\documentclass[11pt]{article}

\let\fn\footnote
\renewcommand{\footnote}[1]{\linespread{1.1}\fn{#1}\linespread{1.29}}

\makeatletter\renewcommand{\section}{\@startsection
{section}{1}{\z@}{-3.5ex plus -1ex minus
    -.2ex}{2.3ex plus .2ex}{\bf }}
\makeatletter\renewcommand{\subsection}{\@startsection{subsection}{2}{\z@}{-3.25ex
plus -1ex minus
   -.2ex}{1.5ex plus .2ex}{\it }}
\makeatletter\renewcommand{\subsubsection}{\@startsection{subsubsection}{3}{-2.45ex}{-3.25ex
plus -1ex minus -.2ex}{1.5ex plus .2ex}{\it }}
\renewcommand{\thesection}{\arabic{section}.}
\renewcommand{\thesubsection}{\arabic{section}.\arabic{subsection}.}

\renewcommand{\theequation}{\thesection\arabic{equation}}
\makeatletter \@addtoreset{equation}{section}

\renewenvironment{thebibliography}[1]
     {\baselineskip=16pt plus 2pt minus 1pt
      \section*{\large\refname
        \@mkboth{\MakeUppercase\refname}{\MakeUppercase\refname}}%
     \list{\@biblabel{\@arabic\c@enumiv}}%
           {\settowidth\labelwidth{\@biblabel{#1}}%
            \leftmargin\labelwidth
            \advance\leftmargin\labelsep
            \@openbib@code
            \usecounter{enumiv}%
            \let\p@enumiv\@empty
            \renewcommand\theenumiv{\@arabic\c@enumiv}}%
      \sloppy
      \clubpenalty4000
      \@clubpenalty \clubpenalty
      \widowpenalty4000%
      \sfcode`\.\@m}

\newcommand{\acknowledgements}{\section*{Acknowledgements}
\addcontentsline{toc}{section}{\hspace{0.6cm}{\bf Acknowledgements}}}

\newcommand{\appendices}{\section*{Appendix}\setcounter{subsection}{0}\setcounter{equation}{0}\renewcommand{\thesubsection}{\Alph{subsection}.}
\renewcommand{\theequation}{\thesubsection\arabic{equation}}
\makeatletter
\@addtoreset{equation}{subsection}
\makeatother
\addtocontents{toc}{\vspace{0.2cm}

{\bf Appendices}}
}

\usepackage{epsfig}
\usepackage{amsmath,amssymb}
\usepackage{amsfonts}
\usepackage{mathrsfs}
\usepackage{bbm}
\usepackage{bm}
\usepackage[vcentermath,enableskew]{youngtab}

\def\slasha#1{\setbox0=\hbox{$#1$}#1\hskip-\wd0\hbox to\wd0{\hss\sl/\/\hss}}

\def\periodb#1{\setbox0=\hbox{$#1$}#1\hskip-\wd0\hbox to\wd0{-}}

\newcommand{\bfa}{\mathbf{a}}				% bold letters
\newcommand{\bfA}{\mathbf{A}}
\newcommand{\bfB}{\mathbf{B}}
\newcommand{\bfb}{\mathbf{b}}
\newcommand{\bfc}{\mathbf{c}}
\newcommand{\bfe}{\mathbf{e}}
\newcommand{\bfeta}{\boldsymbol{\eta}}
\newcommand{\bff}{\mathbf{f}}

\newcommand{\bfphi}{\boldsymbol{\phi}}
\newcommand{\bfG}{\mathbf{G}}
\newcommand{\bfg}{\mathbf{g}}
\newcommand{\bfH}{\mathbf{H}}
\newcommand{\bfL}{\mathbf{L}}
\newcommand{\bflambda}{\boldsymbol{\lambda}}
\newcommand{\bfM}{\mathbf{M}}
\newcommand{\bfN}{\mathbf{N}}

\newcommand{\bfV}{\mathbf{V}}
\newcommand{\bfv}{\mathbf{v}}
\newcommand{\bfw}{\mathbf{w}}
\newcommand{\bfx}{\mathbf{x}}
\newcommand{\unit}{\mathbbm{1}}   			% identity map/matrix
\newcommand{\lsc}{\{\hspace{-0.1cm}[}			% new supercommutators
\newcommand{\rsc}{]\hspace{-0.1cm}\}}
\newcommand{\CA}{\mathcal{A}}    			% cal-letters

\newcommand{\CCH}{\mathscr{H}}
\newcommand{\CI}{\mathcal{I}}

\newcommand{\CL}{\mathcal{L}}

\newcommand{\CN}{\mathcal{N}}
\newcommand{\CO}{\mathcal{O}}

\newcommand{\CP}{\mathcal{P}}
\newcommand{\CCP}{\mathscr{P}}

\newcommand{\FR}{\mathbbm{R}}     			% field of real numbers
\newcommand{\FC}{\mathbbm{C}}     			% field of complex numbers
\newcommand{\NN}{\mathbbm{N}}     			% set of natural numbers
\newcommand{\RZ}{\mathbbm{Z}}     			% ring of integers
\newcommand{\CPP}{{\mathbbm{C}P}}    			% complex projective plane
\newcommand{\PP}{{\mathbbm{P}}}    			% complex projective plane
\newcommand{\dd}{\mathrm{d}}     			% total differential
\newcommand{\dpar}{\partial}     			% partial differential
\newcommand{\diag}{{\mathrm{diag}}}     		% diagonal matrix
\newcommand{\di}{\mathrm{i}}     			% imaginary unit
\newcommand{\eps}{{\varepsilon}}			% antisymmetric tensors
\newcommand{\etab}{{\bar{\eta}}}

\newcommand{\eand}{{~~~\mbox{and}~~~}}     		% and etc. in equations
\newcommand{\ewith}{{~~~\mbox{with}~~~}}
\newcommand{\efor}{{~~~\mbox{for}~~~}}

\newcommand{\der}[1]{\frac{\dpar}{\dpar #1}}   		% partielle ableitung, 1 argument
\newcommand{\tr}{\,\mathrm{tr}\,}     			% trace
\newcommand{\str}{\,\mathrm{str}\,}     		% supertrace
\newcommand{\ad}{\mathrm{ad}}     			% adjoint action
\newcommand{\Ad}{\mathrm{Ad}}     			% adjoint action

\newcommand{\dual}{^\vee}     				% dual space
\newcommand{\au}{\mathfrak{u}}
\newcommand{\asu}{\mathfrak{su}}

\newcommand{\sU}{\mathsf{U}}     			% groups

\newcommand{\sSU}{\mathsf{SU}}
\newcommand{\sSL}{\mathsf{SL}}
\newcommand{\sGL}{\mathsf{GL}}

\newcommand{\sS}{\mathsf{S}}

\newcommand{\remark}[1]{}     				% remark

\newcommand{\contra}{{\diagup\hspace{-0.27cm}\bullet}}     
\newcommand{\co}{{\diagup}}     
\newcommand{\inner}{\mathrm{int}}	% zero component of supernumber
\def\tyng(#1){\hbox{\tiny$\yng(#1)$}}			% small Young diagram

\begin{document}
\begin{titlepage}
\begin{flushright}
  hep-th/0611328\\
  DIAS-STP-06-21
\end{flushright}
\vskip 2.0cm
\begin{center}
{\LARGE \bf Quantization of Flag Manifolds\\[5mm] and their Supersymmetric Extensions}
\vskip 1.5cm
{\Large Se{\'a}n Murray$^{1,2}$ and Christian S{\"a}mann$^1$}
\setcounter{footnote}{0}
\renewcommand{\thefootnote}{\arabic{thefootnote}}
\vskip 1cm
{\em ${}^1$ School of Theoretical Physics\\
Dublin Institute for Advanced Studies\\
10 Burlington Road, Dublin 4, Ireland}\\[5mm]
{\em ${}^2$ Department of Mathematical Physics\\
NUI Maynooth\\
Maynooth, Co. Kildare, Ireland}\\[5mm]
{Email: {\ttfamily smury, csamann@stp.dias.ie}} \vskip
1.1cm
\end{center}
\vskip 1.0cm
\begin{center}
{\bf Abstract}
\end{center}
\begin{quote}
We first review the description of flag manifolds in terms of Pl{\"u}cker coordinates and coherent states. Using this description, we construct fuzzy versions of the algebra of functions on these spaces in both operatorial and star product language. Our main focus is here on flag manifolds appearing in the double fibration underlying the most common twistor correspondences. After extending the Pl{\"u}cker description to certain supersymmetric cases, we also obtain the appropriate deformed algebra of functions on a number of fuzzy flag supermanifolds. In particular, fuzzy versions of Calabi-Yau supermanifolds are found.
\end{quote}
\end{titlepage}

\tableofcontents

\section{Introduction and results}

Quite often in physics, approximation methods like perturbation theory are necessary for explicit computations. In particular, non-perturbative methods, which typically involve the reduction of the field theory to a model with a finite number of degrees of freedom, are required to access the physics of field theories in the strong coupling regime. The standard method of this type is lattice field theory. It has been very successful in the study of confinement in quantum chromodynamics and for non-perturbative regularization of quantum field theories.

Lattice discretizations do have some disadvantages, however. They do not retain the symmetries of the exact theory except in some rough sense. By limiting the couplings to nearest neighbor, the topology and differential geometry of the underlying manifolds are treated only indirectly. Furthermore, the description of fermions in this context leads to the well-known fermion doubling problem.

Fortunately, the lattice is not the only method of reducing a field theory to a finite number of degrees of freedom. An alternative is what has become known as the fuzzy approach \cite{Madore:1991bw,Hoppe:Diss,Grosse:1995ar,Baez:1998he,Balachandran:2000wp,Balachandran:2001dd,Ydri:2001pv,Dolan:2001mi,Vaidya:2001bt,Ramgoolam:2001zx}, see \cite{Balachandran:2005ew} for a detailed review. The basic idea is here to take a classical phase space of finite volume, quantize it and thus obtain a space carrying a function algebra with a finite number of degrees of freedom. There are certain limitations to this approach, such as the even dimensionality of the parent manifold, that can be avoided when the phase space is a co-adjoint orbit of a Lie group. The functions on the resultant fuzzy spaces are described by linear operators on irreducible representations of the group. The simplest such example is the two sphere $S^2$, with the resulting phase space known as the fuzzy sphere \cite{Madore:1991bw}.  Field theory models on the fuzzy sphere then possess only a finite number of modes. The simplest such field theory with $\phi^4$ interaction was proposed in \cite{Grosse:1995ar}.

There are other reasons to consider fuzzy spaces. They lead to matrix models, which have seen much interest by string theorists especially in describing D-branes: When considering D-branes on group manifolds \cite{Klimcik:1996hp}, turning on background fields can render the target space geometry fuzzy \cite{Alekseev:1999bs}. Similarly, a system of D0-branes in a nontrivial background can form the fuzzy sphere \cite{Myers:1999ps}; see also \cite{Trivedi:2000mq}. 

The spaces we chose for deformation play a prominent r{\^o}le in various geometrical areas. Flag manifolds, i.e.\ the spaces of sequences of nested subvector spaces in a given vector space, are generalizations of Gra{\ss}mannians (and thus of complex projective spaces) and serve as non-trivial examples in algebraic geometry. They are special cases of coset spaces, and in particular coset superspaces received growing attention recently \cite{Kleppe:2006ys,Gotz:2006qp}. Moreover, flag manifolds arise naturally in the theory of characteristic classes of vector bundles, in representation theory, in mirror symmetry and in twistor theory. It is therefore clear that studying fuzzy versions of flag manifolds may lead to a deeper understanding of both differential and algebraic geometry on fuzzy spaces. Although the results presented in this paper generalize to arbitrary flag manifolds, we will restrict our attention to those which appear naturally in the double fibrations of twistor theory described e.g.\ in \cite{Howe:1995md}.

We start our discussion by giving a detailed description of flag manifolds in terms of Pl{\"u}cker coordinates and the geometric structures on these spaces. The latter is induced from a canonical embedding of the flag manifolds into Euclidean space. We continue with the description of the correspondence between flag manifolds and coherent states in various representations of the Lie group $\sSU(n)$. In particular, a relationship between the patches covering a flag manifold and dominant weight states in the corresponding representation is established. 

With the appropriate representations found in the coherent state picture together with the Pl{\"u}cker description, the discussion of fuzzy flag manifolds is rather straightforward. We present the matrix algebras corresponding to the algebra of functions on these spaces together with the equivalent star product picture. The latter is used to translate derivatives, which contain information about the geometry of the flag manifolds, into the operator language. In particular, the Laplacian turns into the second order Casimir operator in the considered representation. It is also shown that the constructed matrix algebras converge towards the algebra of continuous functions on flag manifolds in the limit of infinite-dimensional representations.

To prepare the fuzzification of flag supermanifolds, we develop the superanalogue to the Pl{\"u}cker embedding, which is novel, as far as we know. Also, the embedding of these flag supermanifolds into Euclidean superspaces is discussed. A relation between supercoherent states and points on flag supermanifolds is found, which is closely related to the corresponding picture in the case of ordinary flag manifolds. The fuzzification can then be obtained in a rather straightforward way. We give a series of matrix algebras, which approximate functions on the flag supermanifolds and present the equivalent star product formulation. All derivatives can again be translated into the operator language and encode geometric information about the spaces.

The results we obtain may find several applications. First, it is desirable to see whether the Penrose-Ward transform (see e.g.\ \cite{Popov:2004rb} for a review) can be carried over to an analogous construction built on a double fibration of fuzzy spaces. This, however, would demand a clearer understanding of the various gauge theories (i.e.\ holomorphic Chern-Simons and Yang-Mills theory) on the involved fuzzy geometries together with an explicit notion of holomorphic vector bundles over fuzzy spaces, see \cite{Dolan:2006tx} for progress in this direction. Second, the Gra{\ss}mannian $G_{2;4}$ is the conformal compactification of complex Minkowski space and after imposing reality conditions, one arrives at the compactified form of four-dimensional space-times with all possible signatures. Fuzzy versions of these spaces would certainly be very useful; unfortunately, it is not clear, how to impose the corresponding reality conditions in the fuzzy case. The main purpose of constructing fuzzy flag manifolds and in particular their supersymmetric counterparts is, however, to have at hand fuzzy versions of Calabi-Yau supermanifolds. These spaces, as e.g.\ the fuzzy version of the complex projective superspace $\CPP^{3|4}$ discussed in this paper, might be used for the construction of first examples of interacting supersymmetric field theories on fuzzy spaces that can be simulated numerically. Furthermore, there is a conjectured mirror symmetry \cite{Aganagic:2004yh} between two of the flag supermanifolds we describe in this paper ($\CPP^{3|4}$ and $F_{(1|0)(3|3);4|3}$), and trying to understand this mirror symmetry in terms of fuzzy spaces seems very promising.

\section{Pl{\"u}cker coordinates and the geometry of flag manifolds}

\subsection{Flag manifolds of $\sU(4)$}

Consider the vector space $\FC^n$. A {\em flag} $f_{k_1\ldots k_r;n}$ in $\FC^n$ is a sequence of nested vector subspaces $V_{k_1}\subsetneq\ldots \subsetneq V_{k_r}\subset\FC^n$ such that $\dim_\FC V_j=j$. A {\em flag manifold} $F_{k_1\ldots k_r;n}$ is the set of all flags $f_{k_1\ldots k_r;n}$.

The simplest example of a flag manifold is $F_{1;n}$, which is the complex projective space $\CPP^{n-1}$. Furthermore, the Gra{\ss}mannian $G_{k;n}$, the space of $k$-dimensional vector subspaces of $\FC^n$, is the flag manifold $F_{k;n}$. A flag $f_{k;n}=V_k$ is obviously invariant under the subgroup $H=\sU(n-k)\times \sU(k)\subset \sU(n)$, as the elements of $\sU(n-k)$ do not change vectors in $V_k$, while the elements of $\sU(k)$ are just the unitary maps $V_k\rightarrow V_k$. Therefore, the group $H$ defines (maximal) equivalence classes of flags in $\sU(n)$ and we can write $F_{k;n}=\sU(n)/H$. This can be generalized to
\begin{equation}\label{defFlag}
\begin{aligned}
F_{k_1\ldots k_r;n}&\ =\
\sU(n)/(\sU(n-k_r)\times\sU(k_r-k_{r-1})\ldots \times
\sU(k_1))\\&\ =\
\sSU(n)/\sS(\sU(n-k_r)\times\sU(k_r-k_{r-1})\ldots \times
\sU(k_1))~,
\end{aligned}
\end{equation}
and thus the dimension of this flag manifold is $n^2-(n-k_r)^2-(k_r-k_{r-1})^2-\ldots -(k_1)^2$. Note that the above equation cannot be used as a defining relation, as the embedding of the subgroup factored out is not specified. The flag manifolds of $\sSU(4)$ can also be obtained as coset spaces of $\sSL(4,\FC)$, the complexification of $\sSU(4)$, see e.g.\ \cite{Picken:1988fw}. Here, one factors out the group of certain upper block triangular matrices and from this complexified description it follows that flag manifolds are complex manifolds. They are in fact K{\"a}hler manifolds and we will construct their K{\"a}hler structure explicitly later on. We will also see that flag manifolds are adjoint orbits $\{g\CP g^{-1}|g\in \sSU(4)\}$ of certain projectors $\CP$ and therefore carry a natural symplectic structure. Furthermore, a flag manifold is a {\em homogeneous space}.

The flag manifolds of $\sU(n)$ split naturally into {\em irreducible} and {\em reducible} ones, where the irreducible flag manifolds are the Gra{\ss}mannians $F_{k_1;n}=G_{k_1;n}$. In their case, the compact subgroup $H$ consists of two factors. These flag manifolds form {\em hermitian symmetric spaces}, i.e.\ the commutators of two elements of  $\au(n)/(\au(n-k_1)\times\au(k_1))$ is an element of $\au(n-k_1)\times\au(k_1)$.

In the following, we will be exclusively interested\footnote{Nevertheless, all of our discussion trivially translates into the case of flag manifolds of $\sU(n)$.} in flag manifolds of $\sU(4)$, which naturally appear in the double fibrations underlying the most important twistor correspondences, see e.g.\ \cite{Howe:1995md}. These fibrations are obtained by truncating the flags in an obvious manner, e.g.\ there is a projection $F_{12;4}\rightarrow F_{2;4}$. All the twistor double fibrations are included in the following diagram:
\begin{equation}\label{dblfibrationfourself}
\begin{aligned}
\begin{picture}(90,90)
\put(40.0,0.0){\makebox(0,0)[c]{$F_{1;4}$}}
\put(40.0,40.0){\makebox(0,0)[c]{$F_{123;4}$}}
\put(40.0,80.0){\makebox(0,0)[c]{$F_{23;4}$}}
\put(0.0,60.0){\makebox(0,0)[c]{$F_{2;4}$}}
\put(0.0,20.0){\makebox(0,0)[c]{$F_{12;4}$}}
\put(80.0,60.0){\makebox(0,0)[c]{$F_{3;4}$}}
\put(80,20.0){\makebox(0,0)[c]{$F_{13;4}$}}
\put(40.0,50.0){\vector(0,1){20}}
\put(40.0,30.0){\vector(0,-1){20}}
\put(50.0,48.0){\vector(2,1){16}}
\put(30.0,48.0){\vector(-2,1){16}}
\put(50.0,32.0){\vector(2,-1){16}}
\put(30.0,32.0){\vector(-2,-1){16}}
\put(0.0,30.0){\vector(0,1){20}}
\put(80.0,30.0){\vector(0,1){20}}
\put(50.0,48.0){\vector(2,1){16}}
\put(30.0,48.0){\vector(-2,1){16}}
\put(50.0,32.0){\vector(2,-1){16}}
\put(25.0,75.0){\vector(-2,-1){16}}
\put(55.0,75.0){\vector(2,-1){16}}
\put(70.0,10.0){\vector(-2,-1){16}}
\put(10.0,10.0){\vector(2,-1){16}}
\end{picture}
\end{aligned}
\end{equation}
In the twistor context, the space $F_{2;4}$ is the conformal compactification of complexified Minkowski space $\hat{M}$ and the spaces $F_{1;4}$, $F_{3;4}$, $F_{13;4}$ are the spaces of self-dual null planes in $\hat{M}$ (twistor space), anti-self-dual null planes in $\hat{M}$ (dual twistor space) and null geodesics in $\hat{M}$ (a thickening of which is the ambitwistor space), respectively. One can also consider affine (non-compact) subspaces of all the above spaces and the corresponding double fibrations. For more details on this point, see \cite{Popov:2004rb}.

The dimensions of the involved spaces are easily calculated from the formula given below the defining equation \eqref{defFlag}. The minimal number of patches in a covering of the flag manifolds can be calculated inductively in the following way. The minimal number of patches covering all of $G_{k;n}$ is $\binom{n}{k}$; in particular, we have $n$ as the minimal number of patches for $\CPP^{n-1}$. The number of patches needed for a flag manifold is then obtained by multiplying the patches of the contained subflags. For example, to cover $F_{12;4}$, one needs at least $6$ for $F_{2;4}$ times $2$ for $F_{1;2}$ equals $12$ patches. We summarize the results of these calculations in the following table:
\begin{equation*}
\begin{aligned}
\begin{tabular}[t]{|l|c|c|c|c|c|c|c|}
\hline
Flag manifold & $F_{1;4}$ & $F_{2;4}$ & $F_{3;4}$ & $F_{12;4}$ & $F_{13;4}$ & $F_{23;4}$ & $F_{123;4}$ \\
\hline
complex dimension & $3$ & $4$ & $3$ & $5$ & $5$ & $5$ & $6$\\
minimal $\#$ patches & $4$ & $6$ & $4$ & $12$ & $12$ & $12$ & $24$\\
\hline
\end{tabular}
\end{aligned}
\end{equation*}

\subsection{Description of $\CPP^3$}

There are various aspects of the classical description of flag manifolds that we will use for their fuzzification. In particular, we need a description in terms of homogeneous coordinates, a description in terms of projectors and the link between both of them. We will first discuss the simple example of $F_{1;4}=\CPP^3=\sU(4)/(\sU(3)\times\sU(1))$ in detail before going over to the more complicated spaces.

A normalized vector in $\FC^4$ clearly spans a one-dimensional vector subspace of $\FC^4$ and thus corresponds to a flag $f_{1;4}$. There is, however, a redundancy in the total phase of the vector, which needs to be factored out. One is thus naturally led to consider the generalized Hopf fibration defined by the short
exact sequence
\begin{equation}
1\ \longrightarrow\  \sU(1)\ \longrightarrow\  S^{2n-1}\
\longrightarrow\ \CPP^{n-1}\ \longrightarrow\  1
\end{equation}
for the case $n=4$. In coordinates $a^i$ on $\FC^4$, the projection down to $S^7$ amounts to imposing the condition $a^i\bar{a}^j\delta_{ij}=1$ and the subsequent projection down to $\CPP^3$ is performed by considering the auxiliary coordinates
\begin{equation}\label{auxcoords}
x^a_{1;4}\ :=\ \remark{\frac{1}{\bar{a}^k
a^l\delta_{kl}}}\bar{a}^i\lambda^a_{ij}a^j~,
\end{equation}
where $\lambda^a_{ij}$, $a=1,\ldots ,15$ are the Gell-Mann matrices\footnote{We shall adopt the following convention throughout:
\begin{align} 
\tr(\lambda^a \lambda^b )\ =\ \delta^{ab}~,~~~[\lambda^a,\,\lambda^b]\ =\  \sqrt{2}\di f^{ab}{}_{c}\lambda^c ~.\nonumber
\end{align}} of $\sSU(4)$. These coordinates describe an embedding of $\CPP^3$ in $\FR^{15}$. Note that we factored out only the $\sU(1)$ (internal) part from the invariance group of the flag $f_{1;4}$ acting non-trivially on the one-dimensional subspace of $\FC^4$ spanned by the vector $a$. The remaining $\sU(3)$ (external) part acts orthogonally to this vector and therefore leaves it invariant. In an equivalent construction \cite{Dolan:2006tx}, the action of this external group appears more explicitly.

The homogeneous coordinates $a^i$ are a special case of the so-called Pl{\"u}cker coordinates, which we will discuss in the next section. Before, however, let us give a second description of $\CPP^3$ in terms of projectors, see e.g.\ \cite{Balachandran:2001dd}.

A projector $\CP$ is a hermitian $4\times 4$ matrix satisfying $\CP^2=\CP$. The {\em rank} of the projector $\CP$, $\tr(\CP)$, is equal to the dimension of the subspace it projects onto. It is therefore evident that every point on an irreducible flag manifold $F_{k;4}$ corresponds to a rank $k$ projector $\CP_{k;4}(x)$; in particular, $\CPP^3$ is isomorphic to the space of rank one projectors $\CP_{1;4}(x)$. 

The space of projectors acting on $\FC^4$ is spanned by the Gell-Mann matrices of $\sSU(4)$ and the identity. We can write
\begin{equation}
\CP\ =\ x^{\hat{a}}\lambda_{\hat{a}}\ =\  x^0\lambda_0+ x^a
\lambda_a~,
\end{equation}
where $\hat{a}=0,\ldots ,15$ and $a=1,\ldots ,15$. We use $\lambda_0=\unit/\sqrt{4}$, which implies that $x^0=\tr(\CP)/\sqrt{4}$ and
\begin{equation}
\lambda_a \lambda_b\ =\
\frac{\delta_{ab}}{\sqrt{4}}\lambda_0+\frac{1}{\sqrt{2}}(d_{ab}{}^c+\di
f_{ab}{}^c)\lambda_c~,
\end{equation}
where $d_{ab}{}^c$ and $f_{ab}{}^c$ are the (traceless) symmetric invariant tensor and the structure constants of $\sSU(4)$, respectively. Recall that the Lie algebra indices are raised and lowered with the Killing metric $\delta_{ab}$.

As stated above, the irreducible flag manifolds $F_{k;4}$ correspond to the space of projector $\CP_{k;4}$ of rank $k$ and the condition $(\CP_{k;4})^2=\CP_{k;4}$ defines a set of quadratic constraints, embedding the flag manifolds in $\FR^{16}$ (or $\FR^{15}$, if one considers $x^0_{k;4}$  already fixed by the condition on the trace of $\CP_{k;4}$). Explicitly, they read as
\begin{equation}\label{singleprojcond}
 x^a_{k;4} x^a_{k;4}\ =\ \frac{4k-k^2}{4}\eand
 x^a_{k;4} x^b_{k;4}d_{ab}{}^c\ =\ \sqrt{2}\frac{4-2k}{4} x^c_{k;4}~.
\end{equation}

Given a projector $\CP^0_{k;4}$ of rank $k$, all of the space $F_{k;4}$ is obtained by its orbit $g\CP^0_{k;4}g^{-1}$, $g\in\sU(4)$. However, two elements $g$ and $g'$ related by $g=g'h$, where $h\in H=\sU(4-k)\times\sU(k)$, will rotate to the same element $g\CP^0_{k;4}g^{-1}=g'\CP^0_{k;4}g'{}^{-1}$. This simply reflects the definition \eqref{defFlag} of $F_{k;4}$ as a coset space.

There is evidently a relation between $F_{1;4}$ and $F_{3;4}$ since the coordinates $x^a_{1;4}$ of a projector $\CP_{1;4}$ yield the coordinates of a projector $\CP_{3;4}$ by $x^a_{3;4}=-x^a_{1;4}$, as one easily checks using \eqref{singleprojcond}. Furthermore, the coordinates $x^a_{2;4}$ of a projector $\CP_{2;4}$ yield a second projector $\check{\CP}_{2;4}$ with coordinates $\check{x}^a_{2;4}=-x^a_{2;4}$. The meaning of these dualities will become clear in the next section.

For $\CPP^3= F_{1;4}$, the projector is obtained by extending the definition \eqref{auxcoords} of the auxiliary coordinates to $x^{\hat{a}}_{1;4}=\bar{a}^i\lambda^{\hat{a}}_{ij}a^j$, which satisfy the constraints \eqref{singleprojcond}. Due to $\lambda^{\hat{a}}_{ij}\lambda^{\hat{a}}_{kl}=\delta_{il}\delta_{kj}$, the resulting projector is then explicitly given by the matrix $\CP_{1;4}=a\bar{a}^T$ and one easily verifies $(\CP_{1;4})^2=\CP_{1;4}$.

\subsection{Pl{\"u}cker coordinates and projectors describing irreducible flag manifolds}\label{secGrassmann}

To define a two-plane in $\FC^4$, we can use two normalized vectors $a,b\in\FC^4$ antisymmetrized to $A_2:=a\wedge b=\frac{1}{2}\left( a\otimes b -b\otimes a \right) =(A^{ij}_2):=a^{[i}b^{j]}$. As one easily observes, the antisymmetrization projects on the mutually orthogonal components of $a$ and $b$. The $A^{ij}_2$ are so-called {\em Pl{\"u}cker coordinates} on $F_{2;4}=\sU(4)/(\sU(2)\times\sU(2))$ and satisfy by construction the identity 
\begin{equation}\label{plucker1}
\eps_{ijkl}A^{ij}_2A^{kl}_2\ =\ 0~.
\end{equation}
As there are six projective Pl{\"u}cker coordinates, we learn that the Gra{\ss}mannian $F_{2;4}$ is a quadric in $\CPP^5$, the so-called {\em Klein quadric}, see e.g.\ \cite{Mason:1991rf}. Equation \eqref{plucker1} is an example of the {\em Pl{\"u}cker relations}, which describe an embedding of a Gra{\ss}mannian $G_{k;n}$ in $\PP(\Lambda^k\FC^n)$. Although the Pl{\"u}cker relation is straightforward in the present case of $F_{2;4}=G_{2;4}$, we will need more nontrivial such relations when discussing flag supermanifold and we will present a more explicit discussion in section \ref{secPlucker} There are certainly other approaches to coordinatizing flag manifolds; see e.g.\ \cite{Picken:1988fw} for ``Bruhat coordinates'' and more background material on flag manifolds.

Let us now consider the space of hyperplanes in $\FC^4$, i.e.\ $F_{3;4}=\sU(4)/(\sU(1)\times\sU(3))$. Analogous to the case of the two-plane, a three-plane is spanned by three antisymmetrized vectors $a\wedge b\wedge c$, which are naturally dual to a single vector $\check{d}=(\check{d}_i)=(\eps_{ijkl}a^j b^k c^l)$, which in turn spans the orthogonal complement to the hyperplane. However, the non-dualized picture will be useful later on and therefore let us also introduce the Pl{\"u}cker coordinates $A^{ijk}_3=a^{[i}b^jc^{k]}$.

We can contract these new Pl{\"u}cker coordinates with tensor products of the Gell-Mann matrices, which yields auxiliary coordinates describing an embedding of the Gra{\ss}mannians in Euclidean space. In the case of $F_{2;4}$, we have 
\begin{equation}\label{contractionG2}
x^{\hat{a}\hat{b}}_{2;4}\ =\
\bar{A}^{i_1i_2}_2(\lambda^{\hat{a}}\wedge\lambda^{\hat{b}})_{i_1i_2,j_1j_2}A^{j_1j_2}_2
\end{equation}
with the antisymmetrized tensor product $\wedge$ defined in components as
\begin{equation}\label{defantisym2}
(A\wedge B)_{ij;kl}\ =\
\tfrac{1}{4}\left(A_{ik} B_{jl}-A_{jk} B_{il}-A_{il} B_{jk}+A_{jl} B_{ik}\right)~.
\end{equation}
The choice of this contraction, which again factors out a phase, will become obvious after discussing the description of $F_{2;4}$ in terms of projectors. Note that $x^{\hat{a}\hat{b}}_{2;4}$ is {\em symmetric} in its indices. The above contraction is in agreement with the generalized Hopf fibration\footnote{After imposing a certain reality condition, this fibration reduces naturally to 
\begin{equation*}
1\ \rightarrow\  S^1\times S^1\ \rightarrow\  S^3\times S^3\ \rightarrow\  S^2\times S^2\ \cong\  G^\FR_{2;4}\ \rightarrow\  1~.
\end{equation*}}
\begin{equation}
1\ \longrightarrow\  \sU(2)\ \longrightarrow\  S^7\times S^5\
\longrightarrow\  F_{2;4}\ =\ G_{2;4}\ \longrightarrow\  1~.
\end{equation}
As before, a normalized complex vector in $\FC^4$ defines a point on $S^7$, and we choose $a$ to be this point. In the combination $A_2=a\wedge b$, the component of $b$ parallel to $a$ vanishes trivially, and thus the relevant component of $b$ is a point on $S^5$. Factoring out the internal $\sU(2)$ which describes rotations in the plane $a\wedge b$, one obtains $G_{2;4}$. The other $\sU(2)$ factor is again trivially factored out, since it does not affect $A^{ij}_2$. To see that the contraction \eqref{contractionG2} indeed factors out an $\sU(2)$, note that the action
\begin{equation}
\left(\begin{array}{cccc} a^1 & \ldots  & a^4\\ b^1 & \ldots  & b^4
\end{array}\right)\ \mapsto\  g
\left(\begin{array}{cccc} a^1 & \ldots  & a^4\\ b^1 & \ldots  & b^4
\end{array}\right)\ewith g\in \sU(2)~,
\end{equation}
leaves invariant both $\bar{A}^{ij}_2$ and $A_2^{kl}$ up to a phase and therefore \eqref{contractionG2} is indeed invariant. Correspondingly, one can discuss the isotropy groups for all the flag manifolds we construct in the following. We refrain from doing this, but present a more detailed discussion in the quantized picture.

In the case of $F_{3;4}$, we choose the auxiliary coordinates 
\begin{equation}
x_{3;4}^{\hat{a}\hat{b}\hat{c}}\ =\ \bar{A}^{i_1i_2i_3}_3(\lambda^{\hat{a}}\wedge
\lambda^{\hat{b}}\wedge \lambda^{\hat{c}})_{i_1i_2i_3,j_1j_2j_3}A^{j_1j_2j_3}_3~.
\end{equation}
In the dual picture, this corresponds to
\begin{equation}
\check{x}_{3;4}^{\hat{a}}\ =\
\bar{\check{d}}^k\check{\lambda}_{kl}^{\hat{a}}\check{d}^l \ewith
\check{\lambda}^{\hat{a}}_{kl}\sim
\eps_{ki_1i_2i_3}\eps_{lj_1j_2j_3}(\lambda^{\hat{b}}\wedge \lambda^{\hat{c}}\wedge
\lambda^{\hat{d}})_{i_1i_2i_3,j_1j_2j_3}~,
\end{equation}
and the implied map of the Lie algebra indices $(\hat{b}\hat{c}\hat{d})\mapsto \hat{a}$ can easily be calculated. This contraction corresponds to the generalized Hopf fibration
\begin{equation}
1\ \longrightarrow\  \sU(3)\ \longrightarrow\  S^7\times S^5\times
S^3\ \longrightarrow\  F_{3;4}\ \longrightarrow\ 1~.
\end{equation}

Although we already gave a description of the Gra{\ss}mannians $F_{k;4}$ in terms of projectors in the previous section, it will be more convenient to switch to certain rank 1 projectors $\CCP_{k;4}$ acting on the representation spaces of the $\mathbf{6}$ and $\bar{\mathbf{4}}$ of\footnote{Recall that these representations carry two and three antisymmetrized indices of the fundamental of $\au(4)$, respectively.} $\au(4)$ in the cases $F_{2;4}$ and $F_{3;4}$, respectively. This can be done in three equivalent ways. In the first one, one chooses two or three rank one projectors and antisymmetrizes them
\begin{equation}
\CCP_{2;4}\ =\ 2\CP^{1}\wedge \CP^{2}\eand \CCP_{3;4}\ =\
3\CP^{1}\wedge \CP^2\wedge\CP^{3}~,
\end{equation}
where $\CP^r=x^{\hat{a}}_r \lambda_{\hat{a}}$ are some rank one projectors. Besides the usual conditions \eqref{singleprojcond} on rank one projectors, additional conditions between the coordinate vectors $x_r$ and $x_s$ arise to guarantee that $\CCP_{2;4}$ and $\CCP_{3;4}$ are projectors. These conditions state, e.g.\ for $\CCP_{2;4}$ that $\CP^1+\CP^2$ is again a projector, which amounts to $\CP^1\CP^2+\CP^2\CP^1=0$. In terms of coordinates, the first projector is constructed from a complex vector $a^i$ by $x_{1;4}^a=\bar{a}^i\lambda^a_{ij}a^j$, while the second one is constructed from an orthonormalized vector $b_\perp^i$ with $b_\perp^i\sim b^i-(\bar{a}^jb^j)a^i$ by $x_{2;4}^a=\bar{b}^i_\perp\lambda^a_{ij}b^j_\perp$. The sum of these two rank one projectors will automatically yield a rank two projector.

Alternatively, we can also antisymmetrize our previous rank two and rank three projectors $\CP_{2;4}$ and $\CP_{3;4}$
\begin{equation}
\CCP'_{2;4}\ =\ \CP_{2;4}\wedge \CP_{2;4}\eand \CCP'_{3;4}\ =\
\CP_{3;4}\wedge \CP_{3;4}\wedge\CP_{3;4}~,
\end{equation}
as discussed in \cite{Dolan:2001mi}. Note that both approaches are equivalent, the latter, however, is slightly more economical in the use of parameters. Furthermore, due to the formula
\begin{equation}
\tr(A\wedge B)\ =\ \tfrac{1}{2}(\tr(A)\tr(B)-\tr(AB))~,
\end{equation}
which is easily verified using \eqref{defantisym2}, and a similar one for the antisymmetrization of three projectors, all of the above projectors have unit trace and therefore indeed rank 1.

Here, we choose to work in the first approach, embedding the Gra{\ss}mannians in the space of symmetrized products of vectors in $\FR^{16}$. This will eventually lead to simpler expressions for the star product on all the Gra{\ss}mannians. It is also linked to the contractions we obtained from the various Hopf fibrations in the previous section. Let us first introduce the shorthand notation
\begin{equation}\label{shorthandLambda}
\lambda^{\hat{a}_1\ldots \hat{a}_n}\ =\ \lambda^{\hat{a}_1}\wedge\ldots \wedge\lambda^{\hat{a}_n}~.
\end{equation}
Recall that $\lambda^{\hat{a}_1\ldots \hat{a}_n}$ turns out to be totally symmetric in its indices. Using this notation, we can easily define the appropriate projectors in the $\mathbf{6}$ and $\bar{\mathbf{4}}$ of $\au(4)$ as
\begin{equation}
\begin{aligned}
\CCP_{2;4}\ =\ 2x^{(\hat{a}\hat{b})}_{2;4}\lambda^{\hat{a}\hat{b}}&\ewith
x^{(\hat{a}\hat{b})}_{2;4}\ =\ \bar{a}^{i_1}\bar{b}^{i_2}(\lambda^{\hat{a}\hat{b}})_{i_1i_2,j_1j_2}a^{j_1}b^{j_2}~,\\
\CCP_{3;4}\ =\ 3x^{(\hat{a}\hat{b}\hat{c})}_{3;4}\lambda^{\hat{a}\hat{b}\hat{c}}&\ewith
x^{(\hat{a}\hat{b}\hat{c})}_{3;4}\ =\
\bar{a}^{i_1}\bar{b}^{i_2}\bar{c}^{i_3}(\lambda^{\hat{a}\hat{b}\hat{c}})_{i_1i_2i_3,j_1j_2j_3}a^{j_1}b^{j_2}c^{j_3}~.
\end{aligned}
\end{equation}
Here, the subspaces are spanned by complex vectors $a,b$ and $a,b,c$, respectively, and $x^{(\hat{a}\hat{b})}_{2;4}$ and $x^{(\hat{a}\hat{b}\hat{c})}_{3;4}$ describe embeddings of $F_{2;4}$ and $F_{3;4}$ in $\FR^{16\cdot17/2-1}$ and 
$\FR^{16\cdot17\cdot18/(2\cdot3)-1}$, respectively. Note that the coordinates $x^{00}_{2;4}$ and $x^{000}_{3;4}$ are fixed by the ranks of the projectors $\CCP_{2;4}$ and $\CCP_{3;4}$. 

To check that these operators are indeed projectors, one uses identities like
\begin{equation}
(A\wedge B)(C\wedge D)\ =\ \tfrac{1}{2}((AC\wedge BD)+(AD\wedge BC))
\end{equation}
yielding the Fierz identities discussed in appendix B. For example, $\CCP_{2;4}$ can be shown to read as
\begin{equation}
(\CCP_{2;4})_{ij;kl}\ =\ a^{[i}b^{j]}\bar{a}^{[k}\bar{b}^{l]}~,
\end{equation}
where we have chosen $a$ and $b$ orthogonal to each other. It then follows immediately that
\begin{equation}
(\CCP_{2;4})_{ij;kl}(\CCP_{2;4})_{kl;mn}\ =\ (\CCP_{2;4})_{ij;mn}~.
\end{equation}

Note that the na{\"i}ve contraction to obtain the auxiliary coordinates for $F_{2;4}$
\begin{equation}
x^{\hat{a}}_{2;4}\ =\
\bar{A}^{ij}_2(\lambda^{\hat{a}}\wedge\unit)_{ij,kl}A^{kl}_2
\end{equation}
does not yield a projector since $x^{\hat{a}}_{2;4}(\lambda^{\hat{a}}\wedge\unit)_{ij,kl}$ is not idemquadratic.

\subsection{The description of reducible flag manifolds}

The construction of the Pl{\"u}cker coordinates for the reducible flag manifolds is performed in successive steps. For the {\em complete flag manifold}\/\footnote{The manifolds consisting of non-maximal flags are called {\em partial flag manifolds}.} $F_{123;4}=\sU(4)/(\sU(1))^4$, we start from the Pl{\"u}cker coordinates for a line in $\FC^4$, $a^i$, and add a plane containing this line, $A_2^{ij}=a^{[i}b^{j]}$ as well as a hyperplane containing this plane, $A_3^{ijk}=a^{[i}b^jc^{k]}$. We arrive at the set of coordinates
\begin{equation}
a^{[i}b^jc^{k]}~,~~~a^{[i}b^{j]}~,~~~a^i~,
\end{equation}
from which we can construct the auxiliary coordinates
\begin{equation}
\begin{aligned}
x_{123;4}^{\hat{w}_1\ldots \hat{w}_6}\ =\ &\bar{a}^{[i_1}\bar{b}^{i_2}\bar{c}^{i_3]}(\lambda^{\hat{w}_1\hat{w}_2\hat{w}_3})_{i_1i_2i_3,j_1j_2j_3}a^{[j_1}b^{j_2}c^{j_3]}\times \\
&\times \bar{a}^{[i_4}\bar{b}^{i_5]}(\lambda^{\hat{w}_4\hat{w}_5})_{i_4i_5,j_4j_5}a^{[j_4}b^{j_5]}~~
\bar{a}^{i_6}(\lambda^{\hat{w}_6})_{i_6j_6}a^{j_6}~.
\end{aligned}
\end{equation}
The Hopf fibration underlying this contraction reads as 
\begin{equation}
1\ \longrightarrow\  \sU(1)\times\sU(1)\times \sU(1) \ \longrightarrow\  S^7\times S^5\times
S^3\ \longrightarrow\  F_{123;4}\ \longrightarrow\ 1~,
\end{equation}
and the three $\sU(1)$ factors leave invariant the three factors in $x_{123;4}^{\hat{w}_1\ldots \hat{w}_6}$.

On the remaining flag manifolds, the Pl{\"u}cker coordinates are given by subsets of the coordinates for $F_{123;4}$. For example, on $F_{12;4}=\sU(4)/(\sU(2)\times\sU(1)\times\sU(1))$, we have the Pl{\"u}cker coordinates
\begin{equation}
a^{[i}b^{j]}~,~~~a^i
\end{equation}
with obvious auxiliary coordinates. The Hopf fibration reads as 
\begin{subequations}
\begin{equation}
1\ \longrightarrow\  \sU(1)\times \sU(1)\ \longrightarrow\
S^7\times S^5\ \longrightarrow\  F_{12;4}\ \longrightarrow\ 1~.
\end{equation}
This fibration is a reduction of the Hopf fibration for $F_{2;4}$ to the case in which the (internal) isotropy group of the flags is merely $\sU(1)\times \sU(1)$.

The construction of $F_{13;4}=\sU(4)/(\sU(1)\times\sU(2)\times\sU(1))$ and $F_{23;4}=\sU(4)/(\sU(1)\times\sU(1)\times\sU(2))$ follows the same line of argument, and the two Hopf fibrations read as 
\begin{equation}
1\ \longrightarrow\  \sU(1)\times\sU(2) \ \longrightarrow\  S^7\times S^5\times
S^3\ \longrightarrow\  F_{13;4}\ \longrightarrow\ 1~,
\end{equation}
\begin{equation}
1\ \longrightarrow\  \sU(2)\times\sU(1) \ \longrightarrow\  S^7\times S^5\times
S^3\ \longrightarrow\  F_{23;4}\ \longrightarrow\ 1~,
\end{equation}
\end{subequations}
respectively. In particular, the flag manifold $F_{13;4}$ is described by the set of Pl{\"u}cker coordinates $(a^{[i}b^jc^{k]}, a^i)$. In twistor theory, the common description of this space is in terms of a quadric in the space $\CPP^3\times \CPP^3_*$ with coordinates $a^i$ and $a_i^*$. The quadric condition reads as $a^i a_i^*=0$ and with the identification $a_i^*=\eps_{ijkl}a^jb^kc^l$, its relation to the Pl{\"u}cker description becomes clear.

Also the reducible flag manifolds can be mapped to the space of certain tensor products of projectors. For example in the case $F_{12;4}$, we combine $\CCP_{2;4}=\CP^1_{1;4}\wedge \CP^2_{1;4}$ with an additional rank one projector $\CP^3_{1;4}$ given by a linear combination of $\CP^1_{1;4}$ and $\CP^2_{1;4}$:
\begin{equation}
\CP^3_{1;4}\ =\ \alpha\CP^1_{1;4}+\beta\CP^2_{1;4}~,~~~\alpha^2+\beta^2\ =\ 1.
\end{equation}
Thus, $\CP^3_{1;4}$ projects onto a one-dimensional subspace of the plane which $\CCP_{2;4}$ projects onto. The definition of $\CP^3_{1;4}$ implies that the coordinates are linear combinations:
\begin{equation}\label{doubleprojcond}
x^{\hat{a}}_3\ =\ \alpha x^{\hat{a}}_1+\beta x^{\hat{a}}_2~,
\end{equation}
and together with the constraints on the projectors and the antisymmetrization of $\CP^1_{1;4}$ and $\CP^2_{1;4}$ in $\CCP_{2;4}$, this equation describes an embedding of the flag manifold $F_{12;4}$ in Euclidean space.

For the complete flag manifold $F_{123;4}$, we use altogether six rank one projectors, combined in $\CCP^{123}_{3;4}$, $\CCP^{45}_{2;4}$ and $\CP^6_{1;4}$, each satisfying equation \eqref{singleprojcond} and furthermore fulfilling conditions corresponding to \eqref{doubleprojcond}. The coordinates $x^{\hat{a}\hat{b}\hat{c}}_{123}$, $x^{\hat{a}\hat{b}}_{45}$ and $x^{\hat{a}}_6$ form an over-complete set of coordinates on $F_{123;4}$, and the restrictions we impose are an embedding of $F_{123;4}$ in Euclidean space.

All of the reducible flag manifolds can again be described in terms of rank one projectors $\CCP$. Explicitly, these projectors read as:
\begin{equation}
\begin{aligned}
\CCP_{12;4}&\ =\ x^{\hat{w}_1\hat{w}_2\hat{w}_3}\lambda^{\hat{w}_1\hat{w}_2}\otimes \lambda^{\hat{w}_3}~,\\
\CCP_{13;4}&\ =\ x^{\hat{w}_1\ldots \hat{w}_4}\lambda^{\hat{w}_1\ldots \hat{w}_3}\otimes \lambda^{\hat{w}_4}~,\\
\CCP_{23;4}&\ =\ x^{\hat{w}_1\ldots \hat{w}_5}\lambda^{\hat{w}_1\ldots \hat{w}_3}\otimes \lambda^{\hat{w}_4\hat{w}_5}~,\\
\CCP_{123;4}&\ =\ x^{\hat{w}_1\ldots \hat{w}_6}\lambda^{\hat{w}_1\ldots \hat{w}_3}\otimes
\lambda^{\hat{w}_4\hat{w}_5}\otimes \lambda^{\hat{w}_6}~,\\
\end{aligned}
\end{equation}
where the $x^{\hat{w}_1\ldots \hat{w}_k}$ are the auxiliary coordinates constructed from the (independent) Pl{\"u}cker coordinates on the various flag manifolds. Note that in all cases the number of generators in the projector corresponds to the sum of the dimensions of the nested vector spaces in the flag. Even though this description contains a vast redundancy, it turns out to be rather convenient for describing the geometric structures on the flag manifolds inherited from the embedding in Euclidean space, which is the purpose of the next section.

\subsection{Geometric structures on the flag manifolds}

In this section, we will develop expressions for the complex structure, the metric and the symplectic structure on the flag manifolds introduced above, following closely \cite{Balachandran:2001dd}. Given a projector $\CP^0$, which describes a point on the flag manifold $M=F_{i_1\ldots i_k;4}$, all of $M$ is obtained by an appropriate action of $\sU(4)$ on $\CP^0$. That is, the tangent directions are given by infinitesimal actions of $\sU(4)$ and thus the space of tangent vectors is
\begin{equation}
T_{\CP^0} M\ =\ \{R(\Lambda)\CP^0|\Lambda\in \asu(4)\}~.
\end{equation}
Here, we have to distinguish the different representations $R$ of $\Lambda$ for the different projectors $\CP^0$ used for the various flag manifolds. For projectors consisting of $k$-fold antisymmetric combinations of rank 1 projectors, $R(\Lambda)$ is the sum of a $k$-fold tensor product with all entries $\unit$ but one, which is $\ad_\Lambda:=\di [\Lambda,\cdot]$. In particular, we have 
\begin{equation}
\begin{aligned}
&R(\Lambda)\ :=\  \ad_\Lambda~~~\mbox{for }\CP_{1;4}~,\\
&R(\Lambda)\ :=\ \ad_\Lambda\otimes \unit+\unit\otimes \ad_\Lambda~~~\mbox{for }\CCP_{2;4}~,\\
&R(\Lambda)\ :=\ \ad_\Lambda\otimes \unit\otimes \unit+\unit\otimes \ad_\Lambda\otimes \unit+\unit\otimes \unit\otimes \ad_\Lambda~~~\mbox{for }\CCP_{3;4}~.
\end{aligned}
\end{equation}
for the irreducible flag manifolds. The representations in the case of reducible flag manifolds are constructed in an obvious manner, and one has e.g.\
\begin{equation}
\begin{aligned}
&R(\Lambda)\ :=\ \ad_\Lambda\otimes \unit\otimes \ad_\Lambda+\unit\otimes\ad\Lambda\otimes\unit~~~\mbox{for }\CCP_{12;4}~,\\
&R(\Lambda)\ :=\ \ad_\Lambda\otimes \unit\otimes \unit\otimes\ad_\Lambda+\unit\otimes \ad_\Lambda\otimes \unit\otimes\unit+\unit\otimes \unit\otimes \ad_\Lambda\otimes\unit~~~\mbox{for }\CCP_{13;4}~.
\end{aligned}
\end{equation}

If and only if $\Lambda$ is a generator of $H$, $R(\Lambda)\CP^0$ vanishes and therefore $T_{\CP^0} M$ is of the same dimension as $M$. By construction, we have for an element $V\in T_{\CP^0}M$:
\begin{equation}\label{propvectorsCP3}
V^\dagger\ =\ V~,~~~\{\CP^0,V\}\ =\ V~,~~~\tr V\ =\ 0~.
\end{equation}

The orthogonal complement of $T_{\CP^0}M$ in the embedding space is spanned by all other actions of $\sU(4)$ onto $\CP^0$. In particular for $\CPP^3$, the generators $\kappa^a$ of the stabilizer subgroup $H=\sU(3)\times\sU(1)$ of $\CP^0$ span the orthogonal complement of $T_{\CP^0}M$ in the embedding space $\FR^{16}$ as they satisfy by definition $[\CP^0,\kappa^a]=0$ and therefore they are orthogonal to any element of $T_{\CP^0}M$: $\di \tr(\kappa^a[\Lambda,\CP^0])=0$.

To define a complex structure $I$, we start from such a structure on the embedding space, which in turn induces a complex structure on the tangent space at $\CP^0$. Consider the generators $\lambda^{\hat{a}}$ of $\au(4)$. We can pair them into $(\lambda^{2p},\lambda^{2p+1})$, $p=0\ldots7$ and define $I(\lambda^{2p},\lambda^{2p+1})=(-\lambda^{2p+1},\lambda^{2p})$, which amounts to the canonical complex structure on $\FR^{16}$. This translates into a complex structure on any general embedding space and the pairing together with the projection onto $T_{\CP^0}M$ is performed by taking the commutator with $\CP^0$:
\begin{equation}
I V\ =\ -\di [\CP^0,V]\ewith V\in T_{\CP^0}M~.
\end{equation}
One easily checks that $I^2=-1$ and $IW=0$ for $W\in T^\perp_{\CP^0}M$. This definition extends from $T_{\CP^ 0}M$ to the full tangent bundle and yields an almost complex structure on $M$, which turns out to be integrable.

Also the metric is induced from the one on the embedding space, which is the Euclidean (Killing) metric and, after translation into matrices, simply given by the trace. To incorporate the projection onto $T_{\CP^0}M$, we can multiply each vector in $T_{\CP^0}M$ by the complex structure before taking the trace which yields the hermitian metric
\begin{equation}
g(V_1,V_2)\ =\ \tr\left(I V_1 I V_2\right)\ =\
-\tr\left([\CP^0,V_1][\CP^0,V_2]\right)~,
\end{equation}
as for elements $V_1,V_2\in T_{\CP^0}M$, we have $g(V_1,V_2)=g(I
V_1,I V_2)$. The continuation of this metric to all of $M$ is
evident. There is furthermore a symplectic structure defined as
\begin{equation}
\Omega(V_1,V_2)\ :=\ g(IV_1,V_2)~,
\end{equation}
which we can combine as usually with the metric into the K{\"a}hler structure $J$ defined as
\begin{equation}
J(V_1,V_2)\ :=\ \tfrac{1}{2}\left(g(V_1,V_2)+\di
\Omega(V_1,V_2)\right)\ =\ \tr(\CP^0 V_1(\unit-\CP^0)V_2)~,
\end{equation}
which also extends globally.

Note that the projectors we use in the description of flag
manifolds are of rank one and therefore, the above formula
simplifies to
\begin{equation}
J(V_1,V_2)\ =\ \tr(\CP^0 V_1V_2)-\tr(\CP^0V_1)\tr(\CP^0V_2)~.
\end{equation}

On $\CPP^3$, $\Lambda$ is a generator of the fundamental
representation of $\sSU(4)$ and we can introduce the components $\Omega^{\hat{a}\hat{b}}_{1;4}=\Omega(\lambda^{\hat{a}},\lambda^{\hat{b}})$ as well as $J^{\hat{a}\hat{b}}_{1;4}=J(\lambda^{\hat{a}},\lambda^{\hat{b}})$. For these, we have the useful identities
\begin{equation}\label{identOmega}
\Omega^{\hat{a}\hat{b}}_{1;4}\ =\ \sqrt{2} f^{ab}{}_c x^c
\end{equation}
and
\begin{equation}\label{identJK}
\begin{aligned}
J^{\hat{a}\hat{b}}_{1;4} \lambda_{\hat{b}}&\ =\ \tr\left(\CP^0
\lambda^{\hat{a}}(\unit-\CP^0)\lambda^{\hat{b}}\right)\lambda_{\hat{b}}\ =\ \CP^0
\lambda^{\hat{a}}(\unit-\CP^0)~~,\\
\lambda_{\hat{a}}J^{\hat{a}\hat{b}}_{1;4}&\ =\ \lambda_{\hat{a}}\tr\left(\CP^0
\lambda^{\hat{a}}(\unit-\CP^0)\lambda^{\hat{b}}\right)\ =\
(\unit-\CP^0)\lambda^{\hat{b}}\CP^0~.
\end{aligned}
\end{equation}
In deriving these identities, one needs the relation
\begin{equation}\label{identDecomposition}
\tr(\lambda^{\hat{b}}\lambda^{\hat{a}})\lambda_{\hat{a}}\ =\ \lambda^{\hat{b}}\efor
\lambda^{\hat{a}},\lambda^{\hat{b}}\in\au(4)~,
\end{equation}
cf.\ formul\ae{} \eqref{identlambdatrace}. Due to $(\CP^0)^2=\CP^0$, the relations \eqref{identJK} remain valid after omitting the hats over the indices.

Let us briefly comment on the explicit form of the structures obtained in the above discussion for the various flag manifolds. We start with the Gra{\ss}mannians. These spaces are described by rank one projectors in the representation $R$, which is here the previously defined $k$-fold $\wedge$-product of the fundamental one. The tangent directions are given by infinitesimal actions of elements of $\sU(4)$:
\begin{equation}
\begin{aligned}
T_{\CP^0_{2;4}}F_{2;4}&\ =\ \left\{(\ad_\Lambda\otimes
\unit+\unit\otimes\ad_\Lambda)\CP^0_{2;4}|\Lambda\in
\asu(4)\right\}~,\\
T_{\CP^0_{3;4}}F_{3;4}&\ =\ \left\{(\ad_\Lambda\otimes
\unit\otimes\unit+\unit\otimes\ad_\Lambda\otimes\unit+
\unit\otimes\unit\otimes\ad_\Lambda)\CP^0_{3;4}|\Lambda\in
\asu(4)\right\}~.
\end{aligned}
\end{equation}
Note that the most general action e.g.\ on the projector $\CP^0_{2;4}$ is given by $\Ad_{g_1}\otimes \Ad_{g_2}$. Since $\CP^0_{2;4}$ is the sum of antisymmetrized tensor products of the form $\lambda^{\hat{a}}\wedge\lambda^{\hat{b}}$, only the {\em symmetrized form} of $\Ad_{g_1}\otimes \Ad_{g_2}$ is relevant, which is $\frac{1}{2}((\Ad_{g_1}+\Ad_{g_2})\otimes(\Ad_{g_1}+\Ad_{g_2}))$. At infinitesimal level, this yields the action $\ad_\Lambda\otimes \unit+\unit\otimes\ad_\Lambda$.

All the properties \eqref{propvectorsCP3} are easily verified to hold also for the tangent vectors of all the Gra{\ss}mannians. Furthermore, the definitions of the complex structure, the metric and the K{\"a}hler structure is done in a straightforward manner, since the only essential aspect in their definition on $\CPP^3$ was that $\CP^0$ is a projector.

Using again the shorthand notation \eqref{shorthandLambda}, the appropriate components of the symplectic and the K{\"a}hler structure are given by
\begin{equation}
\begin{aligned}
\Omega^{\hat{a}\hat{b},\hat{c}\hat{d}}_{2;4}\ =\ \Omega(\lambda^{\hat{a}\hat{b}},\lambda^{\hat{c}\hat{d}})
&\eand \Omega^{\hat{a}\hat{b}\hat{c},\hat{d}\hat{e}\hat{f}}_{3;4}\ =\ \Omega(\lambda^{\hat{a}\hat{b}\hat{c}},\lambda^{\hat{d}\hat{e}\hat{f}})~,\\
J^{\hat{a}\hat{b},\hat{c}\hat{d}}_{2;4}\ =\ J(\lambda^{\hat{a}\hat{b}},\lambda^{\hat{c}\hat{d}})&\eand
J^{\hat{a}\hat{b}\hat{c},\hat{d}\hat{e}\hat{f}}_{3;4}\ =\ J(\lambda^{\hat{a}\hat{b}\hat{c}},\lambda^{\hat{d}\hat{e}\hat{f}})~.
\end{aligned}
\end{equation}
The corresponding versions of the identities \eqref{identOmega}, \eqref{identJK} read as
\begin{equation}\label{identOmegaStar}
\Omega^{\hat{a}\hat{0},\hat{c}\hat{d}}_{2;4}\ =\ _{(cd)}\sqrt{2}f^{acb}x^{bd}\eand
\Omega^{\hat{a}\hat{0}\hat{0},\hat{d}\hat{e}\hat{f}}_{3;4}\ =\ _{(def)}\sqrt{2}f^{adb}x^{bef}~,
\end{equation}
where  these relations only hold for the components symmetric in $(cd)$ and $(def)$, respectively, and
\begin{equation}
\begin{aligned}
J^{\hat{a}\hat{b},\hat{c}\hat{d}}_{2;4}\lambda^{\hat{c}\hat{d}}&\ =\
\tr(\CP^0\lambda^{\hat{a}\hat{b}}(\unit\otimes\unit-\CP^0)
\lambda^{\hat{c}\hat{d}})\lambda^{\hat{c}\hat{d}}\ =\ \CP^0\lambda^{\hat{a}\hat{b}}(\unit\otimes\unit-\CP^0)~,\\
J^{\hat{a}\hat{b}\hat{c},\hat{d}\hat{e}\hat{f}}_{3;4}\lambda^{\hat{d}\hat{e}\hat{f}}&\ =\
\tr(\CP^0\lambda^{\hat{a}\hat{b}\hat{c}}(\unit\otimes\unit\otimes\unit-\CP^0)
\lambda^{\hat{d}\hat{e}\hat{f}})\lambda^{\hat{d}\hat{e}\hat{f}}\ =\
\CP^0\lambda^{\hat{a}\hat{b}\hat{c}}(\unit\otimes\unit\otimes\unit-\CP^0)~,
\end{aligned}
\end{equation}
which follow from the relations in \eqref{identlambdatrace}. 

To describe the tangent space to the reducible flag manifolds $M=F_{k_1k_2;4}$ at a point $\CP^0_{k_1k_2;4}=\CP^0_{k_2}\otimes\CP^0_{k_1(k_2);4}$, one proceeds completely analogously to above and defines
\begin{equation}
T_{\CP^0_{k_1k_2;4}}M\ :=\ \{(\ad_\Lambda\otimes
\unit\otimes\ldots +\unit\otimes\ad_\Lambda\otimes\ldots +
\ldots \otimes\unit\otimes\ad_\Lambda)(\CP^0_{k_2;4}\otimes
\CP^0_{k_1(k_2);4})|\Lambda\in\asu(4)\}~.
\end{equation}
It immediately follows that the elements of $T_{\CP^0_{k_1k_2;4}}M$ all satisfy \eqref{propvectorsCP3}. The stabilizer subgroup $H$ of $\CP^0_{k_1k_2;4}$ is indeed $\sU(4-k_2)\times \sU(k_2-k_1)\times \sU(k_1)$.

The definition of the complex structure, the metric, the symplectic and the K{\"a}hler structure are again straightforward. For the latter, we introduce components, e.g.\ for $F_{12;4}$:
\begin{equation}
J^{\hat{a}\hat{b},\hat{c}\hat{d};\hat{e}\hat{f}}_{12;4}\ =\ J_{2;4}^{\hat{a}\hat{b},\hat{c}\hat{d}}\otimes J_{1;4}^{\hat{e}\hat{f}}~,
\end{equation}
and one has again obvious identities corresponding to \eqref{identJK}.

\subsection{Spherical functions on the flag manifolds}

Before discussing the fuzzification of functions on the flag manifolds, let us briefly review some aspects of harmonic analysis on these spaces. That is, we want to describe the construction of spherical functions on flag manifolds, which form a complete orthonormal basis on these spaces and are simultaneously eigenfunctions of the Laplace operator. For this, we will extend the standard discussion of spherical functions using the various generalized Hopf fibrations described above. (A detailed discussion for $\CPP^n=F_{1;n}$ is found e.g.\ in \cite{Cucchieri:1998hi}.)

The spherical functions on the sphere $S^m$ are simply the restrictions of the homogeneous harmonic polynomials on $\FR^{m+1}$ to $S^m$ \cite{Gilkey:1998} and the dimension of the eigenspaces $H^j$, $j\in\NN$, corresponding to the eigenvalue $j(j+m-1)$ are $\binom{m+j}{m}$. For $m=2n-1$, the eigenspaces $H^j$ are spanned by homogeneous polynomials in the complex coordinates $a^i$ on $\FC^n$ plus their complex conjugate minus all terms containing contractions $a^i\bar{a}^i$, as these terms belong to spaces $H^k$ with $k<j$.

The complex projective space $\CPP^3$ is obtained from the generalized Hopf fibration $\sU(1)\rightarrow S^7\rightarrow \CPP^3$. The eigenfunctions of the Laplace operator on this space are the subset of the corresponding eigenfunctions on $S^7$ which are invariant under $\sU(1)$. These functions are obviously the product of a homogeneous polynomial of order $k$ in $a^i$ and another such polynomial in $\bar{a}^j$ minus all the possible contractions. We thus get the ``hyperspherical harmonics''
\begin{equation}
Y^{k}_{1;4}{}^{i_1\ldots i_kj_1\ldots j_k}\ :=\ a^{i_1}\ldots a^{i_k}\bar{a}^{j_1}\ldots \bar{a}^{j_k}-\mbox{contraction terms}~,
\end{equation}
which have eigenvalues $\lambda_k=k(k+3)$ and their eigenspaces $H^k_{1;4}$ have dimensions
\begin{equation}
\dim H^k_{1;4}\ =\ \binom{k+3}{k}^2-\binom{k+2}{k-1}^2~,
\end{equation}
where the last term subtracts the dimensions of the contraction terms.

The Gra{\ss}mannian $G_{2;4}$ is obtained from the Hopf fibration $\sU(2)\rightarrow S^7\times S^5\rightarrow G_{2;4}$. The spherical functions on $S^7\times S^5$ are constructed from homogeneous polynomials in $A^{ij}$ and $\bar{A}^{kl}$, where we add again the complex conjugate polynomial to render the expression real and finally subtract all terms containing contractions. The subspace of $\sU(2)$-invariant functions is now spanned by those polynomials, which have an equal number of $A^{ij}$s and $\bar{A}^{kl}$s, and we thus have:
\begin{equation}
Y^{k}_{2;4}{}^{i_1j_1\ldots i_kj_kl_1m_1\ldots l_km_k}\ :=\ A^{i_1j_1}\ldots A^{i_kj_k}\bar{A}^{l_1m_1}\ldots \bar{A}^{l_km_k}-\mbox{contraction terms}~.
\end{equation}

From the generalized Hopf fibrations discussed in the preceding sections, the construction of the remaining flag manifolds is obvious. For our purposes, it is more important to note that these spherical functions on $G/H$ are in one-to-one correspondence with so-called {\em spherical representations} of $G/H$, i.e.\ (finite dimensional) representations of $G$ with non-trivial $H$-invariant vectors, see e.g. \cite{Helgason:1984,dijkhuizen-1995-}. This allows us to associate each set of eigenfunctions of the Laplace operator on the flag manifolds with certain sets of irreducible representations of $G$, which we will do in the next section. Later on, we will use these representations to construct the algebra of fuzzy functions on the flag manifolds. From this construction, one can also read off the eigenvalues of the eigenspaces of the Laplace operator on the various flag manifolds.

\section{Flag manifolds and coherent states}

To quantize the flag manifolds, we would like to connect every point on these spaces to a state in a Hilbert space. A function then automatically becomes an operator on this space. To establish this connection, recall that every point on a flag manifold which is a coset space of $G=\sU(n)$ is in one-to-one correspondence to a generalized coherent state in a specific representation of $G$. We will review this relation in the next section and partly follow the discussion of Perelomov \cite{Perelomov:1986tf}, see \cite{Chaichian:1990bx,Trivedi:2000mq} for quantization using coherent states.

\subsection{Representations of $\sSU(n)$ and coherent states}

Consider the Dynkin diagram of $\sSU(n)$ for simple roots $\alpha_1 , \ldots ,\alpha_{n-1}$
\begin{equation}
\begin{picture}(180,30)(0,-10)
\put(20.0,0.0){\circle{10}}
\put(25.0,0.0){\line(1,0){40}}
\put(70.0,0.0){\circle{10}}
\put(75.0,0.0){\line(1,0){40}}
\put(125.0,0.0){\makebox(0,0)[c]{$\ldots $}}
\put(135.0,0.0){\line(1,0){40}}
\put(180.0,0.0){\circle{10}}
\put(20.0,12.0){\makebox(0,0)[c]{$a_1$}}
\put(70.0,12.0){\makebox(0,0)[c]{$a_2$}}
\put(180.0,12.0){\makebox(0,0)[c]{$a_{n-1}$}}
\put(20.0,-12.0){\makebox(0,0)[c]{$\alpha_1$}}
\put(70.0,-12.0){\makebox(0,0)[c]{$\alpha_2$}}
\put(180.0,-12.0){\makebox(0,0)[c]{$\alpha_{n-1}$}}
\end{picture}
\end{equation}

An irreducible representation $T^\Lambda$ with highest weight $\Lambda$ can be labeled by the Dynkin indices\footnote{See appendix B for more details.} $a_1,\ldots ,a_{n-1}$. In the representation (Hilbert) space $\mathscr{H}^\Lambda$, there is a corresponding highest weight vector $|\Lambda \rangle$ \remark{, i.e.\ a vector satisfying the conditions
\begin{align} 
E_\alpha |\Lambda \rangle \ =\ 0 \qquad H_j |\Lambda \rangle \ =\  \Lambda_j |\Lambda \rangle~,
\end{align}
where $\alpha \in \Sigma_+$ and $\Sigma_+$ is the set of the $\frac{1}{2}N(N-1)$ positive roots.} and $\mathscr{H}^\Lambda$ has a basis of weight vectors\footnote{which can be chosen to be orthonormal} $\{ |\mu \rangle \}$ i.e. $H_j |\mu\rangle = \mu_j |\mu \rangle$.

The isotropy subgroup $H_\mu$ for any weight vector $|\mu\rangle$ contains the Cartan subgroup $H$ of $\sSU(n)$, which is isomorphic to the maximal torus $T^{n-1}=\sU(1)^{\times n-1}=\sU(1)\times\ldots\times\sU(1)$ and for general weight vectors the subgroup $H_\mu$ coincides with $T^{n-1}$. For degenerate representation, where the highest weight $\Lambda$ is orthogonal to some simple root $\alpha_i$, i.e.
\begin{align} 
(\Lambda, \alpha_i)\ =\ 0\ =\ a_i~,
\end{align}
the isotropy subgroup $H_\mu$  may be larger than $T^{n-1}$ for some weight vectors. This is evident since, as explained in appendix B, the $a_i$ indicate the highest power of $E_{-\alpha}$, whose action is still nontrivial on $|\Lambda\rangle$. 

A helpful picture arises, when enlarging $\sSU(n)$ to $\sU(n)$. The Cartan subalgebra consists now of $n$ factors of $\sU(1)$, which one can imagine sitting around the $a_i$. Every $a_i$ which is zero combines the $\sU(m_1)$ left of it with the $\sU(m_2)$ right to it to a $\sU(m_1+m_2)$. This allows us to construct all the isotropy groups one encounters in flag manifolds, and we will be more explicit in the next section.

To construct a coherent state system, one has to choose an initial vector $|\psi_0\rangle$ in $\mathscr{H}^\Lambda$. Then the system of states $\{ | \psi_g \rangle=T^{\Lambda} (g)|\psi_0 \rangle \}$ is called the coherent state system $\{ T^{\Lambda} ,\, |\psi_0 \rangle \}$. Let $H_0$ be the isotropy subgroup for the state $|\psi_0 \rangle$. Then a coherent state $|\psi_g \rangle$ is determined by a point $x=x(g)$ in the coset space $G/H_0$, corresponding to the element $g$ by $|\psi_g \rangle= \exp(\di \alpha)| x \rangle$ up to a phase, $|\psi_0 \rangle=|0\rangle$. 
The isotropy subgroup for a linear combination of weight vectors is, in general, a subgroup of the Cartan subgroup. Therefore it is convenient to choose a weight vector $|\mu\rangle$ as an initial state.
For non-degenerate representations, the isotropy subgroup $H_\mu$ is isomorphic to the Cartan subgroup $H$, and the coherent state $|x\rangle$ is characterized by a point of $G/H$, or equivalently by a point of the orbit of the adjoint representation
\begin{align} H'_j |x\rangle\ =\ T(g)H_j T^{-1}(g)|x\rangle,\quad  |x\rangle\ =\ T(g)|\mu\rangle~.
\end{align}
In a representation with some Dynkin labels vanishing, the isotropy subgroup $H_\mu$ is larger than $H$ for some weight vectors $|\mu\rangle$, in which case the orbit may be degenerate.

There is considerable choice in the selection of the initial state $|0\rangle$, even on restriction to weight vectors. Perelomov \cite{Perelomov:1979sk} has shown that the state $|0\rangle$ must be $|\mu\rangle$, where $\mu$ is a dominant weight, if it is to be closest to classical. That is, $|\mu\rangle$ is obtained from the highest weight by the Weyl reflection group. Then the coherent states minimize the invariant uncertainty relation
\begin{equation}
 \Delta C_2 \ =\  \textrm{min}~, 
\end{equation}
where
\begin{equation}
C_2 \ =\  \sum_j (H_j)^2+\sum_{\alpha \in \Sigma_+}(E_\alpha E_{-\alpha}+E_{-\alpha} E_\alpha) 
\end{equation}
is the quadratic Casimir operator and
\begin{equation}
\Delta C_2\ =\ \langle C_2 \rangle-\left(\sum_j \langle H_j \rangle^2+2\sum_{\alpha \in \Sigma_+}\langle E_\alpha \rangle \langle E_{-\alpha} \rangle \right)~.
\end{equation}

Let us take the initial state to be the highest weight vector $|\Lambda\rangle$. The coherent state system is then more explicitly defined by 
\begin{align}
|a\rangle &\ =\ N ~T^\Lambda (g)|\Lambda\rangle~, \quad N^{-2}\ =\ \langle \Lambda|T^\Lambda (g)|\Lambda\rangle~,\\
|a\rangle &\ =\ N~ \exp\left( \sum_{\alpha \in \Sigma_+} a^-_\alpha E_{-\alpha}\right)  \exp\left(  h_i H_i\right)  \exp\left( \sum_{\alpha \in \Sigma_+} a^+_\alpha E_{\alpha}\right) |\Lambda\rangle~ \nonumber\\  
&\ =\ N~ \exp\left(\sum_{\alpha \in \Sigma_+} a^-_\alpha E_{-\alpha}\right)|\Lambda\rangle~, 
\end{align}
where we have restricted ourselves to elements of $G$ which have a Gau{\ss}ian decomposition. Note that if some of the Dynkin labels $a_i$ vanish (degenerate representations), then the corresponding coordinates $a^-_\alpha$ (and possible some others, see the appendix) are no longer independent and can be eliminated from the definition of the group element $g$ and thus the coherent states correspond to points on various flag manifolds. The  number of independent coordinates $a^-_\alpha$ gives the (complex) dimension of the flag manifold. 

Note that our construction of coherent states yields merely one patch of the covering of the flag manifolds. Starting from a different dominant weight state corresponds to working on a different patch, as the state $|w(\Lambda)\rangle$, where $w$ is an element of the Weyl group $W$, is not contained in the set of the coherent states $|a\rangle$ constructed from $|\Lambda\rangle$. This is because we have restricted ourselves to group elements that have a Gau{\ss}ian decomposition. We can assume all the weight states to be orthogonal. In particular, two states $|\Lambda\rangle$ and $|w(\Lambda)\rangle$ have no overlap, i.e.\ $\langle\Lambda|w(\Lambda)\rangle=0$. Consider now the coherent state $|a\rangle$ as constructed above,
\begin{equation}
|a\rangle\ =\ N \left(1+\sum_{n=1}^\infty \frac{1}{n!} \left(\sum_{\alpha \in \Sigma_+} a^-_\alpha E_{-\alpha}\right)^{\hspace{-0.1cm}n~}\right)|\Lambda\rangle~.
\end{equation}
As $E_{-\alpha}$ contains only lowering operators, we have $\langle \Lambda|E_{-\alpha}|\Lambda\rangle=0$, which implies
\begin{equation}
\langle\Lambda|a^-\rangle\ =\ \langle\Lambda|N|\Lambda\rangle+0\ =\ N~.
\end{equation}
It is thus clear that all $|a\rangle$ have a component parallel to $|\Lambda\rangle$ and therefore $|a\rangle$ never equals $|w(\Lambda)\rangle$, which we wanted to prove.

The number of such dominant weight states from which we can start and thus the number of patches is evidently given by the dimension of $\{w(\Lambda),\,w\in W\}$ or equivalently the number of corners of the convex hull of the states in the weight diagram of the representation. This number is just the rank of the Weyl group modulo group elements acting trivially in a certain representation. As the Weyl group for $\sSU(n)$ is essentially the permutation group of $n$ elements with rank $n!$, the minimal number of patches covering the complete flag manifold $F_{123\ldots n-1;n}$ is given by $n!$, while the number of patches for all other flag manifolds of $\sSU(n)$ is smaller. The reason for this is simply that for other flag manifolds, certain Dynkin labels $a_i$ are zero, which implies that the corresponding Weyl reflections 
\begin{equation}
|S_{\alpha_i}\Lambda\rangle\ =\ |\Lambda-\frac{2 \left( \alpha_i,\,\Lambda\right) } {\left( \alpha_i,\alpha_i \right) }\alpha_i  \rangle~,
\end{equation}
act trivially on the highest weight state.

To clarify the above construction, we will first discuss the simple case of flags in $\sSU(3)$, whose weight diagrams are two-dimensional, before presenting the construction for the flag manifolds of $\sSU(4)$.

\subsection{Examples for $\sSU(3)$}

There are two flag manifolds which arise as coset spaces of $\sSU(3)$: The complex projective space $\CPP^2=F_{1;3}\cong F_{2;3}=\CPP^2_*\cong \frac{\sSU(3)}{\sU(2)}$ and the reducible flag manifold $F_{12;3}=\frac{\sSU(3)}{\sU(1)\times\sU(1)}$. The representations $\mathbf{3}$, $\bar{\mathbf{3}}$ and $\mathbf{6}$ of $\sSU(3)$, corresponding to the diagrams
\begin{equation}
\begin{picture}(80,20)
\put(20.0,0.0){\circle{10}}
\put(25.0,0.0){\line(1,0){40}}
\put(70.0,0.0){\circle{10}}
\put(20.0,12.0){\makebox(0,0)[c]{$1$}}
\put(70.0,12.0){\makebox(0,0)[c]{$0$}}
\end{picture}
~\ =\ ~\tyng(1)~,~~~
\begin{picture}(80,20)
\put(20.0,0.0){\circle{10}}
\put(25.0,0.0){\line(1,0){40}}
\put(70.0,0.0){\circle{10}}
\put(20.0,12.0){\makebox(0,0)[c]{$0$}}
\put(70.0,12.0){\makebox(0,0)[c]{$1$}}
\end{picture}
~\ =\ ~\tyng(1,1)~,
\begin{picture}(80,20)
\put(20.0,0.0){\circle{10}}
\put(25.0,0.0){\line(1,0){40}}
\put(70.0,0.0){\circle{10}}
\put(20.0,12.0){\makebox(0,0)[c]{$2$}}
\put(70.0,12.0){\makebox(0,0)[c]{$0$}}
\end{picture}
~\ =\ ~\tyng(2)~,
\end{equation}
all have dominant weight states with isotropy group $\sU(2)$. Thus, the coherent states constructed from these representations are in one-to-one correspondence with points on $\CPP^2$. The adjoint representation $\mathbf{8}$ as well as the $\mathbf{27}$ corresponding to the diagrams
\begin{equation}
\begin{picture}(80,20)
\put(20.0,0.0){\circle{10}}
\put(25.0,0.0){\line(1,0){40}}
\put(70.0,0.0){\circle{10}}
\put(20.0,12.0){\makebox(0,0)[c]{$1$}}
\put(70.0,12.0){\makebox(0,0)[c]{$1$}}
\end{picture}
~\ =\ ~\tyng(2,1)~,~~~
\begin{picture}(80,20)
\put(20.0,0.0){\circle{10}}
\put(25.0,0.0){\line(1,0){40}}
\put(70.0,0.0){\circle{10}}
\put(20.0,12.0){\makebox(0,0)[c]{$2$}}
\put(70.0,12.0){\makebox(0,0)[c]{$1$}}
\end{picture}
~\ =\ ~\tyng(3,1)
\end{equation}
have dominant weight states, whose isotropy group is only the maximal torus $\sU(1)\times \sU(1)$ and therefore the derived coherent states are in one-to-one correspondence with points on the flag manifold $F_{12;3}$. Recall that the number of patches is found to be the number of corners in (the convex hulls of) the weight diagrams. Consider e.g.\ the diagrams
\begin{equation*}
  \epsfig{file=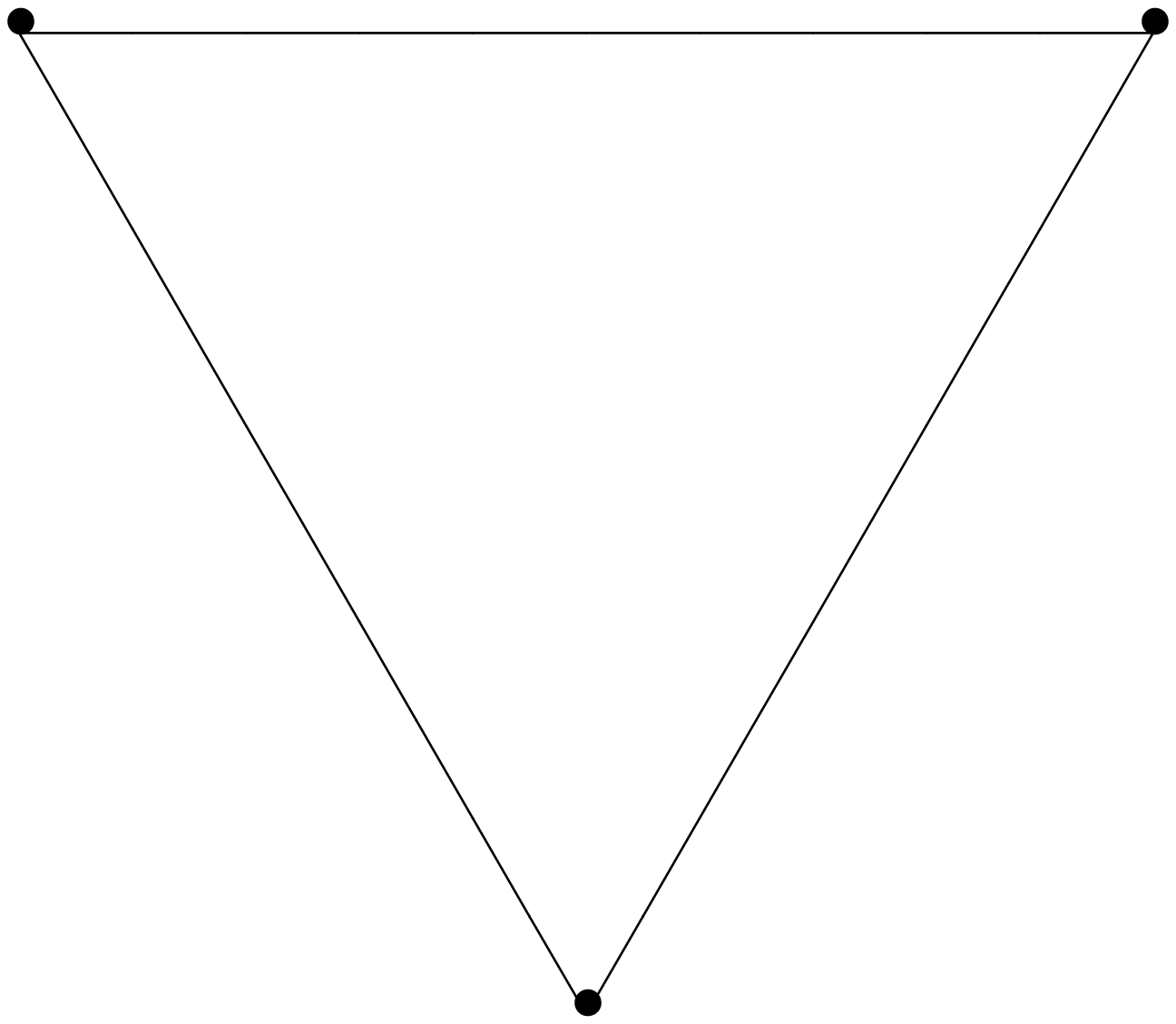,width=3cm,height=3cm}\hspace*{0.8cm}
  \epsfig{file=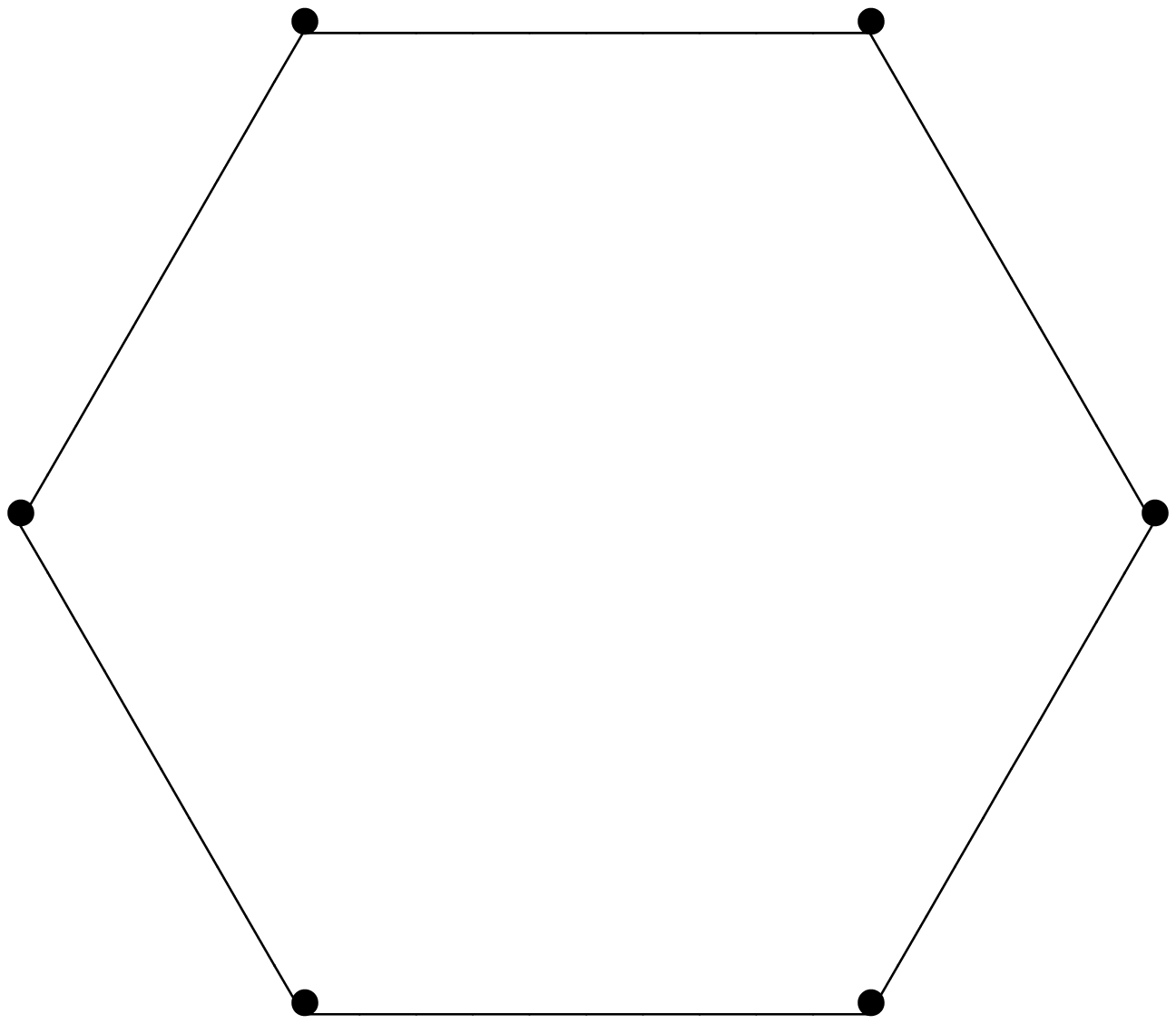,width=3cm,height=3cm}\hspace*{0.8cm}
  \epsfig{file=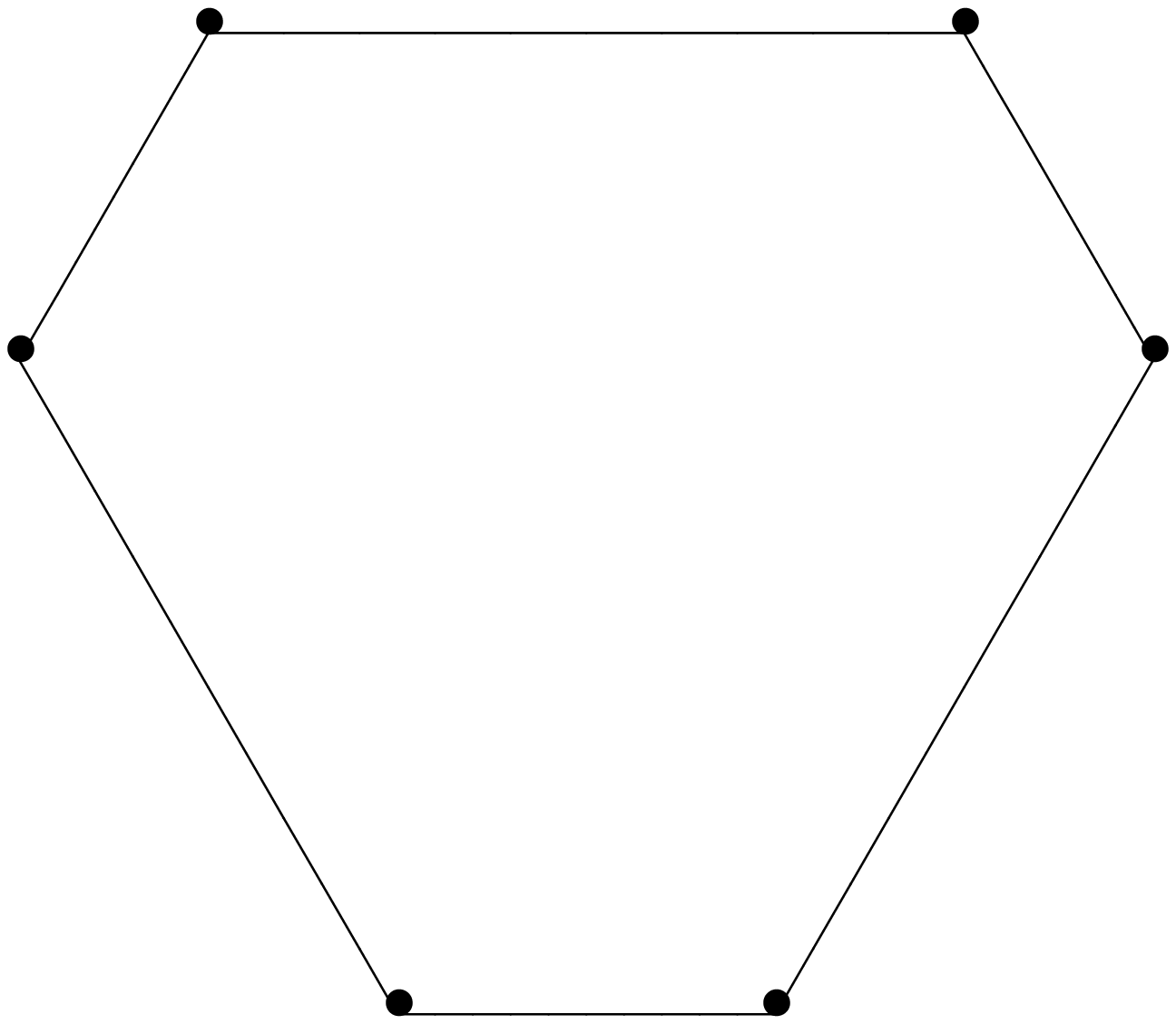,width=3cm,height=3cm}
\end{equation*}
corresponding to the representations with Dynkin labels $(1,0)$, $(1,1)$ and $(2,1)$, respectively. We see that the complex projective space $\CPP^2$ is covered by at least three patches, while the flag manifold $F_{12;3}$ requires a minimum of six patches.

\subsection{The flags in $\FC^4$}

Let us now come to the coherent states which correspond to points on flag manifolds of $\sSU(4)$. For these, the choice of representations as well as the Dynkin diagrams are given in Table 1.

Note that the representations for the reducible flag manifolds are not unique at level $L$. One can choose any representation $(a_1(L),a_2(L),a_3(L))$; however, for considering the limit $L\rightarrow \infty$, the functions $a_i$ should be polynomials of the same order in $L$. 

The minimal numbers of patches for the various flag manifolds are again the numbers of corners in the weight diagrams for the various representations. For example, consider the weight diagrams of the representations $(1,0,0)$, $(1,0,1)$, $(1,1,1)$,
\begin{equation*}
  \epsfig{file=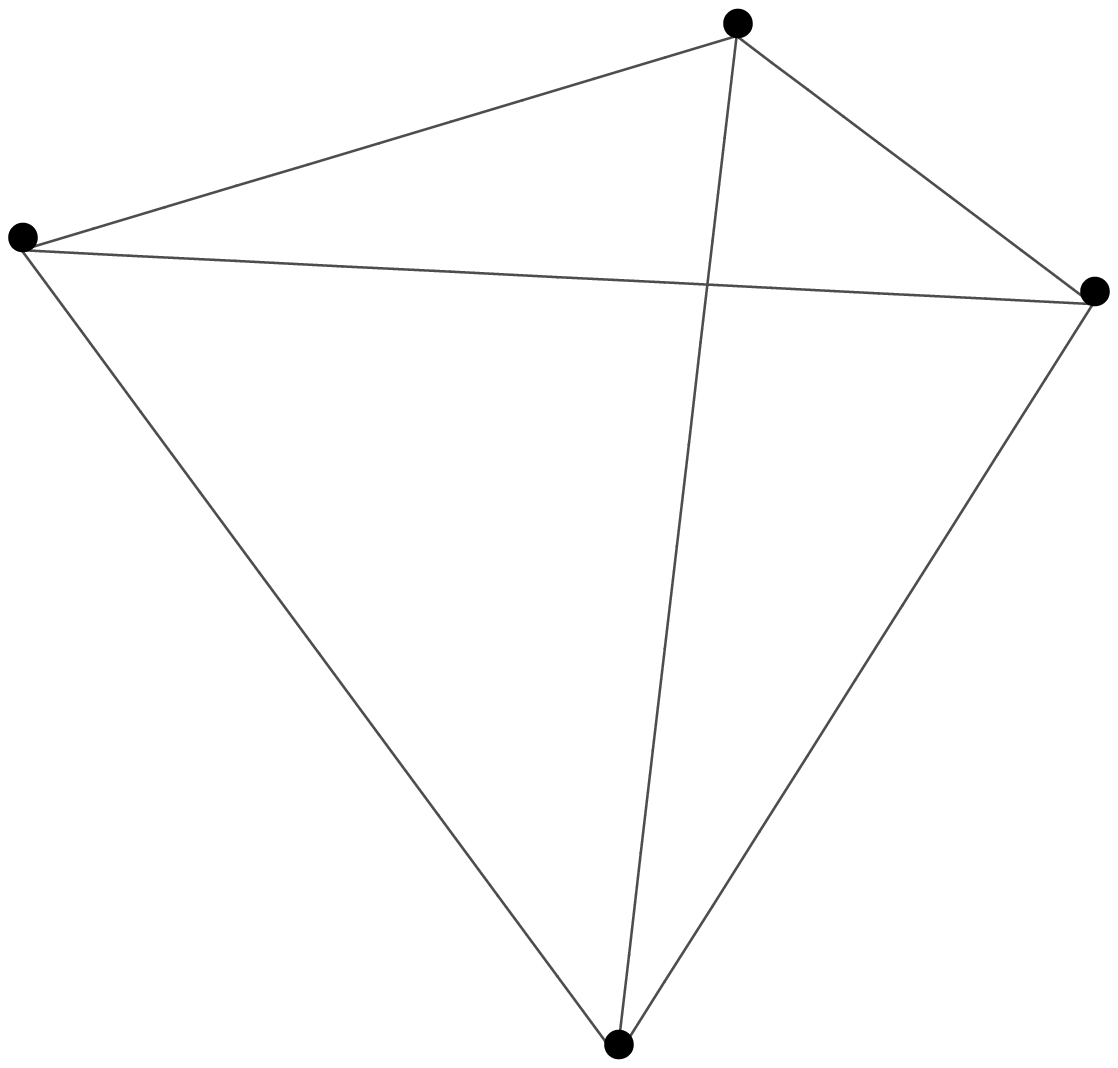,width=3cm,height=3cm}\hspace*{0.8cm}
  \epsfig{file=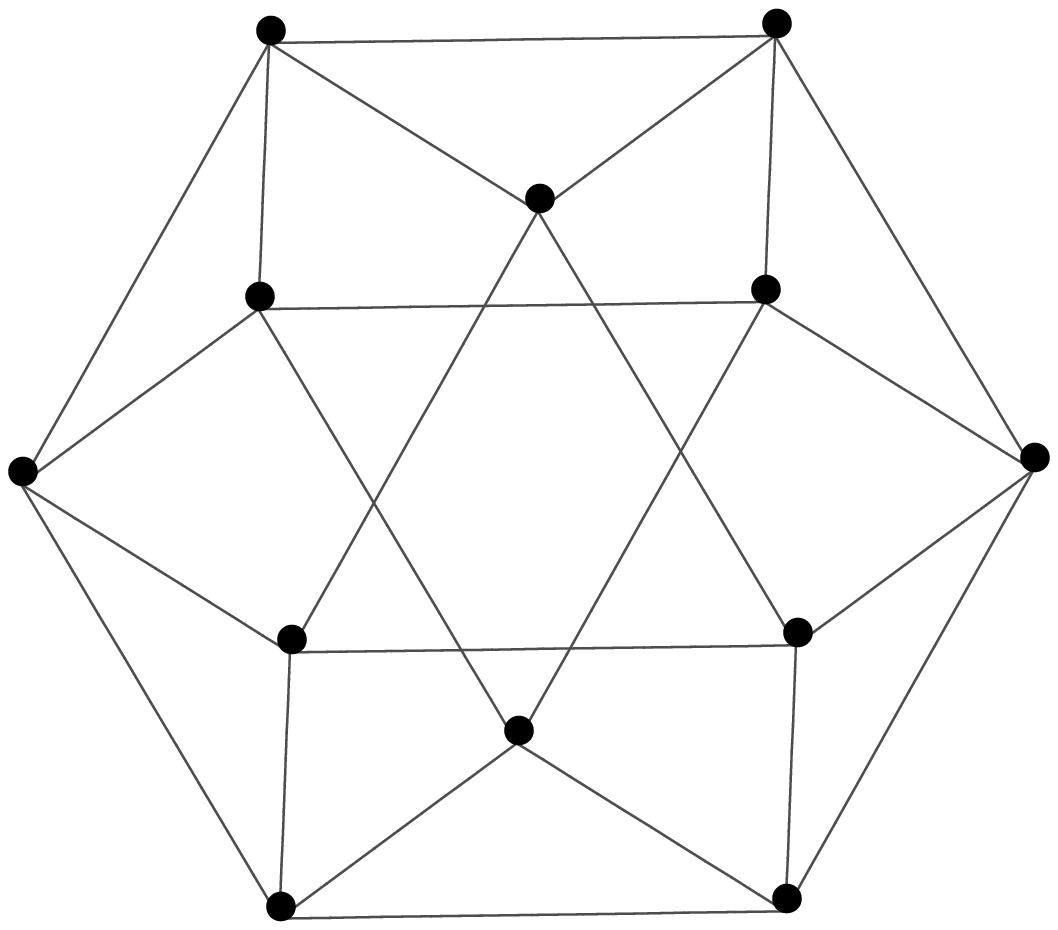,width=3cm,height=3cm}\hspace*{0.8cm}
  \epsfig{file=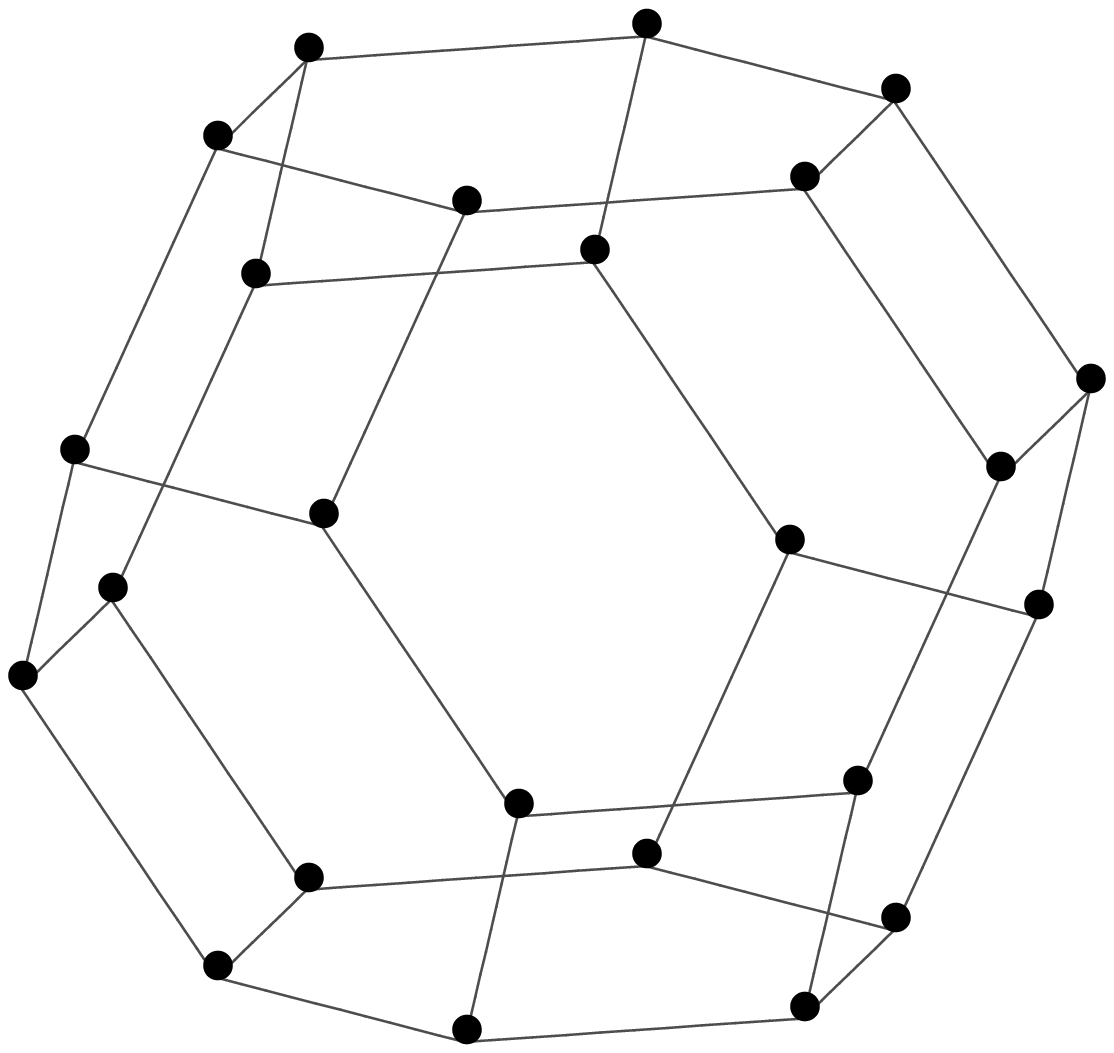,width=3cm,height=3cm}
\end{equation*}
which yield coherent states corresponding to $\CPP^3$, $F_{13;4}$ and $F_{123;4}$ with minimal coverings of $4$, $12$ and $24$ patches, respectively.
\begin{table}[h]
\begin{tabular}[t]{|l|c|c|c|c|c|c|c|}
\hline
 & $\dim_\FC$ & patches & isotropy group in $\sU(4)$ & Dynkin labels & Young diagrams\\
\hline
$~F_{1;4}$\hspace{-0.5cm} $\phantom{\overbrace{{\tyng(1)}}^{\hat{L}}}$ & 3 & 4 &$\sU(1)\times \sU(3)$ & $(L,0,0)$ & $\overbrace{\tyng(4)}^{L}$\\[0.3cm]
$~F_{2;4}$ & 4 & 6 & $\sU(2)\times \sU(2)$ & $(0,L,0)$ & $\overbrace{\tyng(4,4)}^{L}$\\[0.3cm]
$~F_{3;4}$ & 3 & 4 &$\sU(3)\times \sU(1)$ & $(0,0,L)$ & $\overbrace{\tyng(4,4,4)}^{L}$\\[0.3cm]
$~F_{12;4}$ & 5 & 12 & $\sU(1)\times\sU(1)\times \sU(2)$ & $(L,L,0)$ & $\overbrace{\tyng(8,4)}^{L+L}$\\[0.3cm]
$~F_{13;4}$ & 5 & 12 & $\sU(1)\times \sU(2)\times \sU(1)$ & $(L,0,L)$ & $\overbrace{\tyng(8,4,4)}^{L+L}$\\[0.3cm]
$~F_{23;4}$ & 5 & 12 & $\sU(2)\times \sU(1)\times \sU(1)$ & $(0,L,L)$ & $\overbrace{\tyng(8,8,4)}^{L+L}$\\[0.3cm]
$~F_{123;4}$ & 6 & 24 & $\sU(1)\times \sU(1)\times \sU(1)\times \sU(1)$ & $(L,L,L)$ & $\overbrace{\tyng(9,6,3)}^{L+L+L}$\\[0.5cm]
\hline
\end{tabular}
\caption{Representations of $\sSU(4)$ related to the flag manifolds in $\FC^4$.}
\end{table}

\section{Fuzzification of the flag manifolds}\label{SecQuantFlag}

Combining the description of the flag manifolds in terms of Pl{\"u}cker coordinates which we developed in section 2 with the correspondence to coherent states in the previous section, we have an obvious way in which one can truncate the algebra of functions on these spaces to obtain the latter's fuzzy versions. Before we describe the construction of the fuzzy flag manifolds in detail, let us briefly recall the underlying principles.

\subsection{Fuzzification}

By {\em fuzzy geometry}, we mean a truncation of the algebra of functions on a compact space such that the coordinates become noncommutative while all isometries are manifestly preserved. Given a compact Riemannian manifold $M$ without boundary, the spectrum of the Laplace operator is discrete and the eigenfunctions form an orthogonal basis $B$ of $L^2(M)$. A na\"ive guess for a discretization would be to truncate the expansion of a function by using only a finite subset $B^L$ of elements in $B$. Multiplication of functions, however, clearly necessitates a subsequent projection back on to $B^L$, which in turn will render the product non-associative in general.

If the manifold $M=G/H$ is a coadjoint orbit of a Lie group $G$, we can easily circumvent this problem: we can map functions to operators acting as automorphisms\footnote{That is, they can be represented by square matrices.} on the representation space of some representation $R$ of $G$ which admits singlets under $H$ and replace the product between functions by the operator product. In the previous section, we described which representations $R$ are suitable for the various flag manifolds of $\sSU(4)$. We will see that the choice of $R$ corresponds to a choice of the truncation and the closure of multiplication is trivially given. 

Using a projector $\rho_R(x)=|x\rangle\langle x|$ which corresponds to a point $x\in G/H$ and acts on the representation space of $R$, we can establish a map between operators and functions on the coset space by the formula
\begin{equation}\label{deffuzzyfunc}
f_R(x)\ =\ \tr \left(\rho_R(x)\hat{f}\right)~.
\end{equation}
The operator product then induces a star product via
\begin{equation}\label{defstarproduct}
(f_R\star g_R)(x)\ =\ \tr\left(\rho_R(x)\hat{f}\hat{g}\right)~.
\end{equation}
In general, there is an infinite sequence of suitable representations $R_i$ for any coset space and for each of these representations, the star product is different. Choosing higher-dimensional representations amounts to a better approximation of the functions by operators, and there is usually a well-defined limit, in which the complete set of functions on the coset together with the ordinary product is reproduced.

Let us return to equation \eqref{deffuzzyfunc}. To each operator $\hat{f}$ representing a function on $M_R$, assign a corresponding symbol $\tilde{f}(g)$, $g\in G$ by \cite{Dolan:2001mi}
\begin{equation}
\tilde{f}(g)\ =\ n_R \tr\left(D^R(g^{-1})\hat{f}\right)~,
\end{equation}
where $D^R(g)$ is the group element $g$ acting in the representation $R$. The normalization constant $n_R$ is defined by
\begin{equation}
\int \dd \mu(g) D^R(g^{-1})_{ij}D^R(g)_{kl}\ =\
\frac{1}{n_R}\delta_{il}\delta_{jk}
\end{equation}
with $\dd \mu(g)$ being the Haar measure on $G$. Inversely, the
operator $\hat{f}$ corresponding to the symbol $\tilde{f}$ is
therefore obtained from
\begin{equation}
\hat{f}\ =\ \int\dd \mu(g)\tilde{f}(g)D^R(g)~.
\end{equation}
From the symbol $\tilde{f}$ of an operator $\hat{f}$, we can easily calculate
the function defined in \eqref{deffuzzyfunc} using
\begin{equation}
f_R(x)\ =\ \int \dd \mu(g)\omega_R(x,g)\tilde{f}(g) \ewith
\omega_R(x,g)\ :=\ \tr\left(\rho_R(x)D^R(g)\right)~.
\end{equation}
In the definition of the star product \eqref{defstarproduct}, this
translates into
\begin{equation}
(f_R\star g_R)(x)\ =\ \int \dd \mu(g)\int \dd \mu(g')
\omega_R(x,gg')\tilde{f}(g)\tilde{g}(g')~,
\end{equation}
and it is for this formula that we will find explicit expressions
for all the flag manifolds later on.

In the discussion of fuzzy flag manifolds using star products, we can use both of the two equivalent descriptions: either real coordinates describing an embedding of the coset space into flat Euclidean space or the complex homogeneous or Pl{\"u}cker coordinates. In the latter coordinates the star product can be shown to simplify considerably. Moreover, they allow for a direct translation to the operator picture.

Note that so far, we only arrived at an algebra of functions on a topological space. The explicit geometry of this space, i.e.\ its metric structure, has not been described yet. In noncommutative geometry, this information is encoded in a Dirac operator, or -- in a slightly weaker way -- in a Laplacian. Using the above mentioned embedding, we obtain a canonical metric on the coset space and can show that the Laplace operator naturally translates into the second order Casimir in the representation $R$.

For more details on the principle underlying fuzzification, see also \cite{Balachandran:2001dd}.

\subsection{The fuzzy complex projective space $\CPP^3_F$}

The fuzzification of $F_{1;4}$ (and therefore also that of its dual $F_{3;4}$) is well-known \cite{Balachandran:2001dd}, and we follow the usual discussion of the procedure for $\CPP^{N}$. That is, we promote the vector $a$ and its complex conjugate $\bar{a}$ to a four-tuple of annihilation and creation operators satisfying the algebra $[\hat{a}^i,\hat{a}^{j\dagger}]=\delta^{ij}$. The auxiliary coordinates defined in \eqref{auxcoords} also become operators
\begin{equation}
\hat{x}^{\hat{a}}\ :=\ \frac{1}{\hat{a}^{k\dagger}
\hat{a}^l\delta_{kl}}\hat{a}^{i\dagger}\lambda^{\hat{a}}_{ij}\hat{a}^j~,
\end{equation}
which evidently commute with the number operator $\hat{N}=\hat{a}^{k\dagger} \hat{a}^l\delta_{kl}$. Therefore, we can restrict the algebra of functions to the subspace of the Fock space, on which $\hat{N}=L$. This subspace is spanned by
\begin{equation}
\tfrac{1}{C}\hat{a}^{i_1\dagger}\ldots
\hat{a}^{i_L\dagger}|0\rangle\ewith C\ =\ \sqrt{n_1!n_2!n_3!n_4!}~,
\end{equation}
where the $n_i$ are the number of indices being $i$. The truncated algebra of functions $\CA_L$ on this space is the algebra of operators with basis
\begin{equation}\label{functionbasisF1}
\hat{a}^{i_1\dagger}\cdots\hat{a}^{i_L\dagger}|0\rangle\langle0|\hat{a}^{j_1}\cdots
\hat{a}^{j_L}~.
\end{equation}
It is immediately obvious that these operators will commute with the number operator $\hat{N}$, which amounts to factoring out a $\sU(1)$ as implied in the definition of any flag manifold. The coefficients of the expansion of an operator in terms of the basis \eqref{functionbasisF1} form square matrices of dimension $\left(\frac{(3+L)!}{3!L!}\right)^2$ \cite{Balachandran:2001dd}, and in terms of Young diagrams of $\sSU(4)$, we have
\begin{equation}\label{functalgF1}
\overbrace{\overline{\tyng(4)}}^{L}~\otimes~\overbrace{\tyng(4)}^L~\
=\ ~ \overbrace{\tyng(4,4,4)}^{L}~\otimes~\overbrace{\tyng(4)}^L~.
\end{equation}
To expose the underlying $\sSU(4)$ structure and to construct the polarization tensors, we can contract indices from the creation operators with indices from the annihilation operators using the $15$ generators $\lambda^a_{ij}$ of $\sSU(4)$. A contraction with $\lambda^0_{ij}\sim\delta_{ij}$ yields the embedded subalgebra truncated at level $L-1$, since the trace over a fundamental and an antifundamental index corresponds to the determinant over four indices in either the fundamental or antifundamental representation and thus to a column of $4$ boxes in a Young diagram, which is cancelled. The tensor product expansion looks as
\begin{equation}
\overbrace{\tyng(4,4,4)}^L~\otimes~\overbrace{\tyng(4)}^L~\ =\
~\mathbf{1}~\oplus~\tyng(2,1,1)~\oplus~\tyng(4,2,2)~\oplus~
\ldots~\oplus~\overbrace{\tyng(8,4,4)}^{2L}~.
\end{equation}

A representation of the Lie algebra of $\sSU(4)$ is given by the Schwinger construction and we can write
\begin{equation}
\hat{L}^{\hat{a}}\ =\ \hat{a}^{i\dagger} \lambda_{ij}^{\hat{a}} \hat{a}^j~.
\end{equation}
One can easily verify the algebra $[\hat{L}^a,\hat{L}^b]=\di \sqrt{2}f^{ab}{}_c\hat{L}^c$ using $[\hat{a}^i,\hat{a}^{j\dagger}]=\delta^{ij}$. 

In the representations $R=R(L)$ introduced above, the projector corresponding in the case $L=1$ to $\CP_{1;4}=\CP(x_{1;4})$, which describes the embedding of $\CPP^3$ in $\FR^{15}$, is simply the $L$-fold symmetrized tensor product $\rho^L_{1;4}=\rho^L(x_{1;4})=\CP(x_{1;4})\otimes \ldots \otimes \CP(x_{1;4})$, and we can map any operator $\hat{f}$ in the algebra $\CA_L$ to a corresponding function $f_L$ on the embedding of $\CPP^3$ in $\FR^{15}$ by
\begin{equation}\label{functionoperatorF1}
f_L(x_{1;4})\ =\ \tr(\rho^L(x_{1;4}) \hat{f})~.
\end{equation}
Furthermore, this map induces a star product on $\CPP^3$ defined as
\begin{equation}
(f_L\star g_L)(x_{1;4})\ =\ f_L(x_{1;4})\star
g_L(x_{1;4})\ =\ \tr(\rho^L(x_{1;4}) \hat{f} \hat{g})~,
\end{equation}
where $f_L$ and $g_L$ are the functions corresponding to the operators $\hat{f}$ and $\hat{g}$, respectively.

To make the star product more explicit, we calculate $\omega^L(x_{1;4},gg')$ for the fundamental representation $L=1$:
\begin{equation}
\omega^L(x_{1;4},gg')\ =\ \tr(\CP_{1;4}gg')\ =\
\tr(\CP_{1;4}g\CP_{1;4}g')+\tr(\CP_{1;4}g(\unit-\CP_{1;4})g')~.
\end{equation}
Since $\CP_{1;4}$ is a rank one projector, we have
\begin{equation}
\tr(\CP_{1;4}g\CP_{1;4}g')\ =\
\tr(\CP_{1;4}g)\tr(\CP_{1;4}g')\ =\
\omega^L(x_{1;4},g)\omega^L(x_{1;4},g')
\end{equation}
and with the identities \eqref{identJK}, it follows immediately that
\begin{equation}
\begin{aligned}
\tr(\CP_{1;4}g(\unit-\CP_{1;4})g')&\ =\
\tr(\lambda^ag)\tr(\CP_{1;4}\lambda^a(\unit-\CP_{1;4})g')\\&\
=\
\left(\der{x^a}\omega(x_{1;4},g)\right)J^{ab}\left(\der{x^b}\omega(x_{1;4},g')\right)~.
\end{aligned}
\end{equation}
For the representations with $L>1$, we can simply take the $L$-fold tensor product of $\omega^1(x_{1;4},g)$ \cite{Dolan:2001mi}:
\begin{equation}
\omega^L(x_{1;4},g)\ =\ \left(\omega^1(x_{1;4},g)\right)^{\otimes
L}~,
\end{equation}
and the total star product reads as \cite{Balachandran:2001dd}
\begin{equation}
(f_L\star g_L)(x_{1;4})\ =\
\sum_{l=0}^L\frac{(L-l)!}{L!l!}\left(\dpar_{a_1\ldots
a_l}f_L(x_{1;4})\right)J^{a_1b_1}_{1;4}\ldots J^{a_lb_l}_{1;4}
\left(\dpar_{b_1\ldots b_l}g_L(x_{1;4})\right)~.
\end{equation}
In the homogeneous coordinates $a^i, \bar{a}^i$ on $\CPP^3$, the space of functions is spanned by homogeneous polynomials of the form
\begin{equation}
\bar{a}^{i_1}\ldots \bar{a}^{i_L}a^{j_1}\ldots a^{j_L}~,
\end{equation}
which correspond to the operators \eqref{functionbasisF1} under the map \eqref{functionoperatorF1}. In these coordinates, the star product simplifies to \cite{Kurkcuoglu:2006iw}
\begin{equation}\label{StarProductReduced}
(f\star g)\ =\ \mu\left[
\frac{1}{L!}\der{a^{\alpha_1}}\ldots \der{a^{\alpha_L}}\otimes\frac{1}{L!}\der{\bar{a}^{\alpha_1}}\ldots \der{\bar{a}^{\alpha_L}}
(f\otimes g)\right]~,
\end{equation}
where $\mu(a\otimes b)=a\cdot b$. Note that this construction of a star product generalizes in a rather straightforward way to other spaces, as soon as we have a suitable projector $\rho^L(x)$ at hand.

To relate the given matrix algebra to the space $\CPP^3$, we need some additional structure to encode the geometry of this space. For this, consider the vector fields on $\CPP^3$ from the perspective of the embedding space $\FR^{16}$:
\begin{equation}
\CL_a\ =\ -\sqrt{2}f_{ab}{}^cx^b\der{x^c}~,~~~[\CL_a,\CL_b]\ =\ \di\sqrt{2} f_{ab}{}^c \CL_c~,
\end{equation}
where $f_{ab}{}^c$ are the structure constants of $\sSU(4)$. Note here that in the limit $L\rightarrow \infty$, the fuzzy derivatives approach the ones from the continuum in an obvious way. It is now rather straightforward to show that \cite{Balachandran:2001dd} 
\begin{equation}
\CL^a f_L(x)\ =\ \sqrt{2} \Omega^{ab} \der{x^b}f_L(x)\ =\ \frac{L}{\sqrt{2}}(x^a\star f_L(x)-f_L(x)\star x^a)\ =\ \tr\left(\rho^L(x_{1;4})[\hat{L}^a,\hat{f}]\right)~,
\end{equation}
where $[\hat{L}^a,\cdot]$ are the generators of $\sSU(4)$ in the representation $(L,0,0)\otimes(0,0,L)$. It is therefore also clear that the Laplace operator on $\CPP^3$, $\Delta=\CL^a\CL^b\delta_{ab}$, is mapped to the second order Casimir in the adjoint representation $(L,0,0)\otimes(0,0,L)$:
\begin{equation}
\hat{\Delta}_{1;4}\hat{f}\ =\ [\hat{L}^a,[\hat{L}^b,\hat{f}]]\delta_{ab}~.
\end{equation}

\subsection{The fuzzy Gra{\ss}mannian $G^F_{2;4}$}

We proceed analogously to the case of $\CPP^3$, which leads to the results presented in \cite{Dolan:2001mi} in a somewhat simpler form. That is, we take the Pl{\"u}cker description discussed in section 2.3.\ and promote the vector components to creation and annihilation operators $\hat{a}^i,\hat{a}^{i\dagger},\hat{b}^i,\hat{b}^{i\dagger}$. We thus arrive at the algebra
\begin{equation}
[\hat{a}^i,\hat{a}^{j\dagger}]\ =\ [\hat{b}^i,\hat{b}^{j\dagger}]\ =\ \delta^{ij}~,
\end{equation}
and all other commutators vanish. From these operators, we construct the composite creation and annihilation operators
\begin{equation}
\hat{A}_2^{ij}\ =\ \hat{a}^{[i}\hat{b}^{j]}\eand
\hat{A}_2^{ij\dagger}\ =\
\hat{a}^{[i\dagger}\hat{b}^{j]\dagger}~,
\end{equation}
which satisfy
\begin{equation}\label{comx1}
[\hat{A}_2^{ij},\hat{A}_2^{kl\dagger}]\ =\
\left(\delta^{ik}\delta^{jl}+\delta^{jl}\hat{a}^{k\dagger}
\hat{a}^i+\delta^{ik}\hat{b}^{l\dagger}
\hat{b}^j\right)_{[ij][kl]}~,
\end{equation}
where $(\cdot)_{[ij][kl]}$ denotes antisymmetrization of the enclosed components, as well as
\begin{equation}\label{comx2}
[[\hat{A}_2^{ij},\hat{A}_2^{kl\dagger}],\hat{A}_2^{mn\dagger}]\ =\ \left(2\delta^{jl}\delta^{im}\hat{A}_2^{kn\dagger}\right)_{[ij][kl][mn]}~.
\end{equation}
We can now use $\hat{A}_2^{mn\dagger}$ to build an $L$-particle\footnote{Note that a particle is here a composite object consisting of two excitations.} Hilbert space $\CCH^L_{2;4}$. This space is spanned by
\begin{equation}\label{statesG2}
\tfrac{1}{C}\hat{A}_2^{i_1j_1\dagger}\cdots\hat{A}_2^{i_Lj_L\dagger}|0\rangle~,
\end{equation}
where $C$ is the norm of the state. Acting with $\hat{A}_2^{mn}$ on such a state yields a state in $\CCH^{L-1}_{2;4}$ due to \eqref{comx1} and \eqref{comx2}. Recall that in the Pl{\"u}cker description of $G_{2;4}$, we constructed the plane by antisymmetrizing two vectors, which could be chosen orthogonal such that $\bar{a}^ib^i=0$. On the operator level, this translates into
\begin{equation}\label{ComRelOrth}
{}[\hat{a}^{i\dagger} \hat{b}^i,\hat{A}_2^{kl\dagger}]\ =\ 0~,~~~
[\hat{a}^i \hat{b}^{i\dagger},\hat{A}_2^{kl\dagger}]\ =\ 0~,~~~
[\hat{a}^{i\dagger} \hat{b}^i,\hat{A}_2^{kl}]\ =\ 0\eand [\hat{a}^i
\hat{b}^{i\dagger},\hat{A}_2^{kl}]\ =\ 0~,
\end{equation}
and therefore the action of $\hat{a}^{i\dagger} \hat{b}^i$ on any state in $\CCH^L_{2;4}$ vanishes. This implies that we can introduce the number operator
\begin{equation}
\hat{N}\ =\ \hat{a}^{i\dagger}\hat{a}^j\delta_{ij}\ =\
\hat{b}^{i\dagger}\hat{b}^j\delta_{ij}\ =\ L
\end{equation}
and
\begin{equation}
\hat{A}_2^{ij\dagger} \hat{A}_2^{kl}\delta_{ik}\delta_{jl} \ =\  2\hat{N}(\hat{N}-1) \ =\ 2L(L+1)~,
\end{equation}
where the equalities hold only after restriction to $\CCH^L_{2;4}$. Thus, $\CCH^L_{2;4}$ is indeed an $L$-particle Hilbert space. We already know from the discussion in the previous section that the states \eqref{statesG2} are invariant under $\sS(\sU(2)\times \sU(2))$. Let us nevertheless be more explicit on the action of the internal $\sSU(2)$, which acts nontrivially on both $\hat{a}^i$ and $\hat{b}^i$. Its generators $L^r_\inner$ act according to
\begin{equation}
L^r_\inner \ =\ \ad\left(\hat{a}^{i\dagger}_p\lambda^r_{pq}\hat{a}^i_q\right)~,
\end{equation}
where $\hat{a}^i_1=\hat{a}^i$, $\hat{a}^i_2=\hat{b}^i$ and $\lambda^r_{pq}$, $p,q=1,2$, $r=1,2,3$ are the Gell-Mann matrices of $\sSU(2)$. The combinations $A^{ij}_2$ and $A^{ij\dagger}_2$ are now invariant under this action due to the general formula
\begin{equation}\label{oscillatorinternalisotropy}
[a_p^{i\dagger}\lambda^r_{pq}a_q^i,a^{[j_1\dagger}_1\ldots a^{j_k]\dagger}_k]\ =\ 0~,
\end{equation}
where $p=1,\ldots ,k$ and $i,j_p=1,\ldots ,n$; see appendix B for a proof.

The truncated algebra of functions $\CA_L$ is the algebra of operators spanned by
\begin{equation}\label{spanfunctG2}
\hat{A}_2^{i_1j_1\dagger}\cdots\hat{A}^{i_Lj_L\dagger}|0\rangle\langle0|\hat{A}^{k_1l_1}\cdots
\hat{A}_2^{k_Ll_L}~,
\end{equation}
and the coefficients in an expansion in terms of these operators are square matrices of size $\frac{(3+L)!(2+L)!}{3!L!2!(L+1)!}$. Note that these operators again commute with the number operator $\hat{N}$ and in terms of Young diagrams, we have here
\begin{equation}
\overbrace{\overline{\tyng(4,4)}}^{L}~\otimes~\overbrace{\tyng(4,4)}^L ~\ =\ ~\overbrace{\tyng(4,4)}^{L}~\otimes~\overbrace{\tyng(4,4)}^L~.
\end{equation}
Again, each of the tensor product decompositions at level $L$ contains the tensor product decomposition at lower levels:
\begin{equation}
\overbrace{\tyng(4,4)}^{L}~\otimes~\overbrace{\tyng(4,4)}^L~\ =\ ~
\mathbf{1}~\oplus~\tyng(2,1,1)~\oplus~\tyng(2,2)~\oplus~\tyng(4,2,2)~\oplus~\tyng(4,3,1)~\oplus~\tyng(4,4)~\oplus \ldots
\end{equation}
Contrary to the case of $\CPP^3$, where increasing the level $L$ by one yielded precisely one new type of Young tableau in the sum, we here obtain $L+1$ new diagrams in each step, which consist of three rows with $a+b+a$, $a+b$ and $a$ boxes respectively. The new diagrams at level $L$ are the ones for which $a=n$ and $b=2L-2n$ for $n=1\ldots L$.

All these notions readily translate for arbitrary Gra{\ss}mannians. 

To find a representation of the Lie algebra of $\sSU(4)$, we use a generalized Schwinger construction
\begin{equation}\label{Schwingerx}
\hat{L}^a\ =\ \frac{1}{\hat{N}+1}\hat{A}_2^{ij\dagger} \Lambda_{ijkl}^a \hat{A}_2^{kl}\ewith \Lambda_{ijkl}^a\ =\ (\lambda^a\wedge \unit)_{ijkl}~.
\end{equation}
It is important to stress that the $\hat{L}^a$ by themselves do not form a representation of $\sSU(4)$, but again after having them act on a state in $\CCH^L_{2;4}$, they do. This is simply due to the fact that because of \eqref{ComRelOrth}, $\hat{L}^a$ reduces when acting on a state in $\CCH^L_{2;4}$ to
\begin{equation}
\hat{L}^a\ =\
\hat{a}^{i\dagger}\lambda^a_{ij}\hat{a}^j
+\hat{b}^{i\dagger}\lambda^a_{ij}\hat{b}^j~.
\end{equation}

The projector yielding a star product on this space is the symmetrized $L$-fold tensor product $\rho^L_{2;4}=\rho^L(x_{2;4})=\CCP_{2;4}\otimes \ldots \otimes \CCP_{2;4}$. Proceeding precisely along the lines of the
discussion of the star product on $\CPP^3$, we find that 
\begin{equation}
\omega^1(x^{\hat{a}\hat{b}},gg')\ =\
\omega^1(x,g)\left(1+\overleftarrow{\dpar}_{\hat{a}\hat{b}}J^{\hat{a}\hat{b},\hat{c}\hat{d}}\overrightarrow{\dpar}_{\hat{c}\hat{d}}\right)\omega^1(x,g')~,
\end{equation}
where at least one of the indices in each $\hat{a}\hat{b}$ or $\hat{c}\hat{d}$ is nonzero. That is, the component $x^{00}$ plays a similar r\^{o}le to the component $x^0$ in the case of $\CPP^3$. Furthermore, one should stress that as usual for the derivatives on spaces with symmetrized tensors as coordinates, one has
\begin{equation}
\dpar_{\hat{a}\hat{b}}\ =\ \left\{\begin{array}{ll} \der{x^{\hat{a}\hat{b}}}&\mbox{ for $\hat{a} \ =\  \hat{b}\neq 0$}\\
\frac{1}{2}\der{x^{\hat{a}\hat{b}}}&\mbox{ for $\hat{a}\neq \hat{b}$}
\end{array}\right.~.
\end{equation}
It is then quite obvious that the star product is given by
\begin{equation}
\begin{aligned}
(f_L\star g_L)(x_{2;4})\ =\
\sum_{l=0}^L\frac{(L-l)!}{L!l!}&\left(\dpar_{(\hat{a}_1\hat{b}_1)\ldots
(\hat{a}_l\hat{b}_l)}f_L(x_{2;4})\right)J^{\hat{a}_1\hat{b}_1;\hat{c}_1\hat{d}_1}_{2;4}\ldots \\
 & J^{\hat{a}_l\hat{b}_l;\hat{c}_l\hat{d}_l}_{2;4} \left(\dpar_{(\hat{c}_1\hat{d}_1)\ldots
(\hat{c}_l\hat{d}_l)}g_L(x_{2;4})\right)~, 
\end{aligned}
\end{equation}
where $\dpar_{(\hat{a}_1\hat{b}_1)\ldots (\hat{a}_l\hat{b}_l)}:=\dpar_{\hat{a}_1\hat{b}_1}\ldots \dpar_{\hat{a}_l\hat{b}_l}$. Note that our choice of embedding $F_{2;4}$ in $\FR^{16\cdot 17/2-1}$ yielded a slightly simpler expression for the star product on this space than the one in \cite{Dolan:2001mi}, which used an embedding in $\FR^{15}$. 

The expression for the star product further simplifies, if we switch again to complex coordinates $a^{[i}b^{j]}$ on $\FC^4\wedge\FC^4$. The functions corresponding to the operators \eqref{spanfunctG2} read as
\begin{equation}
\bar{a}^{[i_1}\bar{b}^{j_1]}\ldots\bar{a}^{[i_L}\bar{b}^{j_L]}a^{[k_1}b^{l_1]}\ldots a^{[k_L}b^{l_L]}
\end{equation}
and the star product is here defined as 
\begin{equation*}
(f\star g)\ =\ \mu\left[
\frac{1}{L!L!}\der{a^{[i_1}}\der{b^{j_1]}}\ldots \der{a^{[i_L}}\der{b^{j_L]}}\otimes\frac{1}{L!L!}\der{\bar{a}^{[i_1}}\der{\bar{b}^{j_1]}}\ldots \der{\bar{a}^{[i_L}}\der{\bar{b}^{j_L]}}
(f\otimes g)\right]~.
\end{equation*}

The natural Laplacian on $\CCH^L_{2;4}$ encoding the geometry of $G_{2;4}$ is derived from the embedding of $G_{2;4}$ in $\FR^{16\cdot 17/2 -1}$. In terms of Pl{\"u}cker coordinates, the generators read as 
\begin{equation}
\CL^{\hat{a}}\ =\ \bar{a}^i\lambda^{\hat{a}}_{ij}\der{\bar{a}^j}+\bar{b}^i\lambda^{\hat{a}}_{ij}\der{\bar{b}^j}-
a^j\lambda^{\hat{a}}_{ij}\der{a^i}-b^j\lambda^{\hat{a}}_{ij}\der{b^i}~.
\end{equation}
From this expression and equation \eqref{contractionG2}, we obtain the following expression in terms of the embedding coordinates:
\begin{equation}
\CL_a\ =\ -\di \sqrt{2}f_{ab}{}^c x^{bd}\der{x^{cd}}~,
\end{equation}
which satisfies the algebra $[\CL^a,\CL^b]=\di\sqrt{2} f^{ab}{}_c\CL^c$, as is easily verified. Using the first relation in \eqref{identOmegaStar}, we can write
\begin{equation}
\CL^a f_L(x)\ =\ \sqrt{2}I^{a0,cd}\der{x^{cd}}f_L(x)\ =\ x^{a0}\star f_L-f_L\star x^{a0}\ =\ -\di\tr(\rho^L(x_{2;4})[\hat{L}^a,\hat{f}_L])~.
\end{equation}
We thus see again that all the derivatives are mapped to the generators of $\sSU(4)$ acting in the adjoint and therefore the Laplacian is given by the second order Casimir of this representation of $\sSU(4)$,
\begin{equation}\label{casimir}
\hat{\Delta}_{2;4}\ =\ \ad_{\hat{L}^{a}}\ad_{\hat{L}^{b}}\delta_{ab}~.
\end{equation}
This observation from the cases $\CPP^3$ and $G_{2;4}$ translates to all flag manifolds and we suppress this calculation in the remaining cases. For the discussion of the eigenvalues of these Casimirs, see appendix B.

\subsection{The fuzzy dual complex projective space $F^F_{3;4}$}

One can infer the fuzzification of $F_{3;4}$ in a straightforward manner from the ones of $F_{1;4}$ and $F_{2;4}$. We start from the Pl{\"u}cker description and promote the vectors $a,b,c$ to a triple of four-tuples of oscillators with creation and annihilation operators $\hat{a}^i,\hat{a}^{i\dagger},\hat{b}^i,\hat{b}^{i\dagger},\hat{c}^i,\hat{c}^{i\dagger}$. We furthermore introduce the composite operators
\begin{equation}
\hat{A}_3^{ijk}\ =\ \hat{a}^{[i}\hat{b}^j\hat{c}^{k]}\eand
\hat{A}_3^{ijk\dagger}\ =\
\hat{a}^{[i^\dagger}\hat{b}^{j\dagger}\hat{c}^{k]\dagger}~,
\end{equation}
satisfying the commutation relations
\begin{equation}
\begin{aligned}
{}[\tilde{\hat{d}}_i,\tilde{\hat{d}}^\dagger_m]\ :=\ &\eps_{ijkl}\eps_{mnrs}[\hat{A}_3^{jkl},\hat{A}_3^{nrs}]\\
\ =\ &\delta_{im}+\eps_{ijkl}\eps_{mnrs}\Big(\hat{a}^{n\dagger}
\hat{a}^j\delta^{rk}\delta^{sl}+\delta^{nj}\hat{b}^{r\dagger}
\hat{b}^k\delta^{sl}+\delta^{nj}\delta^{rk}\hat{c}^{s\dagger} \hat{c}^l+\\
&\hspace{3cm}\hat{a}^{n\dagger} \hat{a}^j\hat{b}^{r\dagger}
\hat{b}^k\delta^{sl}+ \hat{a}^{n\dagger}
\hat{a}^j\delta^{rk}\hat{c}^{s\dagger} \hat{c}^l+
\delta^{jn}\hat{b}^{r\dagger} \hat{b}^k\hat{c}^{s\dagger}
\hat{c}^l\Big)~.
\end{aligned}
\end{equation}
The expression for $[[[\tilde{\hat{d}}_i,\tilde{\hat{d}}^\dagger_j]\tilde{\hat{d}}^\dagger_k]\tilde{\hat{d}}^\dagger_l]$
contains only the combination $\tilde{\hat{d}}^\dagger_m=\eps_{mnrs}\hat{A}_3^{nrs\dagger}$.

The $L$-particle Fock space $\CCH^L_{3;4}$ is evidently spanned by the states 
\begin{equation}
\hat{A}_3^{i_1j_1k_1\dagger}\ldots \hat{A}_3^{i_Lj_Lk_L\dagger}|0\rangle~,
\end{equation}
and this space forms the representation $(0,0,L)$ of $\sSU(4)$. The isotropy subgroup of any state in this representation is thus $\sS(\sU(3)\times \sU(1))$, and the internal $\sSU(3)$ action, affecting all the elementary oscillators $\hat{a},\hat{b},\hat{c}$ is given by  
\begin{equation}
\hat{L}^r_\inner \ =\ \ad\left(\hat{a}^{i\dagger}_p\lambda^r_{pq}\hat{a}^i_q\right)~,
\end{equation}
where $\hat{a}^i_1, \hat{a}^i_2, \hat{a}^i_3$ stand for $\hat{a}^i,\hat{b}^i,\hat{c}^i$, respectively, and $\lambda^r$ are the Gell-Mann matrices of $\sSU(3)$. Invariance of the operators $\hat{A}_3^{ijk}$ and $\hat{A}_3^{ijk\dagger}$ follows from equation \eqref{oscillatorinternalisotropy}. It is this representation which underlies the construction of vector bundles over $\CPP^3_F$ \cite{Dolan:2006tx}.

The algebra of functions truncated at level $L$ is again constructed from two copies of the Fock space and their $\sSU(4)$ transformation property is captured by the diagrams 
\begin{equation}
\overbrace{\overline{\tyng(4,4,4)}}^{L}~\otimes~\overbrace{\tyng(4,4,4)}^L~\
=\ ~ \overbrace{\tyng(4)}^L~\otimes~\overbrace{\tyng(4,4,4)}^L~.
\end{equation}
We clearly see that this algebra is dual to \eqref{functalgF1}, i.e.\ that of $F^F_{1;4}$. For this reason, we will not go into any further details.

The definition of a star product is obvious. The projector in the symmetrized $L$-fold tensor product reads as $\rho^L_{3;4}(x_{3;4})=\CCP_{3;4}\otimes \ldots \otimes \CCP_{3;4}$ and yields
\begin{equation}
\begin{aligned}
(f_L\star g_L)(x_{3;4})\ =\
\sum_{l=0}^L\frac{(L-l)!}{L!l!}&\left(\dpar_{(\hat{a}_1\hat{b}_1\hat{c}_1)\ldots
(\hat{a}_l\hat{b}_l\hat{c}_l)}f_L(x_{3;4})\right)J^{\hat{a}_1\hat{b}_1\hat{c}_1;\hat{d}_1\hat{e}_1\hat{f}_1}_{3;4}\ldots
\\&~~~J^{\hat{a}_l\hat{b}_l\hat{c}_l;\hat{d}_l\hat{e}_l\hat{f}_l}_{3;4}
\left(\dpar_{(\hat{d}_1\hat{e}_1\hat{f}_1)\ldots (\hat{d}_l\hat{e}_l\hat{f}_l)}g_L(x_{3;4})\right)~,
\end{aligned}
\end{equation}
where again the components $x^{000}$ are dropped in the formula and
\begin{equation}
\dpar_{abc}\ =\ \left\{\begin{array}{ll} \der{x^{abc}}&\mbox{ for $\hat{a}\ =\ \hat{b}\ =\ \hat{c}\neq 0$}\\
\frac{1}{2}\der{x^{\hat{a}\hat{b}\hat{c}}}&\mbox{ for $\hat{a}\ =\ \hat{b}\neq \hat{c}$ etc.}\\
\frac{1}{6}\der{x^{\hat{a}\hat{b}\hat{c}}}&\mbox{ for $\hat{a}\neq \hat{b}\neq \hat{c}\neq \hat{a}$}
\end{array}\right.~.
\end{equation}

The discussion of the star product formalism in the complex coordinates $a^{[i}b^{j}c^{k]}$ on $\FC^4\wedge\FC^4\wedge\FC^4$ is trivially deduced from the cases $\CPP^3_F$ and $G_2^F$. The star product here reads as
\begin{equation*}
\begin{aligned}
(f\star g)\ =\ \mu\Big[&
\frac{1}{L!L!L!}\der{a^{[i_1}}\der{b^{j_1}}\der{c^{k_1]}}\ldots \der{a^{[i_L}}\der{b^{j_L}}\der{c^{k_L]}}\otimes\\&\otimes\frac{1}{L!L!L!}\der{\bar{a}^{[i_1}}\der{\bar{b}^{j_1}}\der{\bar{c}^{k_1]}}\ldots \der{\bar{a}^{[i_L}}\der{\bar{b}^{j_L}}\der{\bar{c}^{k_L]}}
(f\otimes g)\Big]~.
\end{aligned}
\end{equation*}

\subsection{The fuzzy reducible flag manifold $F^F_{12;4}$}

In the case of the reducible flag manifolds, one needs a set of composite creation and annihilation operators. These sets are in one-to-one correspondence with the Pl{\"u}cker coordinates. For $F^F_{12;4}$ we thus have
\begin{equation}
\hat{A}_2^{ij}~,~~~ \hat{A}_2^{ij\dagger}\eand \hat{a}^i~,~~~\hat{a}^{i\dagger}~.
\end{equation}
The missing commutation relations are easily found from \eqref{comx1} and \eqref{comx2}, e.g.\
\begin{equation}
[\hat{a}^i,\hat{A}^{jk\dagger}_2]\ =\ \delta^{i[j}\hat{b}^{k]\dagger}
\end{equation}
and the $(L_1,L_2)$-particle Hilbert spaces are now constructed as
\begin{equation}
\hat{A}_2^{i_1j_1\dagger}\ldots
\hat{A}_2^{i_{L_1}j_{L_1}\dagger}\hat{a}^{k_1\dagger}\ldots
\hat{a}^{k_{L_2}\dagger}|0\rangle~.
\end{equation}
The corresponding operators 
\begin{equation}
\hat{A}_2^{i_1j_1\dagger}\ldots
\hat{A}_2^{i_{L_1}j_{L_1}\dagger}\hat{a}^{k_1\dagger}\ldots
\hat{a}^{k_{L_2}\dagger}|0\rangle \langle0|\hat{A}_2^{m_1n_1}\ldots
\hat{A}_2^{m_{L_1}n_{L_1}}\hat{a}^{l_1}\ldots \hat{a}^{l_{L_2}}
\end{equation}
evidently form a closed algebra and act on the representation space of the representation given in terms of Young diagrams by
\begin{equation}
\overbrace{\overline{\tyng(8,4)}}^{L_1+L_2}~\otimes~\overbrace{\tyng(8,4)}^{L_1+L_2}
~\ =\ ~\overbrace{\tyng(8,8,4)}^{L_2+L_1}~\otimes~\overbrace{\tyng(8,4)}^{L_1+L_2}
\end{equation}
The internal isotropy subgroup here is $\sU(1)\times \sU(1)$, which follows from the discussion of the underlying coherent states. The explicit action of this subgroup on the elementary oscillators is given by the number operators for $\hat{a}^i$ and $\hat{b}^i$.

Note that the full algebra of functions on the flag manifold is obtained from the Hilbert space,
which is the sum of all representations with $L_1+L_2=L$ for some fixed $L$. This is somewhat evident
as the algebra of functions on $F_{12;4}^F$ should contain both the algebra of functions 
of $F_{1;4}^F$ and $F_{2;4}^F$ at level $L$.

To define a star product on this space, recall that we could describe the flag manifold $F_{12;4}$ in terms of two rank one projectors
$\CP_2$ and $\CP_1$ satisfying $\CP_2\CP_1\CP_2=\CP_1$. Furthermore, note that we can split every operator $\hat{f}$ in this representation as
\begin{equation}
\hat{f}\ =\ \tilde{f}_{IJ}\hat{h}_2^I\otimes \hat{h}_1^J\ewith \hat{h}_2^I \in
\overbrace{\overline{\tyng(4,4)}}^{L_1}~\otimes~\overbrace{\tyng(4,4)}^{L_1}\eand
\hat{h}^J_1\in
\overbrace{\overline{\tyng(4,4,4)}}^{L_2}~\otimes~\overbrace{\tyng(4)}^{L_2}~,
\end{equation}
where $I$ and $J$ are multi-indices. To such an operator, a truncated function is assigned by
\begin{equation}
f(x_{2;4},x_{1(2);4})\ =\ \tilde{f}_{IJ}\tr(\rho^L(x_{2;4})
\hat{h}_2^I)\tr(\rho^L(x_{1(2);4}) \hat{h}_1^J)~,
\end{equation}
where $\rho^L(x_{2;4})$ and $\rho^L(x_{1(2);4})$ are the projectors $\CP(x_{2;4})$ and $\CP(x_{1(2);4})$ in the same representations as $\hat{h}^I_{1,2}$ and $x_{1(2);4}$. Furthermore, $x_{2;4},x_{1(2);4}$ are the coordinates on $F_{12;4}$ embedded in Euclidean space, of the plane and the included line, respectively. On this embedding, the star product of two operators is defined as
\begin{equation}
(f\star
g)(x_{2;4},x_{1(2);4})\ =\ \tilde{f}_{IJ}\tilde{g}_{MN}\tr(\rho^L(x_{2;4})\hat{h}_2^I\hat{h}_2^M)
\tr(\rho^L(x_{1(2);4})\hat{h}_1^J\hat{h}_1^N)~.
\end{equation}
Evidently, the star product between $x_{2;4}$ and $x_{1(2);4}$ is simply the ordinary product. Altogether, the star product on $F_{12;4}$ can be derived from the ones on $\CPP^3$ and $G_{2;4}$, and we have
\begin{equation}
\begin{aligned}
(\hat{f}_L\star g_L)&(x_{12;4})\ =\ 
\sum_{l=0}^L\frac{(L-l)!}{L!l!}\sum_{k=0}^L\frac{(L-k)!}{L!k!}
\left(\dpar_{(\hat{a}_1\hat{b}_1)\ldots (\hat{a}_l\hat{b}_l)}\dpar_{\hat{e}_1\ldots
\hat{e}_k}\hat{f}_L(x_{12;4})\right)\times\\&\times J^{(\hat{a}_1\hat{b}_1),(\hat{c}_1\hat{d}_1);\hat{e}_1\hat{f}_1}_{12;4}\ldots
J^{(\hat{a}_l\hat{b}_l),(\hat{c}_l\hat{d}_l);\hat{e}_k\hat{f}_k}_{12;4}
\left(\dpar_{(\hat{c}_1\hat{d}_1)\ldots (\hat{c}_l\hat{d}_l)}\dpar_{\hat{f}_1\ldots
\hat{f}_k}g_L(x_{12;4})\right)~.
\end{aligned}
\end{equation}

\subsection{The fuzzy reducible flag manifolds $F^F_{23;4}$, $F^F_{13;4}$ and $F^F_{123;4}$}

After the discussion of the fuzzy version of $F_{12;4}$, the corresponding constructions for the remaining reducible flag manifolds are quite straightforward. We first choose sets of oscillators, which in turn yield the generators for the algebra of functions:
\begin{equation}
\begin{aligned}
F_{13;4}: &~~ \hat{A}_3^{ijk}~,~~~\hat{A}_3^{ijk\dagger}~,~~~\hat{a}^{i}~,~~~\hat{a}^{i\dagger}~,\\
F_{23;4}: &~~ \hat{A}_3^{ijk}~,~~~\hat{A}_3^{ijk\dagger}~,~~~\hat{A}_2^{ij}~,~~~\hat{A}_2^{ij\dagger}~,\\
F_{123;4}: &~~ \hat{A}_3^{ijk}~,~~~\hat{A}_3^{ijk\dagger}~,~~~\hat{A}_2^{ij}~,~~~\hat{A}_2^{ij\dagger}
~,~~~\hat{a}^{i}~,~~~\hat{a}^{i\dagger}~.
\end{aligned}
\end{equation}
The underlying representations, on which these operators act are given by the Young diagrams
\begin{equation}
\begin{aligned}
F_{13;4}: &~~\overbrace{\overline{\tyng(8,4,4)}}^{L_1+L_2}~\otimes~\overbrace{\tyng(8,4,4)}^{L_1+L_2}
~\ =\ ~\overbrace{\tyng(8,4,4)}^{L_2+L_1}~\otimes~\overbrace{\tyng(8,4,4)}^{L_1+L_2}~,\\
F_{23;4}: &~~\overbrace{\overline{\tyng(8,8,4)}}^{L_1+L_2}~\otimes~\overbrace{\tyng(8,8,4)}^{L_1+L_2}
~\ =\ ~\overbrace{\tyng(8,4)}^{L_2+L_1}~\otimes~\overbrace{\tyng(8,8,4)}^{L_1+L_2}~,\\
F_{123;4}: &
\overbrace{\overline{\tyng(9,6,3)}}^{L_1+L_2+L_3}~\otimes~\overbrace{\tyng(9,6,3)}^{L_1+L_2+L_3}
~\ =\ ~\overbrace{\tyng(9,6,3)}^{L_3+L_2+L_1}~\otimes~\overbrace{\tyng(9,6,3)}^{L_1+L_2+L_3}~.
\end{aligned}
\end{equation}
From these diagrams, it is evident that $F_{23;4}$ is dual to $F_{12;4}$, and that the spaces $F_{13;4}$ and $F_{123;4}$ are dual to themselves.

Furthermore, the explicit action of the internal isotropy subgroups is easily constructed. For example, in the case $F_{23;4}$, the group acting nontrivial on the elementary oscillators but leaving invariant the states of the Fock space is $\sSU(2)\times \sU(1)$. Its action reads as 
\begin{equation}
L^r_\inner\ =\ \ad\left(\hat{a}^{i\dagger}_p\lambda^r_{pq}\hat{a}^i_q\right)\eand L^0_\inner\ =\ \ad\left(\hat{c}^{i\dagger} \hat{c}^i\right)~,
\end{equation}
where as before $\hat{a}^i_1=\hat{a}^i$, $\hat{a}^i_2=\hat{b}^i$ and $\lambda^r_{pq}$ are the Gell-Mann matrices of $\sSU(2)$. Evidently, the composite operators $A^{ij}_2$ and $A^{ij\dagger}_2$ and $A^{ijk}_3$ and $A^{ijk\dagger}_3$ are invariant under this combination.

The definition of a star product is performed analogously to the case of $F_{12;4}$. We decompose an operator $\hat{f}$ representing a function on a fuzzy flag manifold into
\begin{equation}
\hat{f}\ =\ \tilde{f}_{IJ}\hat{h}^I_{k_1}\otimes \hat{h}_{k_2}^J\eand
\hat{f}\ =\ \tilde{f}_{IJK}\hat{h}^I_3\otimes \hat{h}_2^J\otimes
\hat{h}_1^J
\end{equation}
for $F_{k_1k_2;4}$ and $F_{123;4}$, respectively. The map from operators to functions reads as
\begin{equation}
f(x_{k_2;4},x_{k_1(k_2);4})\ =\ \tilde{f}_{IJ}\tr(\rho^{L_1}(x_{k_2;4})
\hat{h}_{k_2}^I)\tr(\rho^{L_2}(x_{k_1(k_2);4}) \hat{h}_{k_1}^J)
\end{equation}
and
\begin{equation}
f(x_{3;4},x_{2(3);4},x_{1(23);4})\ =\
\tilde{f}_{IJK}\tr(\rho^{L_1}(x_{3;4})
\hat{h}_3^I)\tr(\rho^{L_2}(x_{2(3);4})
\hat{h}_2^J)\tr(\rho^{L_3}(x_{1(23);4}) \hat{h}_1^J)~,
\end{equation}
which naturally induces a star product via the usual formula. The explicit form of the star products are then obtained from the obvious sums over the differential operators
\begin{equation*}
\begin{aligned}
F_{13;4}: & \overleftarrow{\dpar}_{(\hat{a}_1\hat{b}_1\hat{c}_1)\ldots
(\hat{a}_l\hat{b}_l\hat{c}_l)}\overleftarrow{\dpar}_{\hat{g}_1\ldots
\hat{g}_k}J^{(\hat{a}_1\hat{b}_1\hat{c}_1),(\hat{d}_1\hat{e}_1\hat{f}_1);\hat{g}_1\hat{h}_1}_{13;4}\ldots\\&\hspace{2cm}
J^{(\hat{a}_l\hat{b}_l\hat{c}_l),(\hat{d}_l\hat{e}_l\hat{f}_l);\hat{g}_k\hat{h}_k}_{13;4}
\overrightarrow{\dpar}_{(\hat{d}_1\hat{e}_1\hat{f}_1)\ldots
(\hat{d}_l\hat{e}_l\hat{f}_l)}\overrightarrow{\dpar}_{\hat{h}_1\ldots
\hat{h}_k}~,\\
F_{23;4}: & \overleftarrow{\dpar}_{(\hat{a}_1\hat{b}_1\hat{c}_1)\ldots
(\hat{a}_l\hat{b}_l\hat{c}_l)}\overleftarrow{\dpar}_{(\hat{g}_1\hat{h}_1)\ldots
(\hat{g}_k\hat{h}_k)}J^{(\hat{a}_1\hat{b}_1\hat{c}_1),(\hat{d}_1\hat{e}_1\hat{f}_1);(\hat{g}_1\hat{h}_1),(\hat{m}_1\hat{n}_1)}_{23;4}\ldots\\&\hspace{2cm}
J^{(\hat{a}_l\hat{b}_l\hat{c}_l),(\hat{d}_l\hat{e}_l\hat{f}_l);(\hat{g}_k\hat{h}_k),(\hat{m}_k\hat{n}_k)}_{23;4}
\overrightarrow{\dpar}_{(\hat{d}_1\hat{e}_1\hat{f}_1)\ldots
(\hat{d}_l\hat{e}_l\hat{f}_l)}\overrightarrow{\dpar}_{(\hat{m}_1\hat{n}_1)\ldots
(\hat{m}_k\hat{n}_k)}~,\\
F_{123;4}: &\overleftarrow{\dpar}_{(\hat{a}_1\hat{b}_1\hat{c}_1)\ldots
(\hat{a}_l\hat{b}_l\hat{c}_l)}\overleftarrow{\dpar}_{(\hat{g}_1\hat{h}_1)\ldots
(\hat{g}_k\hat{h}_k)}\overleftarrow{\dpar}_{\hat{p}_1\ldots
\hat{p}_r}J^{(\hat{a}_1\hat{b}_1\hat{c}_1),(\hat{d}_1\hat{e}_1\hat{f}_1);(\hat{g}_1\hat{h}_1),(\hat{m}_1\hat{n}_1);\hat{p}_1\hat{q}_1}_{123;4}\ldots\\&\hspace{2cm}
J^{(\hat{a}_l\hat{b}_l\hat{c}_l),(\hat{d}_l\hat{e}_l\hat{f}_l);(\hat{g}_k\hat{h}_k),(\hat{m}_k\hat{n}_k);\hat{p}_r\hat{q}_r}_{123;4}
\overrightarrow{\dpar}_{(\hat{d}_1\hat{e}_1\hat{f}_1)\ldots
(\hat{d}_l\hat{e}_l\hat{f}_l)}\overrightarrow{\dpar}_{(\hat{m}_1\hat{n}_1)\ldots
(\hat{m}_k\hat{n}_k)}\overrightarrow{\dpar}_{\hat{q}_1\ldots \hat{q}_r}~.
\end{aligned}
\end{equation*}

\subsection{Continuous limits of the fuzzy flag manifolds}

As a consistency check, one can calculate the dimension of the continuous flag manifolds from considering the $L\rightarrow \infty$ limit of the various representations used in describing the fuzzy algebra of functions on them. The underlying idea is simply that given a cutoff $L$, the number of eigenvalues should be proportional to $L^d$ on a $d$-dimensional manifold. The number of degrees of freedom in the matrix algebras was the square of the dimension $d(a_1(L),a_2(L),a_3(L))$ of the representation $(a_1(L),a_2(L),a_3(L))$, and we can thus deduce that 
\begin{equation}
d=\lim_{L\rightarrow \infty}\frac{\ln(d(a_1(L),a_2(L),a_3(L)))}{\ln L}~.
\end{equation}
A trivial calculation shows that the representations and the matrix algebras we have chosen in the previous sections indeed reproduce the right dimensions for the various flag manifolds:
\begin{equation*}
\begin{aligned}
\begin{tabular}[t]{|l|c|c|c|c|c|c|c|}
\hline
Flag manifold & $F_{1;4}$ & $F_{2;4}$ & $F_{3;4}$ & $F_{12;4}$ & $F_{13;4}$ & $F_{23;4}$ & $F_{123;4}$ \\
\hline
representations & $(L,0,0)$ & $(0,L,0)$ & $(0,0,L)$ & $(L,L,0)$ & $(L,0,L)$ & $(0,L,L)$ & $(L,L,L)$\\
real dimension & $6$ & $8$ & $6$ & $10$ & $10$ & $10$ & $12$\\
\hline
\end{tabular}
\end{aligned}
\end{equation*}

Considering the expressions for the star products on the various flag manifolds found in the previous sections, it is also evident that the star or operator product will go over to the commutative product in the limit $L\rightarrow \infty$. As discussed in \cite{Kurkcuoglu:2006iw}, this limit is, however, not clearly observable in the simplified formul\ae{} using the complex coordinates $a^i,b^i,c^i$. Furthermore, the derivatives in the fuzzy case containing the geometric information are evidently approaching the derivatives in the continuum for $L\rightarrow \infty$.

\section{Super Pl{\"u}cker embeddings and flag supermanifolds}

\subsection{Flag supermanifolds of $\sU(4|n)$}

Flag supermanifolds can be defined analogously to bosonic flag manifolds by considering the supervector space $\FC^{m|n}$ \cite{Howe:1995md}; see also appendix A for more details. A {\em superflag} is a sequence of superspaces $V_{D_1}\subsetneq\ldots \subsetneq V_{D_k}\subset\FC^{m|n}$ such that $\dim_\FC V_D=D=d|\delta$. Note that inclusion requires that $d_i\leq d_j$ and $\delta_i\leq \delta_j$ for $i<j$ with at least one inequality being strict. A {\em flag supermanifold} $F_{D_1\ldots D_k;m|n}$ is correspondingly the set of all superflags $f_{D_1\ldots D_k;m|n}$. One can again write a flag supermanifold as a coset space:
\begin{equation}\label{definitionflagsuper}
F_{D_1\ldots
D_k;m|n}\ =\ \sU(m|n)/(\sU(m-d_k|n-\delta_k)\times\ldots \times\sU(d_1|\delta_1))~.
\end{equation}
The special unitary supergroups $\sSU(m|n)$ are not useful here, as for $m=n$, one has to exclude the identity matrix, which is a central element in this case, from the set of generators of $\asu(m|n)$. See appendix A for more details on this point.

As before, we will be interested in the flag manifolds arising naturally in the double fibration underlying well-known supertwistor correspondences. These spaces are flags in the superspace $\FC^{4|n}$, and because there are again natural projections, they fit into the following diagram:
\begin{equation}\label{dblfibrationsflagsuper}
\begin{aligned}
\begin{picture}(80,95)
\put(40.0,0.0){\makebox(0,0)[c]{$F_{(1|0);(4|n)}$}}
\put(40.0,40.0){\makebox(0,0)[c]{$F_{(1|0)(2|0)(2|n)(3|n);(4|n)}$}}
\put(40.0,83.0){\makebox(0,0)[c]{$F_{(2|0)(2|n)(3|n);(4|n)}$}}
\put(-40.0,60.0){\makebox(0,0)[c]{$F_{(2|0)(2|n);(4|n)}$}}
\put(-40.0,20.0){\makebox(0,0)[c]{$F_{(1|0)(2|0)(2|n);(4|n)}$}}
\put(120.0,60.0){\makebox(0,0)[c]{$F_{(3|n);(4|n)}$}}
\put(120,20.0){\makebox(0,0)[c]{$F_{(1|0)(3|n);(4|n)}$}}
\put(40.0,50.0){\vector(0,1){20}}
\put(40.0,30.0){\vector(0,-1){20}}
\put(50.0,48.0){\vector(4,1){32}}
\put(30.0,48.0){\vector(-4,1){32}}
\put(50.0,32.0){\vector(4,-1){32}}
\put(30.0,32.0){\vector(-4,-1){32}}
\put(-40.0,30.0){\vector(0,1){20}}
\put(120.0,30.0){\vector(0,1){20}}
\put(50.0,48.0){\vector(4,1){32}}
\put(30.0,48.0){\vector(-4,1){32}}
\put(50.0,32.0){\vector(4,-1){32}}
\put(25.0,75.0){\vector(-4,-1){32}}
\put(55.0,75.0){\vector(4,-1){32}}
\put(110.0,10.0){\vector(-4,-1){32}}
\put(-30.0,10.0){\vector(4,-1){32}}
\end{picture}
\end{aligned}
\end{equation}
Here, $F_{(1|0);(4|n)}$ is the superspace $\CPP^{3|n}$ and $F_{(2|0)(2|n);(4|n)}$ is the conformal compactification of super Minkowski space with $n=\CN$ being the number of supersymmetries. Note that $F_{(2|0)(2|n);(4|n)}$ contains the left chiral superspace $F_{(2|0);(4|n)}$ as well as the right chiral superspace $F_{(2|n);(4|n)}$ \cite{Howe:1995md}. Since in the twistor correspondences involving $F_{(1|0);(4|n)}$ and $F_{(3|n);(4|n)}$, only these chiral subspaces play a r{\^o}le, we will also restrict the correspondence spaces and only consider $F_{(1|0)(2|0);(4|n)}$ and $F_{(2|n)(3|n);(4|n)}$ instead of $F_{(1|0)(2|0)(2|n);(4|n)}$ and $F_{(2|0)(2|n)(3|n);(4|n)}$.

To get a reliable handle on the geometry of the flag supermanifolds, it is useful to introduce local coordinates. For simplicity, we will first consider the ordinary Gra{\ss}mannians and then discuss the super case. The extension to reducible flag supermanifolds will be straightforward. On a Gra{\ss}mannian $G_{k;n}$, a patch corresponds to a subset $I\subset\{1,\ldots ,n\}$ with $k$ elements, which selects $k$ columns of a $k\times n$ matrix $Z$. We identify these columns with the columns of a $k\times k$ unit matrix; they fix parts of the vectors spanning the $k$-dimensional vector subspaces. The remaining columns are filled by the local coordinates on the patch $I$. As an example, consider $\CPP^2=G_{1;3}$, where there are three patches
\begin{equation}
Z_1\ =\ (1~z^1_1~z^2_1)~,~~~Z_2\ =\ (z^1_2~1~z^2_2)\eand
Z_3\ =\ (z^1_3~z^2_3~1)~.
\end{equation}
Transition functions are elements of a finite subgroup of $\sGL(n,\FC)$, permuting the $k\times k$ unit matrix to different columns. We also see that on every patch we have the $\sU(k)\times \sU(n-k)\subset \sU(n)$ invariance manifest.

For a super Gra{\ss}mannian $G_{k|\kappa;n|\nu}$, we consider a $(k|\kappa)\times(k+(n-k)|\kappa+(\nu-\kappa))$-dimensional supermatrix $Z_I$, into which we insert the columns of the $k|\kappa\times k|\kappa$-dimensional unit matrix, preserving the grading of the matrix. The set $I$ of the columns which we selected corresponds again to a patch. The discussion of the transition functions is the same as in the bosonic case and the symmetries factored out are again manifest.

A first observation is that for the super Gra{\ss}mannians we are interested in, the fermionic dimension of the subspace is always either maximal or minimal. This leads e.g.\ to matrices $Z_I$
\begin{equation}
\begin{aligned}
G_{2|0;4|4}:~~~&Z_{34|\cdot}\ =\ \left(\begin{array}{cc|c} x_{2\times
2} & \unit_{2\times 2} & \xi_{2\times 4}\\\hline
\end{array}\right)~,\\
G_{2|0;2|4}:~~~&Z_{12|\cdot}\ =\ \left(\begin{array}{c|c}
\unit_{2\times 2} & \xi_{2\times 4}\\\hline
\end{array}\right)~,\\
G_{2|4;4|4}:~~~&Z_{34|1234}\ =\ \left(\begin{array}{cc|c} x_{2\times 2}
&
\unit_{2\times 2} & 0_{2\times 4}\\
\hline
\xi_{4\times 2} & 0_{4\times 2} & \unit_{4\times 4}
\end{array}\right)~,
\end{aligned}
\end{equation}
where the lines indicate the boundaries of the four canonical blocks in the supermatrix $Z$. The first and the third space are the compactified, complexified chiral and anti-chiral $\CN=4$ superspaces, respectively. It is easy to convince oneself that (the bosonic part of) the transition functions are the same as in the purely bosonic case. Therefore, the super Gra{\ss}mannian (and also the flag supermanifolds) we are dealing with are simply certain fermionic vector bundles over their bodies, i.e.\ the embedded ordinary flag manifolds. As an explicit example, consider the space $\CPP^{1|2}$. Following our discussion of local coordinates, we introduce the two patches $U_\pm$, corresponding to the matrices
\begin{equation}
Z_+\ =\ (z_+~1~\zeta^1_+~\zeta^2_+)\eand Z_-\ =\ (1~z_-~\zeta^1_-~\zeta^2_-)~.
\end{equation}
The transition function between both patches is evidently $f_{+-}=z_+$ and therefore we have $\zeta^{1,2}_+=z_+\zeta^{1,2}_-$. Thus, the space $\CPP^{1|2}$ is the total space of the rank $0|2$ vector bundle
\begin{equation}
\Pi\CO(1)\oplus\Pi\CO(1)\ \rightarrow\  \CPP^1~.
\end{equation}
For more details on the definition of super Gra{\ss}mannians and flag supermanifolds, see \cite{Manin:1988ds}.

For simplicity, we restrict our attention to the case $n=\CN=4$ in the following. Using the description of super Gra{\ss}mannians given above, it is easy to determine the dimensions of the various flag supermanifolds in \eqref{dblfibrationsflagsuper} for $n=4$. We have
\begin{equation}
\begin{aligned}
F_{(1|0);(4|4)} : 3|4 ~~~~~&
F_{(2|0);(4|4)} : 4|8\\
F_{(2|4);(4|4)} : 4|8~~~~~&
F_{(2|0)(2|4);(4|4)} : 4|16\\
F_{(1|0)(2|0);(4|4)} : 5|8 ~~~~~&
F_{(3|4);(4|4)} : 3|4 \\
F_{(2|4)(3|4);(4|4)} : 5|8 ~~~~~&
F_{(1|0)(3|4);(4|4)} : 5|16 \\
F_{(1|0)(2|0)(2|4)(3|4);(4|4)} : 6|16~.\,~~~&
\end{aligned}
\end{equation}
The minimal numbers of patches covering the flag supermanifolds are the same as the ones for their bodies.

\subsection{The Pl{\"u}cker and super Pl{\"u}cker embeddings}\label{secPlucker}

Before discussing the super variant of the Pl{\"u}cker embedding, let us briefly recall its common form; see also the review article \cite{Kleiman:1972}. A Gra{\ss}mannian $G_{k;n}$ is the space of all $k$-dimensional vector subspaces in $V=\FC^n$. Each such space is spanned by a basis $f_1,\ldots ,f_k$ and we can combine this basis into an element of the $k$-th exterior power of $V$:
\begin{equation}\label{decomposed}
 A\ :=\ f_1\wedge\ldots\wedge f_k~,~~~A\in\Lambda^kV~.
\end{equation}
Each element of this form describes a point on $G_{k;n}$, however, not every element of $\Lambda^kV$ is of this form. In particular, a sum of two different such elements will not in general be decomposable into a single wedge product of $k$ vectors. One thus needs additional conditions to decide, whether an element of $\Lambda^kV$ is fully decomposable. Physically, this question is completely analogous to the question, whether a $k$-fermion state can be decomposed into a product of $k$ single particle states.

It is well known that the necessary and sufficient condition for $A\in\Lambda^kV$ to be of the form \eqref{decomposed} is
\begin{equation}\label{pluckercond}
 (B\lrcorner\, A)\wedge A\ =\ 0~~~\mbox{for all }B\in\Lambda^{k-1}V\dual~,
\end{equation}
where $V\dual$ denotes the space dual to $V$. The proof of this statement is a simplification of the one in the graded case, which we will give below. The equations arising from \eqref{pluckercond} are called the {\em Pl{\"u}cker relations}. In the case $k=2$, this condition simplifies to $(B\lrcorner\, A)\wedge A=\frac{1}{2}B\lrcorner\,(A\wedge A)$ and thus to $A\wedge A=\eps_{ijkl}A^{ij}A^{kl}=0$, the condition we used in section \ref{secGrassmann} 

Consider now a basis $(e_1,\ldots ,e_n)$ of $V$ with a dual basis $(e^1_\vee,\ldots ,e^n_\vee)$ of $V^*$, $\langle e_i,e^j_\vee\rangle=\delta_i^j$. It is obviously sufficient to consider \eqref{pluckercond} only for elements $B$ of the form $e^{i_1}_\vee\wedge\ldots \wedge e^{i_k}_\vee$. Writing $A=A^{i_1\ldots i_k}e_{i_1}\wedge\ldots \wedge e_{i_{k-1}}$, \eqref{pluckercond} reduces to
\begin{equation}\label{pluckerred}
 \sum_{t=0}^k(-1)^tA^{j_1\ldots j_{k-1}i_t}A^{i_1\ldots \widehat{i_t}\ldots i_{k+1}}\ =\ A^{j_1\ldots j_{k-1}[i_1}A^{i_2\ldots i_{k+1}]}\ =\ 0~,
\end{equation}
where $\hat{\cdot}$ indicates an omission as usual.

Note that not all of these equations are independent, but one can easily read off the number of independent ones. As the Pl{\"u}cker relations are evidently projective, one can fix a nonvanishing component to unity, e.g.\ $A^{p_1\ldots p_k}=1$. Then it follows that one can solve for all coordinates with $m\geq 2$ indices different from all of the $p_r$ \cite{Kleiman:1972}: Consider a sequence $q_1\ldots q_k$ of indices, $m$ of which are not in the sequence $p_1\ldots p_k$. From \eqref{pluckerred}, we obtain for $(j_k)=(q_1\ldots \widehat{q_r}\ldots q_k)$ and $(i_k)=(q_r p_1\ldots p_k)$ the following equation:
\begin{equation}
A^{q_1\ldots \widehat{q_r}\ldots q_kq_r}A^{p_1\ldots p_k}\ =\ \sum_{t=1}^k(-1)^tA^{q_1\ldots \widehat{q_r}\ldots q_kp_t}A^{p_1\ldots \widehat{p_t}\ldots p_k}~.
\end{equation}
If $p_r$ is contained in the sequence $q_1\ldots q_k$, the right hand side vanishes, otherwise exactly $m-1$ of the elements in the sequence $q_1\ldots \widehat{q_r}\ldots q_kp_t$ are not in the sequence $p_1\ldots p_k$. Iterating this prescription, we find that all Pl{\"u}cker coordinates can be expressed in terms of $A^{q_1\ldots q_k}$ with at most one $q_s$ not in the sequence $p_1\ldots p_k$. That is, of the $\binom{n}{k}$ coordinates on $\Lambda^k\FC^n$, only $1+k(n-k)$ are relevant, and this is the number of (homogeneous) coordinates on the Gra{\ss}mannian $G_{k;n}$.

For the discussion of flag supermanifolds, we need a similar picture at hands, and we will find that for Gra{\ss}mannians $G_{k|\kappa;n|\nu}$ for which $\kappa\in\{0,\nu\}$, the Pl{\"u}cker embedding can be straightforwardly extended. This fact is mentioned in \cite{Manin:1988ds}, but beyond this, we are not aware of any explicit discussion of super Pl{\"u}cker embeddings in the literature.

The Gra{\ss}mannian $G_{k|\kappa;n|\nu}$ consists of spaces spanned by $k$ even and $\kappa$ odd supervectors in $\bfV=\FC^{n|\nu}$. Given a basis $\bfe^I=(e^i,\eps^\alpha)$, we can write a point\footnote{A discussion of wedge products of supervector spaces and their duals is found in \cite{Cartier:0202026}.} $\bfA\in\Lambda^{k|\kappa}\FC^{n|\nu}\subset G_{k|\kappa;n|\nu}$ as
\begin{equation}
\bfA\ =\ \bfA^{I_1\ldots I_k\Upsilon_1\ldots \Upsilon_\kappa}\bfe^{I_1}\wedge\ldots \wedge \bfe^{I_k}\wedge \bfe^{\Upsilon_1}\wedge\ldots \wedge \bfe^{\Upsilon_\kappa}~, 
\end{equation}
where $I_r$ and $\Upsilon_\rho$ range each from $1$ to $n+\nu$. The Pl{\"u}cker relations we are looking for are supposed to be the necessary and sufficient conditions that 
\begin{equation}
\bfA\ =\ \bff_1\wedge\ldots \wedge \bff_k\wedge \bfphi_1\wedge\ldots \wedge \bfphi_\kappa~,
\end{equation}
where $\bff_\alpha$ and $\bfphi_\alpha$ are linearly independent\footnote{See appendix A for a discussion of linear independence of supervectors.} even and odd supervectors, respectively. We now claim that for $\kappa\in\{0,\nu\}$ (and these are the only cases we are interested in), the necessary and sufficient condition reads as
\begin{equation}\label{superPlucker}
 (\bfB\lrcorner\, \bfA)\wedge \bfA\ =\ 0~~~\mbox{for any }\bfB\in\Lambda^{k-1|\kappa}\bfV\dual~.
\end{equation}
First, note that an even supervector $\bfv$ with nonvanishing body $\bfv^\circ$ divides an element $\bfA$ of $\Lambda^{k-1|\kappa}\bfV\dual$ with $\bfA^\circ\neq 0$, i.e.\ $\bfA=\bfv\wedge \bfw$ for some $\bfw\in\Lambda^{k-2|\kappa}$ with $\bfw^\circ\neq 0$, if and only if $\bfv\wedge \bfA=0$. Let us assume that $\kappa=0$. If $\bfA$ is of the form $\bff_1\wedge\ldots \wedge \bff_k$, then we can complete the $\bff_\alpha$ to an (orthonormal) basis of $\bfV$ by even supervectors $\bff_{k+1}\ldots \bff_{n+\nu}$. If $\bfB$ is composed only of $\bff_\alpha\dual$ with $\alpha\leq k$, \eqref{superPlucker} is satisfied. If there is one or more of the $\bff_\alpha$ with $\alpha>k$, \eqref{superPlucker} is also true. Since \eqref{superPlucker} is linear in $\bfB$ and the span of the cases we discussed comprises all of $\Lambda^{k-1|\kappa}\bfV\dual$, \eqref{superPlucker} is true in general.

To prove the remaining direction, we follow the proof in the bosonic case \cite{Kleiman:1972} and explicitly construct the Gra{\ss}mannian $G_{k|\kappa;n|\nu}$ from the coordinates $\bfA^{I_1\ldots I_k}$. First, note that we can again fix a bosonic coordinate which has nonvanishing body (at least one such component exists, if $\bfA$ has nonvanishing body), say $\bfA^{i_1\ldots i_k}=1$, since the equations are projective. Furthermore, all coordinates $\bfA^{J_1\ldots J_k}$ with a sequence $J_1\ldots J_k$ of more than one index different from $i_1\ldots i_k$ can again be written in terms of the remaining coordinates; the proof is the same as in the bosonic case. We now construct $k$ vectors spanning a $k$-plane in $\Lambda^{k|0}\FC^{n|\nu}$ by putting $p_m(J)=A^{i_1\ldots i_{m-1}Ji_{m+1}\ldots i_k}$, $m=1\ldots k$, $J=1\ldots n+\nu$. This evidently yields $k$ linearly independent vectors with nonvanishing bodies, as they differ in the $k$ components $J=i_l$: it is $p_m(i_l)=0$ for $m\neq l$ and $p_l(i_l)=1$. It remains to show that the plane corresponding to these vectors is indeed compatible with all the Pl{\"u}cker coordinates. First, it is straightforward to see that the components $A^{i_1\ldots i_{m-1}Ji_{m+1}\ldots i_k}$ all are compatible with our definition. As shown above, the remaining coordinates are derived from these and thus all the Pl{\"u}cker coordinates are the ones corresponding to the Gra{\ss}mannian we constructed. Altogether, if the equation is satisfied, then the multivector $\bfA$ describes a Gra{\ss}mannian, which completes our proof for $\kappa=0$.

For $\kappa=\nu$, it is sufficient to note that the wedge product of the $\nu$ odd supervectors in $\bfA$ either vanishes, if they are linearly dependent, or spans all of the odd supspace. Therefore, it suffices again to focus on the even supervectors in $\bfA$, which is done by contracting with a dual multivector $\bfB\in\Lambda^{k-1|\nu}\bfV\dual$, and there is nothing left to prove.

\subsection{Pl{\"u}cker coordinates and projector description of irreducible flag supermanifolds}\label{SecTensorSuperflags}

From the discussion in section 5.1., the description of $\CPP^{3|4}$ is evident: homogeneous coordinates on $\CPP^{3|4}$ are provided by the components of an even supervector (normalized and in a representation of type I, i.e.\ in a pure even basis) $\bfa^I=(a^i,\eta^\alpha)$. From the discussion above, we know that the space $\CPP^{3|4}$ is the fermionic rank 4 vector bundle 
\begin{equation}
 \FC^4\otimes \Pi\CO(1)\ \rightarrow\  \CPP^3~,
\end{equation}
and its sections are given by homogeneous polynomials of degree one in the coordinates on $\CPP^3$. We can therefore rewrite the fermionic components of the even supervector $\bfa^I$ as
\begin{equation}
\eta^\alpha\ =\ \eta^\alpha_i a^i\ewith \eta^\alpha_i\in\FC^{0|8}~.
\end{equation}
In the following, however, we will not be interested in sections of this bundle but rather in the algebra of functions on its total space.

We therefore continue along the lines of the bosonic case and construct a projector using the supervector $\bfa$ according to
\begin{equation}
\CP_{1|0}\ =\ \bfa\bar{\bfa}^T~.
\end{equation}

Note that the description of a flag manifold using projectors trivially generalizes to the supercase. First of all, the bodies of even and odd supervectors of dimension $4|4$ have non-zero components in the first and the last four components, respectively. This property is preserved by the action of $\sU(4|4)$. Thus, given a projector
\begin{equation}
\CP\ =\ \left(\begin{array}{cc} \CP_A & \CP_B \\ \CP_C & \CP_D
\end{array}\right)~,
\end{equation}
we can read off the dimension $k|\kappa$ of the subspace of $\FC^{4|4}$ onto which it projects to be $\tr(\CP_A^\circ)|\tr(\CP_D^\circ)$, where $\cdot^\circ$ denotes the body of the projector. We will define the rank of such a projector to be $k|\kappa$. 

Also, there is a subgroup $\sU(4-k|4-\kappa)$ of $\sU(4|4)$, which leaves invariant a projector of rank $k|\kappa$. This is easily seen by the usual identification of $\au(4|4)$ with $\au(4+4)$ after combining the odd generators with an odd parameter and the fact that the rank of a projector is left invariant by the action of $\sU(4|4)$.

For $\CP_{1|0}$ to be a projector, we have to demand that
\begin{equation}
(\bfa,\bfa)\ =\ \bar{\bfa}\bfa\ =\ \bar{a}^ia^i+\di\etab^\alpha\eta^\alpha\ =\ 1~,
\end{equation}
see appendix A for a definition of the scalar product of complex supervectors. This condition can also be understood from the construction of $\CPP^{n|n+1}$ via
\begin{equation}
\FC^{n+1|n+1}\ \rightarrow\  S^{2n+1|2n+2}\ \rightarrow\  \CPP^{n|n+1}~,
\end{equation}
as the first projection, see also \cite{Klimcik:1999pr}. We can in fact introduce coordinates $\bfx^{\hat{A}}=\bar{\bfa}^I\bflambda^{\hat{A}}_{IJ}\bfa^J$ in the superspace $\FR^{32|32}$, where the $\bflambda^A$ are the generators of $\sU(4|4)$ as described in the appendix. The coordinates $\bfx^{\hat{A}}$ describe an embedding of $\CPP^{3|4}$ in $\FR^{32|32}$ and due to $\bflambda^{\hat{A}}_{IJ}\bflambda^{\hat{B}}_{KL}g_{\hat{A}\hat{B}}=\delta_{IL}\delta_{JK}$, we have
\begin{equation}
\CP_{1|0}\ =\ \bfx^{\hat{A}}\bflambda^{\hat{B}}g_{\hat{A}\hat{B}}\ =\ \bfa \bar{\bfa}^T~.
\end{equation}
Underlying this construction is again a generalized Hopf fibration
\begin{equation}
1\longrightarrow\sU(1)\ \longrightarrow\  S^{2n+1|2n+2}\ \longrightarrow\ 
\CPP^{n|n+1}\ \longrightarrow\  1~.
\end{equation}

Let us now turn to the Gra{\ss}mannians, whose description is rather straightforward using the super Pl{\"u}cker embedding. We start with $G_{2|0;4|4}$, and the Pl{\"u}cker coordinates are given by 
\begin{equation}
\bfA_{2|0}^{\lsc IJ\rsc}\ =\ \bfa^{\lsc I}\bfb^{J\rsc}\ =\ \bfa^I\bfb^J-(-1)^{\tilde{I}\tilde{J}}\bfa^J\bfb^I\ =\ \bfa^I\bfb^J- \bfb^I\bfa^J~,
\end{equation}
where $\bfa=(a,\eta)$ and $\bfb=(b,\zeta)$ are two even supervectors, $\lsc \cdot,\cdot\rsc$ is the supercommutator and $\tilde{I}$ denotes the parity of the corresponding index. That is, $\tilde{i}=0$ and $\tilde{\alpha}=1$, plus the parity of the supervector under consideration, modulo 2. In more detail, we have
\begin{equation}
\bfA_{2|0}^{ij}\ =\ a^{[i}b^{j]}~,~~~
\bfA_{2|0}^{i\alpha}\ =\ -\bfA_{2|0}^{\alpha
i}\ =\ \tfrac{1}{2}\left(a^i\zeta^\alpha-\eta^\alpha b^i\right)\eand
\bfA_{2|0}^{\alpha\beta}\ =\ \eta^{\{\alpha}\zeta^{\beta\}}~,
\end{equation}
where $\{\cdot\}$ denotes symmetrization, in particular of Gra{\ss}mann-odd quantities. Note that this super-antisymmetrized combination of $\bfa$ and $\bfb$ indeed eliminates all components of $\bfb$ parallel to $\bfa$ and in the following, we will assume that $\bfb$ and $\bfa$ are perpendicular:
\begin{equation}
\bar{\bfa}^I\bfb^I\ =\ \bar{a}^ib^i+\di\bar{\zeta}^\alpha\eta^\alpha\ =\ 0~.
\end{equation}
Together with $\bfa^\circ\neq 0\neq \bfb^\circ$, this equation implies that both supervectors are linearly independent. The (internal) stabilizer subgroup of $\sU(4|4)$ leaving $\bfA_{2|0}^{IJ}$ invariant is $\sU(2|0)$, which rotates the supervector $\bfa$ into $\bfb$ and vice versa; see the discussion of the fuzzy case for more details.

We observed before that the Pl{\"u}cker coordinates $A^{ij}_2$ on $G_{2;4}$ contain some redundancy: first there is a scaling, which renders them effectively coordinates on $\CPP^5$, and second, there is the identity $\eps_{ijkl}A^{ij}_2A^{kl}_2=0$. In the present case, the redundancy is somewhat larger, but from the discussion of the super Pl{\"u}cker embedding, we can be specific about the number of redundant coordinates. First of all, we have $16$ even and $16$ odd homogeneous coordinates and thus we are using a Pl{\"u}cker embedding into $\CPP^{15|16}$. Assuming that $\bfA_{2|0}^{12}=1$ fixes the scaling, only the coordinates
\begin{equation}
\begin{array}{cccccc}
\bfA_{2|0}^{13} & \bfA_{2|0}^{14} & \hspace{1cm} & \bfA_{2|0}^{15} &\ldots & \bfA_{2|0}^{18}\\
\bfA_{2|0}^{23} & \bfA_{2|0}^{24} & & \bfA_{2|0}^{25} &\ldots & \bfA_{2|0}^{28}
\end{array}
\end{equation}
are independent and we thus arrive at a $4|8$-dimensional space, $G_{2|0;4|4}$.

From the bi-supervector $\bfA_{2|0}$, one can again construct a projector:
\begin{equation}
(\CCP_{2|0;4|4})^{IJ;KL}\ =\ \bfA^{IJ}_{2|0}\bar{\bfA}_{2|0}^{KL}
\end{equation}
and with our choice and the orthonormality of $\bfa$ and $\bfb$, $\CCP_{2|0;4|4}$ satisfies indeed $(\CCP_{2|0;4|4})^2=\CCP_{2|0;4|4}$. Underlying the construction of this projector is again a generalized Hopf fibration, which is evidently a superextension of the one for $G_{2;4}$:
\begin{equation}
1 \ \longrightarrow\  \sU(2|0) \ \longrightarrow\  S^{7|8}\times S^{5|8} \ \longrightarrow\  F_{2|0;4|4}\ \longrightarrow\  1~.
\end{equation}
It is now necessary to introduce a super-antisymmetrized tensor product defined as
\begin{equation}\label{defsuperantisym}
\begin{aligned}
(\bfA\Cap \bfB)_{IJ;KL}\ :=\
\tfrac{1}{4}\big(\bfA_{IK}\bfB_{JL}&-(-1)^{\tilde{I}\tilde{J}}\bfA_{JK}\bfB_{IL}-\\
&(-1)^{\tilde{K}\tilde{L}}\bfA_{IL}\bfB_{JK}+(-1)^{\tilde{I}\tilde{J}}(-1)^{\tilde{K}\tilde{L}}\bfA_{JL}\bfB_{IK}\big)~.
\end{aligned}
\end{equation}
As in the case of the antisymmetric product of the generators of $\sU(4)$, we also have here various Fierz identities, see appendix A. Using one of these identities, we can introduce the projector
\begin{equation}
\begin{aligned}
(\CCP_{2|0;4|4})_{IJ;KL}&\ =\ \bar{\bfA}^{KL}_{2|0}\bfA^{IJ}_{2|0}\\
&\ =\ \bar{\bfA}^{K'L'}_{2|0}(\bflambda^{\hat{A}}\Cap\bflambda^{\hat{B}})_{K'L'I'J'}\bfA^{I'J'}_{2|0}
(\bflambda^{\hat{C}}\Cap\bflambda^{\hat{D}})_{IJ;KL}~g_{AC}~ g_{BD}\\
&\ =:\ \mathbf{x}^{\hat{A}\hat{B}}(\bflambda^{\hat{C}}\Cap\bflambda^{\hat{D}})_{IJ;KL}~g_{AC}~g_{BD}~.
\end{aligned}
\end{equation}

The next Gra{\ss}mannian to be described is the space $G_{2|4;4|4}$. This space is certainly ``dual'' to $G_{2|0;4|4}$, as one easily guesses from the description in terms of coordinate matrices $Z_I$ given in the previous section. To describe this space, one needs to take two even and four odd normalized supervectors and then super-antisymmetrize them. A short remark is in order to show that super-antisymmetrizing indeed yields a projection on their mutually orthogonal components. We saw above that this is true for two even supervectors. Since the bodies of normalized even and odd supervectors are non-vanishing exactly in the even and odd indices, respectively, a pair of an even and an odd normalized supervector are always linearly independent; antisupersymmetrization just eliminates redundancies in the description of $1|1$-dimensional subspaces via Pl{\"u}cker relations. Given two odd supervectors, super-antisymmetrization has the same effect on the even and odd components as it had on two even supervectors and thus projects out non-orthogonal components.

Altogether, we have Pl{\"u}cker coordinates
\begin{equation}
 \bfA^{IJ\Upsilon_1\ldots \Upsilon_4}_{2|4}\ =\ \bfa^{\lsc I}\bfb^J\bfeta^{\Upsilon_1}\ldots \bfeta^{\Upsilon_4\rsc}~,
\end{equation}
where $\bfa^I,\bfb^I$ and $\bfeta^\Upsilon$ are even and odd supervectors, respectively. We assume that $\bfA^{1245678}_{2|4}=1$ to fix the scale. From the discussion of the super Pl{\"u}cker embedding, it then follows that the independent coordinates here are given by
\begin{equation}
\begin{array}{cccccc}
\bfA_{2|4}^{135678} & \bfA_{2|4}^{145678} & \hspace{1cm} & \bfA_{2|4}^{155678} &\ldots & \bfA_{2|4}^{185678}\\
\bfA_{2|4}^{235678} & \bfA_{2|4}^{245678} & & \bfA_{2|4}^{255678} &\ldots & \bfA_{2|4}^{285678}
\end{array}~.
\end{equation}

In the case of the dual complex projective superspace $\CPP^{3|4}_*$, we add to this picture another even supervector $\bfc^I$ and obtain the Pl{\"u}cker coordinates
\begin{equation}
 \bfA^{IJK\Upsilon_1\ldots \Upsilon_4}_{2|4}\ =\ \bfa^{\lsc I}\bfb^J\bfc^K\bfeta^{\Upsilon_1}\ldots \bfeta^{\Upsilon_4\rsc}~,
\end{equation}
where $\bfa^I,\bfb^I$ and $\bfeta^\Upsilon$ are even and odd supervectors, respectively. We fix the scale by $\bfA^{12345678}_{3|4}=1$ and the remaining independent coordinates here are
\begin{equation}
\begin{array}{ccccccc}
\bfA_{3|4}^{1245678} & \bfA_{3|4}^{1345678} & \bfA_{3|4}^{2345678}& \hspace{1cm} & \bfA_{3|4}^{155678} &\ldots & \bfA_{2|4}^{185678}
\end{array}~.
\end{equation}
We refrain from going into any further detail at this point and refer to the discussion of the fuzzy pictures of these super Gra{\ss}mannians.

\subsection{The reducible flag supermanifolds}

As in the bosonic case, the description of the reducible flag supermanifolds is merely a combination of the underlying ``elementary'' Gra{\ss}mannians.

The complexified, compactified super Minkowski space $F_{(2|0)(2|4);(4|4)}$ is a reducible flag supermanifold, although its body is an irreducible flag manifold. Combining the sets of Pl{\"u}cker coordinates on $G_{2|0;4|4}$ and $G_{2|4;4|4}$ and factoring out redundancies arising from the fact that the even supervectors in both cases are the same, one arrives at the Pl{\"u}cker coordinates on the space $F_{(2|0)(2|4);4|4}$, which is of superdimension $4|16$.

Similarly, the remaining flag supermanifolds are constructed by combining the coordinates of the Gra{\ss}mannians corresponding to the various subflags. Again, a more detailed discussion will be presented, when we develop the fuzzy versions of these spaces.

\subsection{Geometric structures on the flag supermanifolds}

The description of the flag supermanifolds in terms of projectors allows us to proceed similarly to the case of ordinary flag manifolds in the description of their geometry. That is, we describe a point on a flag supermanifold $\bfM$ again by a projector $\CP_0$, and the action of $\bfG=\sU(4|4)$ on this point produces all of $\bfM$. The coordinates $\bfx^A$ provide an embedding of $\bfM$ in $\FR^{m|n}$ (e.g.\ $\CPP^{3|4}$ is embedded in $\FR^{32|32}$). The tangent vectors at $\CP_0$ are naturally obtained from an appropriate action $R(\Lambda)$ on $\CP_0$ and we define analogously to the bosonic case
\begin{equation}
T_{\CP_0}\bfM\ =\ \{ R(\Lambda)\CP_0|\Lambda\in\au(4|4)\}~.
\end{equation}
The generators of the subgroup $\bfH$ in $\bfM=\bfG/\bfH$ will leave $\CP_0$ invariant, and thus $T_{\CP_0}\bfM$ is of the same dimension as $\bfM$. In the case of $\CPP^{3|4}$, $R(\Lambda)$ is the superadjoint action, and since $\CP_0$ is an even supermatrix, it is thus clear that a vector $\bfV\in T_{\CP_0}\bfM$ satisfies
\begin{equation}
\{\CP_0,\bfV\}\ =\ \bfV\eand \str \bfV\ =\ 0~.
\end{equation}
From this point, it is rather obvious that we can proceed with the definition of the complex structure, the metric, as well as the K{\"a}hler structure exactly as in the bosonic case. We thus go over to an arbitrary projector $\CP$ describing any point on $\bfM$ and define the complex structure as
\begin{equation}
I\bfV\ :=\ -\di\lsc\CP,\bfV\rsc\ =\ -\di[\CP,\bfV]~,
\end{equation}
which trivially satisfies $I^2=-\unit$ and the hermitian supermetric as
\begin{equation}
g(\bfV_1,\bfV_2)\ =\ -\str(I\bfV_1I\bfV_2)~,
\end{equation}
which satisfies $g(I\bfV_1,I\bfV_2)=g(\bfV_1,\bfV_2)$, is invariant under the action of $\sU(4|4)$ on the vectors $\bfV_1$ and $\bfV_2$ and defines a supersymmetric even tensor: $g(\bfV_1,\bfV_2)=(-1)^{\tilde{\bfV}_1\tilde{\bfV}_2}g(\bfV_2,\bfV_1)$. Evidently, there is the supersymplectic structure
\begin{equation}
\Omega(\bfV_1,\bfV_2)\ =\ g(I\bfV_1,\bfV_2)~,
\end{equation}
satisfying $\Omega(\bfV_1,\bfV_2)=-(-1)^{\tilde{\bfV}_1\tilde{\bfV}_2}\Omega(\bfV_2,\bfV_1)$, from which we obtain the super K{\"a}hler structure $J$ as
\begin{equation}
J(\bfV_1,\bfV_2)\ =\ \tfrac{1}{2}(g(\bfV_1,\bfV_2)+\di\Omega(\bfV_1,\bfV_2))\ =\ \str (\CP \bfV_1(\unit-\CP)\bfV_2)~.
\end{equation}

Note that for any normalized supervector $\bfa$, there is a transformation $\bfg\in\sU(4|4)$ mapping it to the vector $\bfa'=(1,0,\ldots )^T$. This implies that by an appropriate action of $\bfg$, one can turn $\CP$ into $\diag(1,0,\ldots )$ and therefore we have again the formula $\str(\CP \bfA\CP \bfB)=\str(\CP \bfA)\str(\CP \bfB)$~, which allows us to write
\begin{equation}
J(\bfV_1,\bfV_2)\ =\ \str (\CP \bfV_1\bfV_2)-\str(\CP \bfV_1)\str(\CP \bfV_2)~.
\end{equation}
Furthermore, we introduce the obvious components $J^{AB}=J(\bflambda^A,\bflambda^B)$ etc., for which we have e.g.\ the identity
\begin{equation}
J^{AB}\bflambda_B\ =\ \str(\CP\bflambda^A(\unit-\CP)\bflambda^B)\bflambda_B\ =\ \CP\bflambda^A(\unit-\CP)~.
\end{equation}

Note that a projection onto the body of all the structures introduced in this section naturally reduces them to their ordinary counterparts on bosonic flag manifolds. These geometric structures are thus (unique) supersymmetric extensions. Furthermore, the above discussion naturally extends to the case of all other flag supermanifolds involving super-antisymmetrized tensor products of the type $\bflambda^A\Cap\bflambda^B\Cap\ldots $.

\section{Fuzzy flag supermanifolds}

Having a description of flag supermanifolds using the Pl{\"u}cker embedding, we obtain quite straightforwardly the description of fuzzy flag supermanifolds. We will be rather concise and essentially stress the differences with the bosonic case. For earlier accounts of quantizing $\CPP^{1|2}$, see e.g.\ \cite{Ivanov:2003qq}.

\subsection{Supercoherent states}

The discussion of supercoherent states is done in close analogy to the case of bosonic coherent states. Consider the generators of the supergroup $\sU(4|4)$. After taking out the matrix $\diag(\unit_4,-\unit_4)$, we are left with 31 bosonic and 32 fermionic generators. There are seven generators of the Cartan subalgebra: the six Cartan generators of the two $\sSU(4)$s contained in $\sU(4|4)$ together with $\unit_8$. We pair the remaining 24 bosonic and 30 fermionic generators into raising and lowering operators. Seven of them are fundamental, the remaining ones are generated by supercommutators of these. All of the fundamental and Cartan generators are of the form
\begin{equation*}
H_i\ =\ \left(\begin{array}{cccc}
\ddots  & 0 & 0 & 0\\
0 & 1 & 0 & 0\\
0 & 0 & -1 & 0\\
0 & 0 & 0 & \ddots
\end{array}\right)~,~
E^+_i\ =\ \left(\begin{array}{cccc}
\ddots  & 0 & 0 & 0\\
0 & 0 & 1 & 0\\
0 & 0 & 0 & 0\\
0 & 0 & 0 & \ddots
\end{array}\right)~,~
E^-_i\ =\ \left(\begin{array}{cccc}
\ddots  & 0 & 0 & 0\\
0 & 0 & 0 & 0\\
0 & 1 & 0 & 0\\
0 & 0 & 0 & \ddots
\end{array}\right)~,~~~
\end{equation*} 
except for the following three:
\begin{equation*}
H_4\ =\ \left(\begin{array}{cc|cc}
\ddots  & 0 & 0 & 0\\
0 & 1 & 0 & 0\\
\hline
0 & 0 & 1 & 0\\
0 & 0 & 0 & \ddots
\end{array}\right)~,~~
E^+_4\ =\ \left(\begin{array}{cc|cc}
\ddots  & 0 & 0 & 0\\
0 & 0 & 1 & 0\\
\hline
0 & 0 & 0 & 0\\
0 & 0 & 0 & \ddots
\end{array}\right)~,~~
E^-_4\ =\ \left(\begin{array}{cc|cc}
\ddots  & 0 & 0 & 0\\
0 & 0 & 0 & 0\\
\hline
0 & 1 & 0 & 0\\
0 & 0 & 0 & \ddots
\end{array}\right)~.
\end{equation*}
The lines mark the boundaries between the four blocks of the supermatrices.

Correspondingly, we can introduce a super Dynkin diagram of the form 
\begin{equation}
\begin{picture}(320,20)
\put(20.0,0.0){\circle{10}}
\put(25.0,0.0){\line(1,0){40}}
\put(70.0,0.0){\circle{10}}
\put(75.0,0.0){\line(1,0){40}}
\put(120.0,0.0){\circle{10}}
\put(125.0,0.0){\line(1,0){40}}
\put(170.0,0.0){\circle{10}}
\put(175.0,0.0){\line(1,0){40}}
\put(220.0,0.0){\circle{10}}
\put(225.0,0.0){\line(1,0){40}}
\put(270.0,0.0){\circle{10}}
\put(275.0,0.0){\line(1,0){40}}
\put(320.0,0.0){\circle{10}}
\put(20.0,12.0){\makebox(0,0)[c]{$a_1$}}
\put(70.0,12.0){\makebox(0,0)[c]{$a_2$}}
\put(120.0,12.0){\makebox(0,0)[c]{$a_3$}}
\put(170.0,12.0){\makebox(0,0)[c]{$q_0$}}
\put(170.0,0.0){\makebox(0,0)[c]{$\times$}}
\put(220.0,12.0){\makebox(0,0)[c]{$a_4$}}
\put(270.0,12.0){\makebox(0,0)[c]{$a_5$}}
\put(320.0,12.0){\makebox(0,0)[c]{$a_6$}}
\end{picture}
\end{equation}
where the Dynkin labels indicate again the number of nontrivial actions of $E_i^-$ on a highest weight state. Setting these labels to zero, we evidently enlarge the isotropy group of the highest weight states in the same way as in the bosonic case. That is, the highest weight state in the representation $(L,0,0,0,0,0,0)$ has isotropy group $\sSU(3|4)$, while the highest weight state in the representation $(0,L,0,0,0,0,0)$ has isotropy group $\sSU(2)\times\sSU(2|4)$. These are thus the representations, in which the coherent states correspond to points on the flag supermanifolds $F_{1|0;4|4}$ and $F_{2|0;4|4}$.

The remaining representations for the flag manifolds $F_{3|4;4|4}$ and $F_{2|4;4|4}$ are given by representations with the Dynkin labels $(0,0,0,0,0,0,L)$ and $(0,0,0,0,0,L,0)$. The representations corresponding to flag manifolds are derived by choosing the Dynkin labels corresponding to all the contained Gra{\ss}mannians to be non-vanishing.

The further discussion of the construction of coherent states as well as the treatment of the various patches (choosing dominant weight states instead of highest weight states) is evident. Instead of going into details, we continue directly with the construction of fuzzy matrix algebras on the flag supermanifolds in the next section.

\subsection{The fuzzy complex projective superspace $\CPP^{3|4}_F$}

The fuzzification of $\CPP^{3|4}$ is obtained by promoting its homogeneous coordinates to creation and annihilation operators of bosonic and fermionic harmonic oscillators:
\begin{equation}
\bfa^I\ =\ (a^i,\eta^\alpha)~~~\ \rightarrow\ ~~~ \hat{\bfa}^I\ =\ (\hat{a}^i,\hat{\eta}^\alpha)\ewith\lsc\bfa^I,\bfa^{J\dagger}\rsc\ =\ \delta^{IJ}~.
\end{equation}
The total number operator reads then as $\hat{\bfN}=\hat{\bfa}^{I\dagger}\hat{\bfa}^I=\hat{a}^{i\dagger}\hat{a}^i+\hat{\eta}^{\alpha\dagger}\hat{\eta}^\alpha$, and commutes with the auxiliary coordinate operator
\begin{equation}
\hat{\bfx}^A\ :=\ \frac{1}{\hat{\bfN}}\hat{\bfa}^{I\dagger}\bflambda^A_{IJ}\hat{\bfa}^J~.
\end{equation}
We can thus restrict the algebra of functions to the $L$-particle Hilbert space $\CCH_{1|0;4|4}^F$ spanned by the states
\begin{equation}
 \hat{\bfa}^{I_1\dagger}\ldots \hat{\bfa}^{I_L\dagger}|0\rangle~~~\mbox{or}~~~\hat{a}^{i_1\dagger}\ldots \hat{a}^{i_{L-k}}\hat{\eta}^{\alpha_1\dagger}\ldots \hat{\eta}^{\alpha_k\dagger}|0\rangle~,
\end{equation}
where, evidently, $k\leq4$. Using the Schwinger construction for Lie superalgebras, we can define an action of $\au(4|4)$ on this space:
\begin{equation}
\hat{\bfL}^A\ =\ \hat{\bfa}^{I\dagger}\bflambda^A_{IJ}\hat{\bfa}^J~.
\end{equation}
In the standard notation\footnote{see e.g.\ \cite{Bars:1982se,Bars:1982ps}} for Young supertableaux, the representations space $\CCH_{1|0;4|4}^F$ corresponds to
\begin{equation}
\overbrace{\young(\contra\contra\contra\contra)}^{L}~. 
\end{equation}
The algebra of functions on $\CPP^{3|4}$ is given by the matrix algebra
\begin{equation}
\hat{\bfa}^{I_1\dagger}\ldots \hat{\bfa}^{I_L\dagger}|0\rangle\langle 0|\hat{\bfa}^{J_1}\ldots \hat{\bfa}^{J_L}~,
\end{equation}
which corresponds to the tensor product
\begin{equation}
\overbrace{\young(\contra\contra\contra\contra)}^{L}~\otimes~\overbrace{\young(\co\co\co\co)}^{L}~,
\end{equation}
where $\young(\co)$ stands for the dual (contragredient) representation of $\young(\contra)$.

For simplicity, let us give the star product on this space only for the complex coordinates $\bfa^I=(a^i,\eta^\alpha)$. The algebra of functions is spanned by the monomials
\begin{equation}
 \bar{\bfa}^{I_1}\ldots \bar{\bfa}^{I_L}\bfa^{J_1}\ldots \bfa^{J_L}~,
\end{equation}
and the star product is defined as
\begin{equation}
 (\bff\star\bfg)\ =\ \mu\left[\frac{1}{L!}\der{\bfa^{I_1}}\ldots \der{\bfa^{I_L}}\otimes\frac{1}{L!}\der{\bar{\bfa}^{I_1}}\ldots \der{\bar{\bfa}^{I_L}}(\bff\otimes\bfg)\right]~.
\end{equation}

Although the second order Casimir still labels representations to some extend, the Laplacian in the continuum does not have any immediate meaning. We are thus more interested in translating all the various derivatives, written in terms of the embedding coordinates to the fuzzy picture. This is easily done using the generators $\CL_A$ described in appendix B. One can show in complete analogy to the case of $\CPP^3$ that
\begin{equation}
\CL_A f_L(\bfx)\ =\ \frac{L}{\sqrt{2}}\left(\bfx_A\star f_L(\bfx)-(-1)^{\tilde{A}\widetilde{f_L}}f_L(\bfx)\star \bfx_A\right)\ =\ \tr\left(\rho^L(\bfx_{1|0;4|4})\lsc\hat{\CL}_A,\hat{f}\rsc\right)~.
\end{equation}
Together with the Killing metric $g_{AB}$, this sufficiently describes the geometry on fuzzy $\CPP^{3|4}$ embedded in $\FR^{32|32}$.

\subsection{The remaining fuzzy Gra{\ss}mannian supermanifolds}

The next flag manifold in our list is $G_{2|0;4|4}$, which is of complex dimension $4|8$. From the super Pl{\"u}cker embedding of this space into $\Lambda^{2|0}\FC^{4|4}$, the fuzzification is immediately obvious. We start from two sets of supersymmetric oscillators
\begin{equation}
\lsc\hat{\bfa}^I,\hat{\bfa}^{J\dagger}\rsc\ =\ \delta^{IJ}~,~~~
\lsc\hat{\bfb}^I,\hat{\bfb}^{J\dagger}\rsc\ =\ \delta^{IJ}
\end{equation}
with components $\hat{\bfa}^I=(\hat{a}^i,\hat{\eta}^\alpha)$ and $\hat{\bfb}^I=(\hat{b}^i,\hat{\theta}^\alpha)$. From these, we construct the composite annihilation and creation operators
\begin{equation}
 \hat{\bfA}^{IJ}_{2|0}\ :=\ \hat{\bfa}^{\lsc I}\hat{\bfb}^{J\rsc}~,~~~\hat{\bfA}^{IJ\dagger}_{2|0}\ :=\ \hat{\bfa}^{\lsc I\dagger}\hat{\bfb}^{J\rsc\dagger}~,
\end{equation}
or, in more detail, 
\begin{equation}
\hat{\bfA}^{ij}_{2|0}\ =\ \hat{a}^{[i}\hat{b}^{j]}~,~~~
\hat{\bfA}^{i\alpha}_{2|0}\ =\ \hat{a}^i\hat{\theta}^\alpha-\hat{\eta}^\alpha\hat{b}^i~,~~~
\hat{\bfA}^{\alpha i}_{2|0}\ =\ -\hat{\bfA}^{i\alpha}_{2|0}~,~~~
\hat{\bfA}^{\alpha\beta}_{2|0}\ :=\ \hat{\eta}^{\{\alpha}\hat{\theta}^{\beta\}}~,
\end{equation}
together with their hermitian conjugates. The construction of the usual $L$-particle Hilbert space is now again straightforward and this space forms a representation of $\sU(4|4)$. Combining this Hilbert space with a dual copy, we obtain the algebra of functions on $G_{2|0;4|4}^F$ as a matrix algebra of the form
\begin{equation}
\overbrace{\young(\contra\contra\contra\contra,\contra\contra\contra\contra)}^{L}~\otimes~\overbrace{\young(\co\co\co\co,\co\co\co\co)}^{L}
\end{equation}

From this construction, the deformed algebra of functions on $G_{2|0;4|4}$ together with the star product are also obvious.

The construction for the remaining two Gra{\ss}mannians $G_{2|4;4|4}$ and $G_{3|4;4|4}=\CPP^{3|4}_*$ uses composite creation and annihilation operators obtained from super-antisymmetrizing 2 even and 4 odd, and 3 even and 4 odd sets of superoscillators, respectively:
\begin{equation}
\hat{\bfA}^{IJ\Upsilon_1\ldots \Upsilon_4}_{2|4}\ =\ \hat{\bfa}^{\lsc I}\hat{\bfb}^J\hat{\bfeta}^{\Upsilon_1}_1\ldots \hat{\bfeta}^{\Upsilon_4\rsc}_4\eand\hat{\bfA}^{IJK\Upsilon_1\ldots \Upsilon_4}_{3|4}\ =\ \hat{\bfa}^{\lsc I}\hat{\bfb}^J\hat{\bfc}^K\hat{\bfeta}^{\Upsilon_1}_1\ldots \hat{\bfeta}^{\Upsilon_4\rsc}_4~,
\end{equation}
where the $\hat{\bfeta}^\Upsilon=(\hat{\eta}^i,\hat{a}^\alpha)$ are sets of odd annihilation operators satisfying with their hermitian conjugate the algebra
\begin{equation}
 \lsc \hat{\bfeta}^{\Upsilon_1},\hat{\bfeta}^{\Upsilon_2\dagger}\rsc\ =\ \delta^{\Upsilon_1\Upsilon_2}~.
\end{equation}
From the discussion above, the representations are clear.

\subsection{The fuzzy reducible flag supermanifolds}

The discussion of the fuzzy reducible flag supermanifolds is now completely obvious. By combining sets of oscillators from the various super Gra{\ss}mannians, we obtain the appropriate sets of oscillators (and thus the relevant Fock spaces from which the algebras of functions are constructed) for the flag supermanifolds naturally projecting on these Gra{\ss}mannians. Instead of repeating the discussion for all the flag supermanifolds, let us merely study the example of $F_{(1|0)(2|0);4|4}$.

For this flag supermanifold, we need the oscillators of $G_{1|0;4|4}=\CPP^{3|4}$ together with the ones of $G_{2|0;4|4}$:
\begin{equation}
 \hat{\bfa}^I~,~~~\hat{\bfa}^{I\dagger}~,~~~\hat{\bfA}_{2|0}^{IJ}\ =\ \hat{\bfa}^{\lsc I}\hat{\bfb}^{J\rsc}~,~~~\hat{\bfA}_{2|0}^{IJ\dagger}\ =\ \hat{\bfa}^{\lsc I\dagger}\hat{\bfb}^{J\rsc\dagger}~.
\end{equation}
Using $L_1$ operators $\hat{\bfa}^{I\dagger}$ and $L_2$ operators $\hat{\bfA}_{2|0}^{IJ\dagger}$ we construct the $(L_1,L_2)$-particle Fock space and its dual. Tensoring them yields the algebra of functions on $F_{(1|0)(2|0);4|4}$ as the matrix algebra
\begin{equation}
\overbrace{\young(\contra\contra\contra\contra\contra\contra\contra\contra,\contra\contra\contra\contra)}^{L_2+L_1}~\otimes~\overbrace{\young(\co\co\co\co\co\co\co\co,\co\co\co\co)}^{L_2+L_1}~.
\end{equation}

\subsection{Fuzzy Calabi-Yau supermanifolds}

Calabi-Yau supermanifolds received much attention recently in twistor string theory \cite{Witten:2003nn}, where $\CPP^{3|4}$ was used as a target space for the topological B-model. The interest in this particular space is due to the fact that $\CPP^{3|4}$ is simultaneously a supertwistor space and a Calabi-Yau supermanifold. The latter spaces are defined as spaces whose canonical bundle is trivial and thus have a nowhere vanishing holomorphic volume form. It was furthermore conjectured \cite{Witten:2003nn,Aganagic:2000gs} that there is a mirror symmetry between $\CPP^{3|4}$ and the superambitwistor space $\CL^{5|6}:=F_{(1|0)(3|3);4|3}$. (Note that in our above constructions, we instead considered the space $F_{(1|0)(3|4);4|4}$.) The space $\CL^{5|6}$ is of real dimension $10|12$ and a coset space of $\sU(4|3)$, as defined in \eqref{definitionflagsuper}. The corresponding fuzzy space is obtained from merging the fuzzy versions of $\CPP^{3|3}=F_{1|0;4|3}$ and $\CPP^{3|3}_*=F_{3|3;4|3}$ in the same way fuzzy flag manifolds are obtained from their sub-Gra{\ss}mannians. Since the construction is again rather trivial, we stop here and postpone the analysis of fuzzy mirror symmetry to future work.

\acknowledgements

We would like to thank Charles Nash and Sachin Vaidya for helpful comments. In particular, we wish to express our gratitude towards Brian Dolan, Denjoe O'Connor and Peter Pre\v{s}najder for numerous useful discussions and for valuable remarks on a draft of this paper.

\appendices

\subsection{Supermathematics, conventions and definitions}

We denote even objects by Latin letters and odd objects by Greek ones. Boldface symbols will represent superobjects. Furthermore, a tilde over an index or an object will denote the naturally assigned parity, and $\cdot^\circ$ attached to an object refers to its body.

\subsubsection{Supernumbers}

A {\em supernumber} is an element of the Gra{\ss}mann algebra $\Lambda_N$, $N\in\NN\cup\{\infty\}$ which has generators $\xi^i$, $i=1,\ldots N$ satisfying $\xi^i\xi^j+\xi^j\xi^i=0$. The Gra{\ss}mann algebra decomposes into an even and an odd part, $\Lambda_{N,0}=\Lambda_{N,c}$, $\Lambda_{N,1}=\Lambda_{N,a}$, which are the subsets of supernumbers built from an even and an odd number of Gra{\ss}mann generators, respectively. The {\em body} of a supernumber is denoted by $z^\circ$ and consists of the purely complex part of $z$ containing no Gra{\ss}mann generator. For {\em complex conjugation of Gra{\ss}mann odd quantities}, there are essentially two conventions used throughout the literature. First, and most commonly, there is
\begin{equation}
(\theta_1\theta_2)^*\ =\ \bar{\theta}_2\bar{\theta}_1\ =\ -\bar{\theta}_1\bar{\theta}_2~,
\end{equation}
which is used e.g.\ in \cite{Bars:1982ps,DeWitt:1992cy}. Second, there is
\begin{equation}
(\theta_1\theta_2)^*\ =\ -\bar{\theta}_2\bar{\theta}_1\ =\ +\bar{\theta}_1\bar{\theta}_2~,
\end{equation}
which is used in \cite{Cartier:0202026,Deligne:1999qp}. The latter convention respects the sign rule that interchanging two Gra{\ss}mann-odd objects in a monomial should always be accompanied by an additional sign. There is a discussion of this issue in \cite{Deligne:1999qp}. Manin in his book \cite{Manin:1988ds} also discusses all of these conventions. In this paper, we use the second convention. 

\subsubsection{Supervectors}

A {\em supervector space} is a free module over a supercommutative ring. We restrict our considerations to supervector spaces which are endowed with a so-called pure basis. In particular, consider a supervector space $V$ with a so-called class I (even) basis $(\bfe^A)$ of $n$ even and $\nu$ odd elements: $(\bfe^A)=(e^1,\ldots ,e^n,\eps^1,\ldots ,\eps^\nu)$. The supervector space $V$ is then said to be of dimension $n|\nu$.

A $n|\nu$-dimensional {\em supervector} consists of $n+\nu$ components. If the first $n$ of the components of a supervector are even and the remaining $\nu$ odd, the supervector is called even. If the inverse statement is true, the supervector is said to be odd:
\begin{equation}
\bfx \ =\ x^a e^a+\xi^\alpha\eps^\alpha~~ \Rightarrow~~ \tilde{\bfx}\ =\ 0\eand
\bfx \ =\ \xi^a e^a+x^\alpha\eps^\alpha~~ \Rightarrow~~ \tilde{\bfx}\ =\ 1~,
\end{equation}
where $x^i$ and $\xi^i$ are complex even and odd supernumbers, respectively. If the supervector is neither odd nor even, it is of mixed parity. 

We will also allow for class II bases, in which the parity of the even and odd basis elements is interchanged: $(\bfe^A)=(\eps^1,\ldots ,\eps^n,e^1,\ldots ,e^\nu)$. Here, the dependence of the parity of a supervector on the parity of its components is evidently inverted.

Two supervectors $(\bfe_1^A), (\bfe^A_2)$ are called {\em linearly independent}, if and only if
\begin{equation}
 \alpha \bfe^A_1+\beta \bfe^A_2\ =\ 0~~~\Rightarrow~~~\alpha\ =\ \beta\ =\ 0~,
\end{equation}
where $\alpha,\beta\in\Lambda_N$. This is equivalent to their bodies $\bfe_1^\circ, \bfe_2^\circ$ being linearly independent. A {\em scalar product} between two complex supervectors is supposed to be graded antilinear, i.e.\
\begin{equation}
(\bfa,\bfb)\ =\ (-1)^{\tilde{\bfa}\tilde{\bfb}}(\bfb,\bfa)^*~.
\end{equation}
For even supervectors with components $\bfa=(a^i,\eta^\alpha)$ and $\bfb=(b^i,\zeta^\alpha)$, where $a^i,b^i\in \Lambda_{N,c}$ and $\eta^\alpha,\zeta^\alpha \in \Lambda_{N,a}$, we can thus define
\begin{equation}
(\bfa,\bfb)\ :=\ \bfa^\dagger\bfb\ :=\ \bar{a}^ib^i+\di \bar{\eta}^\alpha\zeta^\alpha~.
\end{equation}
Two supervectors are perpendicular, if they have nonvanishing bodies and the scalar product between them vanishes. It follows that they are linearly independent.

A general supermatrix acting on the elements of an $n|\nu$-dimensional supervector space is of the block form
\begin{equation}\label{defGenSupermatrix}
\bfM\ =\ \left(\begin{array}{c|c} A & B \\\hline C & D
\end{array}\right)~,
\end{equation}
where $A$ is of dimension $n\times n$ and $D$ of dimension $\nu\times \nu$. The supermatrix $\bfM$ is called even, if $A$ and $D$ have only even components and $B$ and $C$ consist only of odd components; it thus preserves the parity of any supervector it acts on. Furthermore, if it inverts the parity of the supervector, it is called odd.

Note that the space $\FC^{n|\nu}$ is defined in two different ways throughout the literature. Most commonly, it denotes a space described by a set of coordinates consisting of $n$ even and complex numbers and $\nu$ complex Gra{\ss}mann variables. On the other hand, it is a $n|\nu$-dimensional supervector space over a complex supercommutative ring, as e.g.\ the ring of complex supernumbers. For the description of flag supermanifolds, we need the latter definition.

\subsubsection{The supergroup $\sU(4|4)$}

In our conventions for supergroups, we follow essentially
\cite{Bars:1982ps}, see also \cite{Kac:1977hp}. The Lie superalgebra $\au(n|\nu)$ is given by block supermatrices of the form \eqref{defGenSupermatrix}, where $A$ and $D$ are elements of $\au(n)$ and $\au(\nu)$, respectively, while $B$ and $C$ are hermitian conjugates of each other. This algebra is generated by $n^2+\nu^2$ generators for the components $A$ and $C$, which are even supermatrices, as well as $2n\nu$ generators for the components $B$ and $C$, which are odd supermatrices. Exponentiating these generators with even and odd parameters, respectively, yields the supergroup $\sU(n|\nu)$. To obtain the superanalogue of $\asu(n|\nu)$, one linearly combines the identities $\lambda^0_n$ and $\lambda^0_\nu$ of $\au(n)$ and $\au(\nu)$ into
\begin{equation}
\left(\begin{array}{cc}\frac{1}{\sqrt{n}} \unit_n & 0 \\ 0 &\frac{1}{\sqrt{\nu}} \unit_\nu \end{array}\right)\eand
\left(\begin{array}{cc}\frac{1}{\sqrt{n}} \unit_n & 0 \\ 0 &-\frac{1}{\sqrt{\nu}} \unit_\nu \end{array}\right)~.
\end{equation}
Imposing the condition $\str(\cdot)=0$ on the generators, eliminates the second generator. For $n\neq \nu$, this yields a semisimple super Lie algebra $\asu(n|\nu)$. For $n=\nu$, however, the first factor becomes $\frac{1}{n}\unit_{2n}$ and generates an invariant Abelian subgroup. For this reason, one excludes this generator and arrives at $\mathfrak{p}\asu(n|n)$. However, the lowest dimensional representation is the adjoint, see e.g.\ \cite{Bars:1982ps,DeWitt:1992cy} for more details. To avoid these complications, we choose to work with $\au(4|4)$.

The Killing form for $\au(n|n)$ is easily evaluated to be 
\begin{equation}
g_{\hat{A}\hat{B}}\ :=\ \str(\bflambda_{\hat{A}}\bflambda_{\hat{B}})\ =\ \left(\begin{array}{cc|ccc}
\unit_n & 0 & & 0 &\\
0 & \unit_\nu & & &\\
\hline 
& & -\sigma_2 & 0 & 0\\
0&   & 0 & -\sigma_2 & 0\\
& &  0 & 0 & \ddots
\end{array}\right)_{\hat{A}\hat{B}}~,
\end{equation}
and we also define $g^{\hat{A}\hat{B}}$ with $g_{\hat{A}\hat{B}}g^{\hat{B}\hat{C}}=\delta_{\hat{A}}^{\hat{C}}$. The Killing form is furthermore supersymmetric, i.e.\ $g_{\hat{A}\hat{B}}=(-1)^{\tilde{\hat{A}}\tilde{\hat{B}}}g_{\hat{B}\hat{A}}$ and non-degenerate.

\subsubsection{Supermanifolds}

We define a (complex) supermanifold as a topological space $X$ together with a sheaf $\CO_N$ of $\RZ_2$-graded supercommutative rings on $X$, satisfying the following two conditions
\begin{itemize}
\item[(i)] There is a projection on the ``body'' of $X$, which is an ordinary complex manifold of dimension $m$. More explicitly, consider the reduced structure sheaf $\CO^\circ:=\CO_N/\CI$, where $\CI$ is the ideal of nilpotent elements in $\CO_N$. We demand that $(X,\CO^\circ)$ is a complex manifold of dimension $m$. 
\item[(ii)] Locally, the structure sheaf is the structure sheaf of the body with values in a Gra{\ss}mann algebra. That is, for every point $x$ in $X$, there is an open neighborhood $U$ such that
\begin{equation}
\CO_N|_U\ \cong\  \CO_{\mathrm{red}}|_U(\Lambda^*\FC^n)~.
\end{equation}
\end{itemize}
We will define the dimension of such a supermanifold to be $m|n$. For more details on supermanifolds, see \cite{Manin:1988ds,Cartier:0202026,DeWitt:1992cy}.

\subsubsection{Riemannian supergeometry}

A {\em supermetric} on a (real) supermanifold $\bfM:=(M,\CO_N)$ is an $\CO_N$-linear, even map $g:T\bfM\otimes_{\CO_N}T\bfM$ satisfying the following properties:
\begin{itemize}
\item[(i)] $g$ is supersymmetric: $g(X\otimes Y)=(-1)^{\tilde{X}\tilde{Y}}g(Y\otimes X)$.
\item[(ii)] $g$ induces a Riemannian metric on $(M,\CO^\circ)$.
\item[(iii)] $g$ induces a symplectic form on the fermionic tangent directions of $T\bfM$.
\end{itemize}

An {\em almost complex structure} on a (real) supermanifold $\bfM:=(M,\CO_N)$ is an even, smooth map $I:T\bfM\rightarrow T\bfM$, which satisfies $I^2=-\unit$. As in ordinary complex geometry, a real supermanifold underlying a complex supermanifold has a natural almost complex structure. 

A {\em hermitian supermetric} on a supermanifold $\bfM:=(M,\CO_N)$ with almost complex structure $I$ is a supermetric $g$ which satisfies $g(IX\otimes IY)=g(X\otimes Y)$ for all vector fields $X,Y$ in $T\bfM$. A {\em K{\"a}hler supermetric} is a supermetric, the derived {\em K{\"a}hler form} $J(X,Y)=g(X,IY)$ of which is closed: $\dd J=0$. As an example of a K{\"a}hler supermanifold see the discussion of the space $\CPP^{3|4}$ in section 5.5. For more details on supergeometry, see e.g.\ \cite{Varsaie:6701}.

\subsection{Representations of $\asu(4)$ and $\au(4|4)$}

In this appendix, we briefly recall a few facts on the representation theory of $\asu(4)$ and give the necessary background on the superalgebra $\au(4|4)$ and its subalgebra $\asu(4)$.

\subsubsection{Representation of $\asu(4)$ in terms of Pl{\"u}cker and embedding coordinates}

Consider the case $\CPP^3\cong \sSU(4)/\sU(3)\subset \FR^{16}$. A representation of $\sSU(4)$ acting on functions written in terms of complex Pl{\"u}cker coordinates $a^i\in\FC^4$ is given by
\begin{equation}
\CL^{\hat{a}}\ =\ \bar{a}^i\lambda_{ij}^{\hat{a}}\der{\bar{a}^j}-a^j\lambda_{ij}^{\hat{a}}\der{a^i}~,
\end{equation}
where $\lambda_{ij}^{\hat{a}}$ are again the Gell-Mann matrices of $\sSU(4)$ together with the identity. One easily verifies that $[\CL^a,\CL^b]=\di\sqrt{2}f^{ab}{}_c\CL^c$, where $f^{ab}{}_c$ are the structure constants of $\sSU(4)$. In terms of the real coordinates
\begin{equation}
x^{\hat{a}}\ =\ \bar{a}^i\lambda_{ij}^{\hat{a}} a^j
\end{equation}
describing the canonical embedding of $\CPP^3$ in $\FR^{16}$, the above generators read as
\begin{equation}
\CL^a\ =\ -\di \sqrt{2}f^{abc} x_b\der{x^c}~,~~~\CL^0\sim x^a\der{x^a}~.
\end{equation}
as one easily verifies. The representations in terms of the embedding coordinates for $G_{2;4}$ is given in the text, from which also the remaining cases follow.

\subsubsection{Dynkin and Young diagrams for $\asu(4)$}

The 15 generators of $\asu(4)$ split into three generators of the Cartan subalgebra $H_i$, $i=1,\ldots ,3$ and 12 raising and lowering operators $E_{\pm\vec{\alpha}_j}$, $j=1,\ldots 6$ satisfying the commutation relations
\begin{equation}
\begin{aligned}
{}[H_i,H_j]\ =\ 0~,~~~[H_i,&E_{\vec{\alpha}}]\ =\ \alpha_i 
E_{\vec{\alpha}}~,\\
[E_{\vec{\alpha}},E_{-\vec{\alpha}}]\ =\ \sum_i \alpha_i H_i~,~~~ &[E_{\alpha},E_{\beta}]\ =\  N_{\alpha \beta} E_{\alpha+\beta}~. 
\end{aligned}
\end{equation}
Here, $\vec{\alpha}_j$ are the six three-dimensional positive root vectors, $\Sigma_+$, three of which are simple. The irreducible representations of $\asu(4)$ can be labeled by the three eigenvalues $\mu_i$ of a highest weight state $|\mu\rangle$ under the action of the $H_i$. Equally well, one can label them by three integers $a_i$, the Dynkin labels, which are given by
\begin{equation}
a_i\ =\ 2\frac{(\vec{\mu},\vec{\alpha}_i)}{(\vec{\alpha}_i , \vec{\alpha}_i)}~,
\end{equation}
where $\vec{\alpha}_i$ are the three simple roots. The Dynkin diagram labeling irreducible representations of $\asu(4)$ is then
\begin{equation}
\begin{picture}(180,20)
\put(20.0,0.0){\circle{10}}
\put(25.0,0.0){\line(1,0){40}}
\put(70.0,0.0){\circle{10}}
\put(75.0,0.0){\line(1,0){40}}
\put(120.0,0.0){\circle{10}}
\put(20.0,12.0){\makebox(0,0)[c]{$a_1$}}
\put(70.0,12.0){\makebox(0,0)[c]{$a_2$}}
\put(120.0,12.0){\makebox(0,0)[c]{$a_{3}$}}
\end{picture}
\end{equation}
and these representation are of dimension
\begin{equation}
d=\frac{a_1+1}{1}\frac{a_2+1}{1}\frac{a_3+1}{1}\frac{a_1+a_2+2}{2}
\frac{a_2+a_3+2}{2}\frac{a_1+a_2+a_3+3}{3}~.
\end{equation}
The Dynkin labels indicate, how often one can act with a lowering operator on the highest weight state without obtaining a trivial state:
\begin{equation}
(E_{-\vec{\alpha}_i})^{a_i} |\mu\rangle\neq 0~,~~~(E_{-\vec{\alpha}_i})^{a_i+1}|\mu\rangle\ =\  0~.
\end{equation}

On the other hand, the Dynkin labels appear naturally in the Young diagrams of the representation $(a_1,a_2,a_3)$:
\begin{equation}
\overbrace{\tyng(16,8,4)}^{a_3+a_2+a_1}~,
\end{equation}
and $a_i$ counts the number of columns with $i$ boxes.

Due to the existence of the $\eps$-tensor, which is invariant under $\sSU(4)$, four antisymmetrized boxes combine to a singlet. Furthermore, this tensor provides a duality between three antisymmetrized indices and one index as well as two antisymmetric ones and their complement:
\begin{equation}
\check{\phi}_i\ =\ \eps_{ijkl}\phi^{jkl}\eand\check{\phi}_{ij}\ =\ \tfrac{1}{2}\eps_{ijkl}\phi^{kl}~.
\end{equation}

\subsubsection{Schwinger construction for Lie superalgebras}

Consider a Lie superalgebra having the even generators $\bflambda^a$ and the odd generators $\bflambda^\alpha$ satisfying the commutation relations
\begin{equation}
[\bflambda^a,\bflambda^b]\ =\ f^{ab}{}_c\lambda^c~,~~~
[\bflambda^a,\bflambda^\alpha]\ =\ f^{a\alpha}{}_\beta \bflambda^\beta~,~~~
\{\bflambda^\alpha,\bflambda^\beta\}\ =\ f^{\alpha\beta}{}_a \bflambda^a~.
\end{equation}
We summarize the generators into
$\bflambda^A=(\bflambda^a,\bflambda^\alpha)$ and the commutation
relations to
\begin{equation}
\lsc \bflambda^A,\bflambda^B\rsc\ =\ f^{AB}{}_C\bflambda^C~,
\end{equation}
where $\lsc\cdot\rsc$ denotes the supercommutator. Assume furthermore that the generators are in a representation acting on an $(m|n)$-dimensional supervector space. After introducing a set of $m$ bosonic and $n$ fermionic oscillators together with the corresponding annihilation and creation operators
\begin{equation}
\hat{\bfa}^I\ =\ (\hat{a}^i,\hat{\eta}^\alpha)\eand\hat{\bfa}^{I\dagger}\ =\ (\hat{a}^{i\dagger},\hat{\eta}^{\alpha\dagger})~,
\end{equation}
the Schwinger construction yields a representation of the Lie superalgebra by
\begin{equation}
\hat{\bfL}^A\ =\ \hat{\bfa}^{I\dagger}\bflambda^A_{IJ}\hat{\bfa}^J~.
\end{equation}

\subsubsection{Representation of $\au(4|4)$ in terms of Pl{\"u}cker and embedding coordinates}

Completely analogously to the representations of $\au(4)$ in terms of coordinates describing $\CPP^3$, one finds a representation of $\au(4|4)$ in terms of coordinates describing $\CPP^{3|4}$. We have 
\begin{equation}
\CL^{\hat{A}}\ =\ \bar{\bfa}^I\bflambda_{IJ}^{\hat{A}}\der{\bar{\bfa}^J}-\bfa^J\bflambda_{IJ}^{\hat{A}}\der{\bfa^I}~,
\end{equation}
where $\bfa^I=(a^i,\eta^\alpha)\in \FC^{4|4}$ and in terms of the real coordinates
\begin{equation}
\bfx^{\hat{A}}\ =\ \bar{\bfa}^I\bflambda_{IJ}^{\hat{A}} \bfa^J
\end{equation}
describing the embedding of $\CPP^{3|4}$ in $\FR^{32|32}$, the above generators read as
\begin{equation}
\CL^A\ =\ -\di \sqrt{2}f^{ABC} \bfx_B\der{\bfx^C}~,~~~\CL^0\sim \bfx^A\der{\bfx^A}~.
\end{equation}

\subsubsection{Representations and Dynkin diagrams for $\mathfrak{p}\asu(4|4)$}

The representations of supergroups are divided into two classes. In the first class, the representation space is spanned by an even basis, while in the second one, the basis is odd and they can be regarded as dual to each other. Representations are again labelled by highest weights, which are the eigenvalues of the generators of the Cartan subalgebra. The latter is generated by two copies of the Cartan subalgebra of $\asu(4)$ as well as $H_4:=\diag(0,0,0,1,1,0,0,0)$, cf.\ section 6.1. Accordingly, the Dynkin diagram has seven nodes:
\begin{equation}
\begin{picture}(320,20)
\put(20.0,0.0){\circle{10}}
\put(25.0,0.0){\line(1,0){40}}
\put(70.0,0.0){\circle{10}}
\put(75.0,0.0){\line(1,0){40}}
\put(120.0,0.0){\circle{10}}
\put(125.0,0.0){\line(1,0){40}}
\put(170.0,0.0){\circle{10}}
\put(175.0,0.0){\line(1,0){40}}
\put(220.0,0.0){\circle{10}}
\put(225.0,0.0){\line(1,0){40}}
\put(270.0,0.0){\circle{10}}
\put(275.0,0.0){\line(1,0){40}}
\put(320.0,0.0){\circle{10}}
\put(20.0,12.0){\makebox(0,0)[c]{$a_1$}}
\put(70.0,12.0){\makebox(0,0)[c]{$a_2$}}
\put(120.0,12.0){\makebox(0,0)[c]{$a_3$}}
\put(170.0,12.0){\makebox(0,0)[c]{$q_0$}}
\put(170.0,0.0){\makebox(0,0)[c]{$\times$}}
\put(220.0,12.0){\makebox(0,0)[c]{$a_4$}}
\put(270.0,12.0){\makebox(0,0)[c]{$a_5$}}
\put(320.0,12.0){\makebox(0,0)[c]{$a_6$}}
\end{picture}
\end{equation}
It is now evident that a class I representation is dual to a class II representation provided that 
\begin{equation}
a_i^I\ =\ a_{7-i}^{II}\eand q_0^I\ =\ q_0^{II}~.
\end{equation}

Since the superdeterminant is no longer a polynomial of finite degree, there is no invariant totally antisymmetric $\eps$-tensor for the $\sSU(m|n)$ supergroups. However, the duality between class I and II representation takes over the r\^ole of the duality between the representation and the corresponding conjugate representation. The latter arises from the interchange between covariant and contravariant indices corresponding to a contraction with an $\eps$-tensor in the $\sSU(4)$ picture, as we saw above.

\subsubsection{The second order Casimir operator on irreducible representations of $\asu(4)$}

We constructed the algebra of functions on the fuzzy flag manifolds from spherical representations of the groups underlying the flag manifolds. These representations are described by Young diagrams, and given such a Young diagram with $m\leq 4$ rows of $n_1\geq \ldots \geq n_m\geq 0$ boxes, the eigenvalue of the second order Casimir operator on these representations reads as\footnote{This formula is given, e.g., in \cite{Gross:1992tu} with a different normalization of the Lie algebra generators.}
\begin{equation}
\frac{1}{2}\left(4\sum_{i=1}^m n_i+\sum_{i=1}^m n_i(n_i+1-2i)-\frac{(\sum_{i=1}^mn_i)^2}{4}\right)~.
\end{equation}
We will be mostly interested in the representations consisting of a row of $2L$ boxes and $2$ rows of $L$ boxes. For these diagrams, the above formula reduces to $L(L+3)$. The other type of diagrams we will encounter consist of three rows with $a+b+a$, $a+b$ and $a$ boxes, respectively. For them, the above formula reduces to $a^2+a(3+b)+\frac{1}{2}b(4+b)$.

\subsubsection{The second order Casimir operator on representations of $\au(n|n)$}

We can write the second order Casimir operator acting on the Hilbert space $\CCH_{1|0;4|4}^F$ of $\CPP^{3|4}_F$ in terms of one set of oscillators using the Schwinger construction:
\begin{equation*}
\begin{aligned} 
C_2 &\ =\ g_{AB} \hat{\bfL}^A \hat{\bfL}^B\\
&\ =\ \frac{n-1}{n}\hat{\bfN}_b(\hat{\bfN}_b+n)+\frac{1}{n}\hat{\bfN}^2_b -\frac{n+1}{n}\hat{\bfN}_f(n-\hat{\bfN}_f)-\frac{1}{n}\hat{\bfN}_f^2+(2\hat{\bfN}_b\hat{\bfN}_f+n \hat{\bfN}_f-n\hat{\bfN}_b)\\
&\ =\ \frac{1}{2}\hat{\bfN}(\hat{\bfN}-1)~,
\end{aligned}
\end{equation*}
where $\hat{\bfN}_b=\hat{a}^{i\dagger}\hat{a}^i$ and $\hat{\bfN}_f=\hat{\eta}^{\alpha\dagger}\hat{\eta}^\alpha$. For the other flag supermanifold we require more than one set of oscillators and the calculation is more complicated.

\subsubsection{Fierz and super Fierz identities}

Consider the Gell-Mann matrices $\lambda^a$, $a=1,\ldots ,15$ of $\asu(4)$ and extend them to the generators $\lambda^{\hat{a}}$ of $\au(4)$ by adding $\lambda^0=\unit/\sqrt{4}$. We have the Fierz identity
\begin{equation}
\lambda^{\hat{a}}_{ij}\lambda^{\hat{a}}_{kl}\ =\ \delta_{il}\delta_{jk}~,
\end{equation}
which trivially extends in the case of the antisymmetric tensor products $\lambda^{\hat{a}\hat{b}\ldots }:=\lambda^{\hat{a}}\wedge \lambda^{\hat{b}}\wedge\ldots $ we defined in section 2.3 to
\begin{equation}\label{identlambdalambda}
\begin{aligned}
(\lambda^{\hat{a}\hat{b}})_{ij;kl}(\lambda^{\hat{a}\hat{b}})_{mn;pq}&\ =\ 
(\delta_{ip}\delta_{km}\delta_{jq}\delta_{ln})_{[ij][kl][mn][pq]}~,\\
(\lambda^{\hat{a}\hat{b}\hat{c}})_{ijk;lmn}(\lambda^{\hat{a}\hat{b}\hat{c}})_{pqr;stu}&\ =\ 
(\delta_{is}\delta_{lp}\delta_{jt}\delta_{mq}\delta_{ku}\delta_{nr})_{[ijk][lmn][pqr][stu]}~.
\end{aligned}
\end{equation}
Moreover, we can write
\begin{subequations}\label{identlambdatrace}
\begin{equation}
\tr(A_1\lambda^{\hat{a}})\lambda^{\hat{a}}\ =\ A_1\eand 
\tr(A_1\lambda^a)\lambda^a\ =\ A_1-\tfrac{1}{4}\tr(A_1)\unit
\end{equation}
for any hermitian matrix $A_1\in\mathrm{Mat}_4$ as well as
\begin{equation}
\begin{aligned}
\tr(A_2\lambda^{\hat{a}\hat{b}})\lambda^{\hat{a}\hat{b}}&\ =\ A_2\\
\tr(A_3\lambda^{\hat{a}\hat{b}\hat{c}})\lambda^{\hat{a}\hat{b}\hat{c}}&\ =\ A_3\\
\end{aligned}
\end{equation}
\end{subequations}
for hermitian matrices $A_2\in\mathrm{Mat}_4\wedge\mathrm{Mat}_4$ and $A_3\in\mathrm{Mat}_4\wedge\mathrm{Mat}_4\wedge\mathrm{Mat}_4$.

We can use the Killing form on $\sU(4|4)$ to establish the following super Fierz identity:
\begin{equation}
g^{\hat{A}\hat{B}}\bflambda^{IJ}_{\hat{A}}\bflambda^{KL}_{\hat{B}}\ =\ \delta^{IL}\delta^{JK}~.
\end{equation}
To prove this formula quickly, one can e.g.\ extend the one for $\sSU(n|n)$ given in \cite{Bars:1982ps}. From this Fierz identity, we can, as in the purely bosonic case, immediately derive the further identities
\begin{equation*}
\begin{aligned}
(\bflambda^{\hat{A}\hat{B}})_{IJ;KL}(\bflambda^{\hat{A}\hat{B}})_{MN;PQ}&\ =\ 
(\delta_{IP}\delta_{KM}\delta_{JQ}\delta_{LN})_{\lsc IJ\rsc\lsc KL\rsc\lsc MN\rsc \lsc PQ\rsc}~,\\
(\bflambda^{\hat{A}\hat{B}\hat{C}})_{IJK;LMN}(\bflambda^{\hat{A}\hat{B}\hat{C}})_{PQR;STU}&\ =\ 
(\delta_{IS}\delta_{LP}\delta_{JT}\delta_{MQ}\delta_{KU}\delta_{NR})_{\lsc IJK\rsc \lsc LMN\rsc \lsc PQR\rsc \lsc STU\rsc}~,
\end{aligned}
\end{equation*}
where $\bflambda^{\hat{A}\hat{B}}$ denotes again the graded antisymmetric tensor product
\begin{equation}
\bflambda^{\hat{A}\hat{B}}\ =\ \bflambda^{\hat{A}} \Cap \bflambda^{\hat{B}}
\end{equation}
introduced in section 5.3.

\subsubsection{Oscillator representation of internal isotropy subgroups}

In the description of the various Gra{\ss}mannians, we needed the following formula to describe the action of the internal part of the isotropy subgroup acting on the states of certain Fock spaces: 
\begin{equation}
[(a_p^i)^\dagger\lambda^A_{pq}a_q^i,a^{[j_1\dagger}_1\ldots a^{j_k]\dagger}_k]\ =\ 0~.
\end{equation}
To prove this formula, consider the following form of the generators $L_{pq}$ of $U(k)$:
\begin{equation}
L_{pq}\ =\ a^{i \dagger}_p a^i_q ~,
\end{equation}
where $i$ is summed over. The composite creation operators relevant in the description of $G_{k;n}$ can be written as
\begin{equation}
A^{i_1 \cdots i_k\dagger}_k\ =\ \frac{1}{k!}\epsilon_{p_1 \cdots p_k} a^{[i_1 \dagger}_{p_1}\cdots a^{i_k] \dagger}_{p_k}~.
\end{equation}
Then
\begin{align*}
[L_{pq},\,A^{i_1\cdots i_k\dagger}_k]&\ =\ \frac{1}{k!}\epsilon_{p_1 \cdots p_k} a^{i \dagger}_p[a^i_q ,\,a^{[i_1 \dagger}_{p_1}\cdots a^{i_k ]\dagger}_{p_k}]\\
&=\frac{1}{k!}\sum_{j=1}^k \epsilon_{p_1 \cdots p_k} a^{i \dagger}_p a^{[i_1 \dagger}_{p_1}\cdots \delta^{i i_j}\delta_{q p_j} \cdots a^{i_k] \dagger}_{p_k}\\
&=\frac{k}{k!}\epsilon_{q p_2 \cdots p_k} a^{[i_1 \dagger}_{p}a^{i_2 \dagger}_{p_2}\cdots a^{i_k] \dagger}_{p_k}\\
&=\delta_{pq}A^{i_1 \cdots i_k\dagger}_k~,
\end{align*}
and thus the (traceless) generators of $\sSU(k)$ leave the composite creation operators invariant.

In the supersymmetric case, the same statement holds. Here, the creation and annihilation operators satisfy 
\begin{equation}
\lsc\bfa_p^I,\bfa_q^{J\dagger}\rsc\ =\ \delta_{pq}\delta^{IJ}~.
\end{equation}
Using the relations 
\begin{equation}
\begin{aligned}
\lsc\bfa,\bfb\bfc\rsc&\ =\ \lsc\bfa,\bfb\rsc \bfc+(-1)^{\tilde{\bfa}\tilde{\bfb}}\bfb\lsc\bfa,\bfc\rsc~,\\
\lsc \bfa\bfb,\bfc\rsc&\ =\ \bfa\lsc\bfb\bfc\rsc+(-1)^{\tilde{\bfb}\tilde{\bfc}}\lsc\bfa,\bfc\rsc\bfb~,
\end{aligned}
\end{equation}
we obtain
\begin{equation}
\begin{aligned}
\lsc\bfa^I_p,\,\bfA^{JK\dagger}_{2|0}\rsc&\ =\ \lsc\bfa^I_p,\bfa^{J\dagger}_1\rsc\bfa_2^{K\dagger}+(-1)^{\tilde{I}\tilde{J}}\bfa^{J\dagger}_1\lsc\bfa^I_p,\bfa^{K\dagger}_2\rsc-(-1)^{\tilde{J}\tilde{K}}(J\leftrightarrow K)\\
&\ =\ \delta_{p1}\delta^{IJ}\bfa^{K\dagger}_2+\delta_{p2}(-1)^{\tilde{I}\tilde{J}}\delta^{IK}\bfa^{J\dagger}_1-(-1)^{\tilde{J}\tilde{K}}(J\leftrightarrow K)~.
\end{aligned}
\end{equation}
With this relation, we can conclude that
\begin{equation}
 \begin{aligned}
\lsc\bfa^{I\dagger }_p \bfa^I_q,\bfA^{JK\dagger}_{2|0}\rsc\ =\ &\bfa^{I\dagger }_p\lsc \bfa^I_q,\bfA^{JK\dagger}_{2|0}\rsc\\
\ =\ &\delta_{q1}(\bfa^{J\dagger}_p \bfa^{K\dagger}_2-(-1)^{\tilde{J}\tilde{K}}\bfa^{K\dagger}_p \bfa^{J\dagger}_2)+\delta_{q2}((-1)^{\tilde{K}\tilde{J}}\bfa^{K\dagger}_p \bfa^{J\dagger}_1-\bfa^{J\dagger}_p \bfa^{K\dagger}_1)\\
\ =\ &\delta_{p1}\delta_{q1}\bfA^{JK\dagger}_{2|0}+\delta_{p2}\delta_{q2}\bfA^{JK\dagger}_{2|0}\\
\ =\ &\delta_{pq}\bfA^{JK\dagger}_{2|0}~.
 \end{aligned}
\end{equation}
Similar relations can also be proven for the remaining cases of $\bfA_{2|4}$ and $\bfA_{3|4}$ by observing that, for example,
\begin{equation} \bfA^{JKL\dagger}_{3|0} \ =\  \bfA^{JK\dagger}_{2|0}\bfa^{L\dagger}_3 -(-1)^{\tilde{K}\tilde{L}}\bfA^{JL\dagger}_{2|0}\bfa^{K\dagger}_3+(-1)^{\tilde{J}\tilde{K}+{\tilde{J}\tilde{L}}}\bfA^{KL\dagger}_{2|0}\bfa^{J\dagger}_3 \end{equation}
and
\begin{align} 
\lsc\bfa^{I\dagger }_p \bfa^I_q,\bfA^{JK\dagger}_{2|0}\bfa^{L\dagger}_3\rsc\ =\ &\lsc\bfa^{I\dagger }_p \bfa^I_q,\bfA^{JK\dagger}_{2|0}\rsc \bfa^{L\dagger}_3+\bfA^{JK\dagger}_{2|0}\lsc\bfa^{I\dagger }_p \bfa^I_q,\bfa^{L\dagger}_3\rsc \\
\ =\ & (\delta_{pq}-\delta_{p3}\delta_{q3})\bfA^{JK\dagger}_{2|0}\bfa^{L\dagger}_3 + \delta_{q3}\bfA^{JK\dagger}_{2|0} \bfa^{L\dagger}_p~.
\end{align}
Hence,
\begin{equation}
\lsc\bfa^{I\dagger }_p \bfa^I_q,\bfA^{JKL\dagger}_{3|0}\rsc\ =\ \delta_{pq}\bfA^{JKL\dagger}_{3|0} ~. \end{equation}

\end{document}